\documentclass[10pt,a4paper]{article}
\pdfoutput=1
\usepackage{jheppub}
\usepackage{amsmath}
\usepackage{verbatim}
\usepackage{amssymb}
\usepackage{inputenc, array}
\usepackage{textcomp}
\usepackage{appendix}
\usepackage{amsfonts}
\usepackage{amsthm}
\usepackage{tikz}
\usetikzlibrary{decorations.markings}
\usepackage{mathtools}
\usepackage{bm}
\allowdisplaybreaks
\title{Carrollian Approach to $1+3$D Flat Holography}
\author{Amartya Saha}
\affiliation{Indian Institute of Technology Kanpur, Kanpur 208016, INDIA} 
\emailAdd{amartyas@iitk.ac.in}
\preprint{}
\abstract{The isomorphism between the (extended) BMS$_4$ algebra and the $1+2$D Carrollian conformal algebra hints towards a co-dimension one formalism of flat holography with the field theory residing on the null-boundary of the asymptotically flat space-time enjoying a $1+2$D Carrollian conformal symmetry. Motivated by this fact, we study the general symmetry properties of a source-less $1+2$D Carrollian CFT, adopting a purely field-theoretic approach. After deriving the position-space Ward identities, we show how the $1+3$D bulk super-translation and the super-rotation memory effects emerge from them, manifested by the presence of a temporal step-function factor in the same. Temporal-Fourier transforming these memory effect equations, we directly reach the bulk null-momentum-space leading and sub-leading soft graviton theorems. Along the way, we construct six Carrollian fields $S^\pm_0$, $S^\pm_1$, $T$ and $\bar{T}$ corresponding to these soft graviton fields and the Celestial stress-tensors, purely in terms of the Carrollian stress-tensor components. The 2D Celestial shadow-relations and the null-state conditions arise as two natural byproducts of these constructions. We then show that those six fields consist of the modes that implement the super-rotations and a subset of the super-translations on the quantum fields. The temporal step-function allows us to relate the operator product expansions (OPEs) with the operator commutation relations via a complex contour integral prescription. We deduce that not all of those six fields can be taken together to form consistent OPEs. So choosing $S^+_0$, $S^+_1$ and $T$ as the local fields, we form their mutual OPEs using only the OPE-commutativity property, under two general assumptions. The symmetry algebra manifest in these holomorphic-sector OPEs is then shown to be $\text{Vir}\ltimes\hat{\overline{\text{sl}(2,\mathbb{R})}}$ with an abelian ideal.}
\begin{document}
\maketitle
\flushbottom
\section{Introduction}
The goal of the ongoing research program named flat space holography is to understand how the holographic principle \cite{tHooft:1993dmi,Susskind:1994vu} can be extended beyond the framework of the celebrated AdS-CFT correspondence \cite{Maldacena:1997re,Witten:1998qj,Gubser:1998bc} to a more realistic space-time resembling the physical universe, by establishing a (holographic) duality between gravity in the asymptotically flat space-times (AFS) and a lower-dimensional quantum field theory without gravity. 

\medskip

In the AdS-CFT correspondence, the conformal field theory (CFT) lives on the co-dimension one boundary of the bulk AdS space-time. But in the most well-developed formalism till date, of $1+3$D flat holography, called the Celestial holography, the dual Celestial CFT \cite{Pasterski:2016qvg} resides on the co-dimension two celestial sphere $S^2$ at the null-infinity $\mathcal{I}$ of the AFS. The key point of the Celestial holography is that scattering amplitudes of processes taking place in the bulk $1+3$D AFS can be encoded as correlation functions of the Celestial CFT \cite{Pasterski:2016qvg,Pasterski:2017kqt}. Moreover, the Ward identities of the currents of the Celestial CFT are nothing other than the bulk AFS quantum field theory soft theorems of the $S$-matrices written in the boost-eigenstate basis \cite{Pasterski:2016qvg,Pasterski:2017kqt,Banerjee:2018fgd}. For detailed discussions on the properties of the 2D Celestial CFT, some of which are quite surprising from the point of view of the usual 2D relativistic CFT, please see the recent reviews \cite{Pasterski:2021rjz,Raclariu:2021zjz,Travaglini:2022uwo}.

\medskip

It was the fundamental observation that the bulk AFS soft theorems \cite{Weinberg:1965nx,Cachazo:2014fwa} can be interpreted as the Ward identities \cite{Strominger:2013jfa,He:2014laa,Kapec:2014opa,Kapec:2016jld,Cheung:2016iub,He:2017fsb,Donnay:2022hkf} of the (extended) BMS$_4$ symmetries which are the asymptotic symmetries at null infinity of the AFS \cite{Bondi:1962px,Sachs:1962wk,Sachs:1962zza,Barnich:2009se,Barnich:2010ojg,Barnich:2010eb} that led to this insightful approach to $1+3$D flat space holography. But all of the Celestial currents generating these Ward identities were built explicitly from the terms appearing in a $\frac{1}{r}$-expansion (near the null infinity) \cite{Barnich:2011mi} of the asymptotically flat metric in Bondi gauge.

\medskip

To find the quantum symmetry algebra underlying the Celestial CFT, one needs to find the self and mutual operator-product-expansions (OPEs) between the Celestial currents, as is done in usual 2D relativistic CFT \cite{Belavin:1984vu}. In the 2D CFT, the singularities of the OPEs between the energy-momentum (EM) tensor and Kac-Moody currents can be completely determined using only general symmetry arguments and the assumption that there is no local field in the theory with negative scaling dimension \cite{Zamolodchikov:1985wn,Zamolodchikov:1989mz}, without the need to consider any specific action. But, so far, all of the graviton (and gluon) OPEs obtained and used to find the symmetry algebra in Celestial CFT \cite{Distler:2018rwu,Fan:2019emx,Pate:2019mfs,Guevara:2019ypd,Fotopoulos:2019tpe,Pate:2019lpp,Fotopoulos:2019vac,Banerjee:2020kaa,Banerjee:2020zlg,Banerjee:2020vnt,Guevara:2021abz,Banerjee:2021cly,Strominger:2021lvk,Banerjee:2021dlm} were derived from a bulk (tree-level) quantum field theory with a lagrangian that corresponds to the linearized Einstein (and Yang-Mills) action \cite{He:2014laa}. Moreover, since the 2D Celestial CFT has some very different properties than the usual 2D CFT, the powerful 2D CFT techniques can not be readily used for general symmetry arguments to engineer those bulk results back from the gravity-free boundary theory.

\medskip 

In the other approach called Carrollian holography, the Carrollian fields live on a co-dimension one null boundary (one of $\mathcal{I}^\pm$) of the AFS, much like what happens in the co-dimension one AdS/CFT holography. So, unlike the Celestial conformal fields, the Carrollian fields also depend on the retarded or advanced time co-ordinate $u$ or $v$ that generates the null direction $\mathbb{R}$ of $\mathcal{I}^\pm$, in addition to the $S^n$ stereographic co-ordinates. It was shown in \cite{Duval:2014uva,Duval:2014lpa} that the original BMS$_{d}$ group is isomorphic to the (level 2) conformal extension of the Carroll$_{d-1}$ group\footnote{More precisely, the BMS$_{d}$ group is isomorphic to the level 2 conformal transformations of the Carrollian manifold \cite{Duval:2014uoa} $\mathbb{R}\times S^{d-2}$. Following relativistic CFT, one can assume Weyl invariance and use the flat metric on $S^{d-2}\simeq\mathbb{R}^{d-2}\cup\{\infty\}$.} \cite{LL,SG}. Thus, the field theory residing on the co-dimension one null boundary of the AFS is a Carrollian conformal field theory; hence, the name of this holographic approach. Only in $d=3$ and $d=4$, the (local) conformal transformations on the celestial $S^{d-2}$ form an infinite dimensional algebra \cite{Belavin:1984vu}; together with the super-translations, these lead to the extended BMS$_{d}$ symmetry \cite{Barnich:2009se,Barnich:2010ojg,Barnich:2010eb,Barnich:2011mi} of the AFS.

\medskip

Carrollian holography has so far been successful mainly in the case of 3D bulk/2D boundary \cite{Barnich:2012aw,Barnich:2012xq,Bagchi:2012xr,Barnich:2013yka,Bagchi:2014iea,Jiang:2017ecm,Hijano:2017eii,Bagchi:2015wna,Hartong:2015usd} with the case of the $1+3$D AFS gaining attention recently \cite{Bagchi:2016bcd,Donnay:2022aba,Bagchi:2022emh,Donnay:2022wvx}. In \cite{Donnay:2022aba,Donnay:2022wvx}, it was argued that since gravitational radiation escapes from $\mathcal{I}^+$, the dual Carrollian conformal field theory living there must be coupled to some external source-field (residing in the bulk) to account for the non-conservation of charges. Thus, the dual boundary theory is in need of some input from the physics in the bulk.

\medskip

As one of the two major results of this work, we show how Weinberg's soft-graviton theorem \cite{Weinberg:1965nx} and the Cachazo-Strominger sub-leading soft-graviton theorem \cite{Cachazo:2014fwa} in $1+3$D arise directly from the position-space EM tensor Ward identities of an honest (i.e. source-less) Carrollian conformal field theory in $1+2$D, that is derived just from general symmetry principles without needing any input from an underlying bulk AFS theory. Since we work with the Carrollian position-space $(t,z,\bar{z})$ coordinates ($t$ is either $u$ or $v$ and $z,\bar{z}$ are stereographic coordinates on $S^2$), the pole in energy $\omega$ that arises when an external bulk-particle gets soft in the $1+3$D AFS, manifests itself as a temporal step-function (which is the Fourier transformation of complex $\frac{1}{\omega}$) in these Ward identities, thus directly leading to the corresponding memory effects \cite{Strominger:2014pwa,Pasterski:2015tva,Strominger:2017zoo}. The advent of this temporal step-function is a direct consequence of the fact that a Carrollian theory is not a relativistic theory i.e. space and time are treated in different footings here.

\medskip

All of those Carrollian conformal Ward identities are shown to completely follow as a consequence of just the $1+2$D global Carrollian conformal or the $\text{ISL}(2,\mathbb{C})$ Poincare symmetry, if we assume only the original BMS$_4$ \cite{Bondi:1962px,Sachs:1962wk,Sachs:1962zza} and Weyl invariance of the action of the Carrollian field theory. Thus, it is not necessary to extend the BMS$_4$ symmetry of the AFS to unveil the above mentioned connection. As in 2D CFT the vacuum is global conformal invariant, similarly all the Carrollian correlators are expectation values on the global Carrollian conformal invariant vacuum.   

\medskip

That we are getting only upto the sub-leading soft graviton theorem as a consequence of the general Carrollian conformal symmetry is consistent with the findings of \cite{Laddha:2017ygw} that in a generic theory of quantum gravity, there is a non-universal contribution to the sub-subleading soft theorem. Moreover, it was shown in \cite{Freidel:2021dfs} that in Einstein gravity, the sub-subleading soft theorem arises as a consequence of conservation of a spin-2 charge that generates non-local space-time symmetry at null infinity. On the contrary, the Carrollian conformal transformations are local space-time transformations acting on the null infinity. Furthermore, only the leading and the sub-leading soft graviton theorems were shown to arise from a cleverly constructed `AdS radius to infinity' limit of the AdS$_4$/CFT$_3$ correspondence in \cite{deGioia:2023cbd}.   

\medskip

We want to clarify that, for the purpose of this work, it is sufficient to think of the boundary Carrollian theory living only on one of $\mathcal{I}^\pm$ and not on the other so that there is only one Carrollian conformal field theory on one of $\mathcal{I}^\pm$. This scenario is exactly similar to that described in \cite{Banerjee:2018fgd,Banerjee:2018gce}. A Carrollian conformal field insertion $\Phi(t_p,z_p,\bar{z}_p)$ (primary with scaling dimension $\Delta=1$ and transforming under the Carrollian spin-boost irrep, as will be shown) in the Carrollian correlator corresponds, after a temporal Fourier transformation, to either a bulk incoming or an outgoing mass-less particle taking part in a scattering process in the bulk AFS. Just like the opposite Fourier transformation factors appearing for incoming/outgoing particles in the usual relativistic (bulk) LSZ formula, here opposite Fourier transformation (from Carrollian time $t$ to bulk-energy $\omega$) factors will distinguish whether a Carrollian field corresponds to bulk incoming/outgoing particles. It can be readily seen when one `derives' the global (bulk) energy conservation law for mass-less scattering from the following global Carrollian time-translation invariance:
\begin{align*}
\sum_{p=1}^n\partial_{t_p}\left\langle\Phi_1(t_1,z_1,\bar{z}_1)\Phi_2(t_2,z_2,\bar{z}_2)\ldots\Phi_n(t_n,z_n,\bar{z}_n)\right\rangle=0
\end{align*} 
The Carrollian position co-ordinates $(z_p,\bar{z}_p)$ are identified with the parameters describing the null-momentum direction of the initial and final mass-less bulk AFS particles \cite{He:2014laa}. So, by Fourier transforming only $t$ , we go from the Carrollian position-space(-time) to the null-momentum-space describing mass-less scattering in $1+3$D bulk AFS \cite{Donnay:2022wvx}. 

\medskip

The result that the universal soft graviton theorems and the corresponding memory effects follow directly from the honest Carrollian conformal Ward identities avoiding the possibility of explicitly breaking the Carrollian conformal symmetry (by coupling the Carrollian theory to an external source) is consistent with the understanding that the super-translation memory effect arises as a result of a radiation-induced transition between two BMS super-translation inequivalent degenerate vacua which are Poincare invariant \cite{Strominger:2014pwa,Strominger:2017zoo}; this is the spontaneous breaking of the BMS symmetry into a Poincare or global Carrollian conformal symmetry. This may serve as a justification why here we consider only the Poincare invariant vacuum.

\medskip

In this work, the non-conservation of the Carrollian conformal Noether charges \cite{Donnay:2022aba,Donnay:2022wvx} will instead be inferred when we show that finite conserved quantum charges that implement the Carrollian conformal transformations on the quantum fields or the Hilbert space are different from the Noether ones. While we could find the quantum charges generating all the holomorphic and anti-holomorphic super-rotations and super-translations, we report the failure to obtain any finite conserved quantum charges that generate the arbitrary mixed super-translations on the Hilbert space.

\medskip

The temporal step-function in the Ward identities then motivates an $i\epsilon$-form of the latter, following the treatment in the $1+1$D Carrollian CFT in \cite{Saha:2022gjw}; this helps us fix the definition of the quantum conserved charges completely and establish the relation between the operator commutation relations and (time-ordered) OPEs via a complex contour integral. In 2D Euclidean CFT, this is achieved by radial quantization \cite{DiFrancesco:1997nk} by introducing a plane-to-cylinder map which converts time-ordering into radial-ordering. Here we need not introduce radial-ordering at all; the properties of the temporal step-function is all what is required.

\medskip

The Carrollian conformal Ward identity \eqref{a13} led us to the (negative-helicity) sub-leading soft graviton theorem \cite{Cachazo:2014fwa} extracted from the Ward identity of the Carrollian conformal field\footnote{The notation for a Carrollian conformal field is identical to that of a Celestial field with similar action.} $S_1^-$. The same Ward identity \eqref{a13} could be recast into a form resembling the Virasoro holomorphic EM tensor Ward identity \cite{Kapec:2016jld} given as the Ward identity of another Carrollian conformal field ${T}$. We provide purely boundary-theoretic construction of both $S^-_1$($S^+_1$) and ${T}$($\bar{T}$) in terms of the Carrollian EM tensor components. From this, it is easily inferred that the Carrollian fields $S_1^+$($S^-_1$) and $\bar{T}$($T$) are 2D shadow transformations of each other instead of assuming it as is done in Celestial holography \cite{Fotopoulos:2019tpe,Fotopoulos:2019vac}. The field $S^+_0$ whose Ward identity is the positive-helicity soft graviton theorem \cite{Weinberg:1965nx}, is similarly constructed in terms of the Carrollian EM tensor components. It is found to be the 2D shadow transformation of the $S^-_0$ field governing the negative-helicity soft graviton theorem.

\medskip

On the other hand, looking at the Carrollian Ward identity of $S^+_1$ we introduced three currents each with holomorphic weight $h=1$, following \cite{Banerjee:2020zlg,Polyakov:1987zb}; such a decomposition is valid only inside a correlator. The Ward identities of these currents resemble those of the holomorphic $\text{sl}(2,\mathbb{R})$ Kac-Moody currents. But, unlike 2D relativistic CFT, neither these currents nor the $T$ field are classically holomorphic as seen from the above constructions but they are holomorphic inside any Carrollian conformal correlator involving only local fields. 

\medskip

From the above Ward identities, we can readily write the corresponding OPEs from which 2D CFT-like mode-expansions for the six fields $S^\pm_0$, $S^\pm_1$, $T$ and $\bar{T}$ are read off. We then proceed to derive the OPEs between appropriate pairs formed among those six fields using only general symmetry arguments to finally deduce the algebra of the above-defined modes. When choosing the two operators whose product is to be expanded, we need to remember that all the operators involved in an OPE are local (composite) operators and that OPEs are associative. Since a field and its shadow transformation both can not be treated as local operators in a theory \cite{Banerjee:2022wht}, we have to choose one as a candidate local-field from each of the three following pairs: $\{T,S^-_1\}$,$\{\bar{T},S^+_1\}$ and $\{S^+_0,S^-_0\}$, leaving the other as its non-local shadow. As will be explained, only the following two choices can respect the OPE associativity: $\{{T},S^+_1,S^+_0\}$ and $\{\bar{T},S^-_1,S^-_0\}$. That a `$+$' Carrollian field and a `$-$' Carrollian field can not simultaneously give rise to a consistent OPE is the Carrollian manifestation\footnote{In this work, there is no notion of Carrollian particles, let alone Carrollian soft particles.} of the Celestial ambiguity encountered while considering double soft theorems for positive and negative helicity gravitons \cite{Klose:2015xoa}.

\medskip

In this work, we choose to take $\{{T},S^+_1,S^+_0\}$ as the local Carrollian conformal fields. The sector of the $1+2$D Carrollian CFT governed by these three generator-fields is called the holomorphic sector. Demanding these fields transform covariantly under the global Lorentz $\text{SL}(2,\mathbb{C})$ group (there is a similar assumption in Celestial holography \cite{Fotopoulos:2019tpe}) and assuming that no local field in the holomorphic sector possesses negative holomorphic weight, we are able to completely determine the pole-singularities of the mutual OPEs of these three fields from the general structures of their OPEs and using the bosonic (all of them have integer spins) exchange property between them, just as done in 2D relativistic CFT \cite{Zamolodchikov:1985wn,Zamolodchikov:1989mz}. This approach has recently been successfully used to derive the Carrollian EM tensor OPEs in \cite{Saha:2022gjw} and the Carrollian Kac-Moody current OPEs in \cite{Bagchi:2023dzx} in $1+1$D, without resorting to any `ultra-relativistic' limit. Using the prescription to translate the $1+2$D Carrollian conformal OPEs into the language of operator commutation relations, we extract the mode-algebra from these holomorphic sector OPEs to be $\text{Vir}\ltimes\hat{\overline{\text{sl}(2,\mathbb{R})}}$ (along with an abelian super-translation ideal) that is consistent with the results in \cite{Banerjee:2020zlg,Guevara:2021abz,Banerjee:2021dlm}. The modes of the $\hat{\overline{\text{sl}(2,\mathbb{R})}}$ current algebra are moreover shown to generate a special class of Carrollian diffeomorphisms, generalizing the corresponding $S^2$ diffeomorphisms \cite{Campiglia:2014yka,Campiglia:2015yka,Compere:2018ylh,Donnay:2020guq,Campiglia:2020qvc}. This is the other major result of this work. 

\medskip

We reemphasize that to reach these OPEs and the resulting symmetry algebra, we needed no input from the bulk physics. This is to be contrasted with the approaches within the Celestial holography framework where these results were obtained in the context of bulk (tree-level) Einstein-Yang-Mills theory. E.g. in \cite{Pate:2019lpp}, the symmetry argument to restrict the singularity of the OPE of two conformal primaries to a simple pole required intuition from the bulk momentum-space physics of collinear scattering or in \cite{Fan:2019emx,Fotopoulos:2019vac,Banerjee:2020kaa}, this OPE singularity was obtained by a Mellin transformation of the collinear limit of the bulk AFS scattering amplitude. Finally, conformally soft limits of these Celestial OPEs were taken to find the symmetry algebra at the level of the OPEs in e.g. \cite{Banerjee:2020zlg,Guevara:2021abz,Strominger:2021lvk}. 

\medskip

The rest of the paper is organized as follows. Section \ref{s2} contains the study of the classical aspects of the $1+2$D Carrollian CFT. After discussing on the Carrollian conformal transformations themselves, in section \ref{s2.1} we define the Carrollian multiplet fields that transform under the finite dimensional reducible but indecomposible matrix representations of the $1+2$D Carrollian spin-boost subgroup. Then in section \ref{s2.2}, we clarify the transformation properties of the Carrollian conformal primary and quasi-primary multiplet fields. Next in section \ref{s2.3}, we look at the classical properties of the EM tensor of a Carrollian CFT on a $1+2$D flat Carrollian background. In section \ref{s3}, we derive the $1+2$D source-less Carrollian conformal Ward identities and show how the leading and the sub-leading soft graviton theorems in the $1+3$D bulk AFS emerge from the same. In section \ref{s4}, we attempt to construct the finite quantum charges that inflict the $1+2$D Carrollian conformal transformations on the quantum fields. The complete definition of these charges is established in section \ref{s4.3} upon proposing a $j\epsilon$-form of the Carrollian conformal OPEs. We proceed to find the symmetry algebra generated by those quantum charges, manifest at the level of the OPEs, in section \ref{s5} before concluding with a summary in section \ref{s6}.
 
\medskip

\section{Carrollian Fields}\label{s2}
We start by reviewing some properties of the $1+2$D (level 2) Carrollian conformal (CC) transformations.

\medskip

An explicit map between the generators in \cite{Bagchi:2016bcd} showed that the $1+3$D Poincare algebra is isomorphic to the global sub-algebra of the $1+2$D Carrollian conformal algebra. In $1+2$D, the Carroll group \cite{LL,SG} is formed by three space-time translations, a spatial rotation and two Carrollian boosts. Along with these transformations, a dilation, a temporal special CC transformation (TSCT) and two spatial special CC transformations (SSCT) generate the $1+2$D Carrollian conformal group on $\mathbb{R}_t\times S^2$. All of these transformations are collectively expressed as \cite{Oblak:2015qia}:
\begin{align}
z\rightarrow z^\prime=\frac{az+b}{cz+d}\hspace{2.5mm},\hspace{2.5mm}\bar{z}\rightarrow \bar{z}^\prime=\frac{\bar{a}\bar{z}+\bar{b}}{\bar{c}\bar{z}+\bar{d}}\hspace{2.5mm},\hspace{2.5mm}t\rightarrow t^\prime=\frac{t}{{|cz+d|}^2}+\lambda z\bar{z}+\mu z+\bar{\mu}\bar{z}+\nu\label{25}
\end{align}
with $a,b,c,d,\mu\in\mathbb{C}$, $ad-bc=1$ and $\lambda,\nu\in\mathbb{R}$. We recall that $z=x+iy$ is the stereographic coordinate on $S^2$ and $t\in\mathbb{R}$. On the sphere we have $z^*=\bar{z}$ etc; thus, we have the ten parameter group.

\medskip 

In physics, one is usually more interested in local aspects of transformations. The $1+2$D CC transformations, not necessarily globally defined\footnote{For a transformation to be globally defined, we shall also demand globally non-singular behavior of the corresponding generators \cite{Blu:2009zz}. In this sense, the other super-translations not included in \eqref{25} fail to be globally defined on $S^2$.} on $\mathbb{R}\times S^2$, have the following finite form:
\begin{align}
x\rightarrow x^\prime=U_0(x,y)\hspace{2.5mm};\hspace{2.5mm}y\rightarrow y^\prime=V_0(x,y)\hspace{2.5mm};\hspace{2.5mm}t\rightarrow t^\prime=t\left[\left(\frac{\partial U_0}{\partial x}\right)^2+\left(\frac{\partial V_0}{\partial x}\right)^2\right]^{\frac{1}{2}}+H(x,y)
\end{align}
whose infinitesimal version, compactly expressed as $x^{\mu}\rightarrow x^{\mu}+\epsilon^af^{\mu}_{\hspace{1.5mm}a}(\mathbf{x})$, is:
\begin{align}
x\rightarrow x+\epsilon^xU_1+\epsilon^yU_2\hspace{2.5mm};\hspace{2.5mm}y\rightarrow y+\epsilon^xV_1+\epsilon^yV_2\hspace{2.5mm};\hspace{2.5mm}t\rightarrow t\left(1+\epsilon^x\frac{\partial U_1}{\partial x}+\epsilon^y\frac{\partial U_2}{\partial x}\right)+\epsilon^tH\label{eq:1}
\end{align}
where $\epsilon^x, \epsilon^y, \epsilon^t$ are real infinitesimal parameters and the functions $U_i(x,y)$ and $V_i(x,y)$ satisfy the Cauchy-Riemann conditions:
\begin{align}
\frac{\partial U_i}{\partial x}=\frac{\partial V_i}{\partial y}\hspace{2.5mm},\hspace{2.5mm}\frac{\partial U_i}{\partial y}=-\frac{\partial V_i}{\partial x}\hspace{2.5mm}\Longrightarrow\hspace{2.5mm} \nabla^2U_i=0=\nabla^2V_i\nonumber
\end{align}
while the function $H(x,y)\equiv H(z,\bar{z})$ is arbitrary. So, $f_j(z,\bar{z})\equiv U_j+iV_j$ and $\bar{f}_j(z,\bar{z})\equiv U_j-iV_j$ satisfy $\partial_{\bar{z}}f_j=0=\partial_z\bar{f}_j$. Thus, for the finite case:
\begin{align}
z\rightarrow z^\prime=f_0(z)\hspace{2.5mm};\hspace{2.5mm}\bar{z}\rightarrow \bar{z}^\prime=\bar{f}_0(\bar{z})\hspace{2.5mm};\hspace{2.5mm}t\rightarrow t^\prime=t\left(\frac{df_0}{dz}.\frac{d\bar{f}_0}{d\bar{z}}\right)^{\frac{1}{2}}+H(z,\bar{z})\label{34}
\end{align} 
and for the infinitesimal case: 
\begin{align}
z\rightarrow z+\epsilon f(z)\hspace{2.5mm},\hspace{2.5mm}\bar{z}\rightarrow \bar{z}+\bar{\epsilon}\bar{f}(\bar{z})\hspace{2.5mm},\hspace{2.5mm}t\rightarrow t+\epsilon\frac{t}{2}\frac{df}{dz}+\bar{\epsilon}\frac{t}{2}\frac{d\bar{f}}{d\bar{z}}+\epsilon^tH(z,\bar{z})\label{35}
\end{align}
From \eqref{25}, it is clear that the global infinitesimal $1+2$D CC transformations that act on $S^2$ are given by such $f(z)$ and $\bar{f}(\bar{z})$ that are at most quadratic polynomials in $z$ and $\bar{z}$ respectively. In the case of infinitesimal transformations, $z$ and $\bar{z}$ (and similarly, $\epsilon$ and $\bar{\epsilon}$) can be treated as independent variables. 

\medskip

Now we consider a multi-component field transforming as a finite matrix representation under \eqref{eq:1}, schematically as ($i$ denote collection of suitable indices):
\begin{align}
\Phi^i(\mathbf{x})\rightarrow{\tilde{\Phi}}^i(\mathbf{x^\prime})=\Phi^i(\mathbf{x})+\epsilon^a{(\mathcal{F}_a\cdot\Phi)}^i(\mathbf{x})\label{79}
\end{align}
The generators of these transformations are defined as \cite{DiFrancesco:1997nk}:
\begin{align}
\delta_{\bm{\epsilon}}\Phi^i(\mathbf{x})\equiv{\tilde{\Phi}}^i(\mathbf{x})-{{\Phi}}^i(\mathbf{x}):=-i\epsilon^aG_a(\mathbf{x})\Phi^i(\mathbf{x})\equiv -i\epsilon^aG_a\Phi^i(\mathbf{x})
\end{align}
so the generators are given by:
\begin{align}
-iG_a(\mathbf{x})\Phi^i(\mathbf{x})={(\mathcal{F}_a\cdot\Phi)}^i(\mathbf{x})-f^{\mu}_{\hspace{1.5mm}a}(\mathbf{x})\partial_\mu\Phi^i(\mathbf{x})\label{31}
\end{align}

\medskip

Thus, the generator of an infinitesimal space-time transformation $x^{\mu}\rightarrow x^{\mu}+\epsilon^af^{\mu}_{\hspace{1.5mm}a}(\mathbf{x})$ in the space of ordinary functions $\phi(\mathbf{x})$ (i.e. having ${(\mathcal{F}_a\cdot\phi)}(\mathbf{x})=0$) is obtained as:
\begin{align}
-i\epsilon^a\mathcal{G}_a\phi(\mathbf{x})=\phi(\mathbf{x}-\epsilon^a\mathbf{f}_{a}(\mathbf{x}))-\phi(\mathbf{x})\hspace{2.5mm}\Longrightarrow\hspace{2.5mm}\mathcal{G}_a(\mathbf{x})=-if^{\mu}_{\hspace{1.5mm}a}(\mathbf{x})\partial_\mu
\end{align}
So we have the following generators of the $1+2$D CC transformations (with $n,a,b\in\mathbb{Z}$) in the space of functions (we treat $z$ and $\bar{z}$ independently):
\begin{align}
&\text{the $z^{n+1}$ super-rotation is generated by: $\mathcal{L}_n=-iz^{n+1}\partial_z-i\frac{n+1}{2}z^nt\partial_t$}\label{26}\\
&\text{the $\bar{z}^{n+1}$ super-rotation is generated by: $\bar{\mathcal{L}}_n=-i\bar{z}^{n+1}\partial_{\bar{z}}-i\frac{n+1}{2}\bar{z}^nt\partial_t$}\label{27}\\
&\text{the $z^{a+1}\bar{z}^{b+1}$ super-translation is generated by: $\mathcal{P}_{a,b}=-iz^{a+1}\bar{z}^{b+1}\partial_t$}\label{28}
\end{align}

\medskip

These differential generators satisfy the (extended) BMS$_4$ algebra \cite{Barnich:2009se,Barnich:2010ojg,Barnich:2010eb,Barnich:2011mi}:
\begin{align}
&\left[\mathcal{L}_n\hspace{1mm},\hspace{1mm}\mathcal{L}_m\right]=i(n-m)\mathcal{L}_{n+m}\hspace{2.5mm};\hspace{2.5mm}\left[\bar{\mathcal{L}}_n\hspace{1mm},\hspace{1mm}\bar{\mathcal{L}}_m\right]=i(n-m)\bar{\mathcal{L}}_{n+m}\hspace{2.5mm};\hspace{2.5mm}\left[\mathcal{L}_n\hspace{1mm},\hspace{1mm}\bar{\mathcal{L}}_m\right]=0\label{29}\\
&\left[\mathcal{L}_n\hspace{1mm},\hspace{1mm}\mathcal{P}_{a,b}\right]=i\left(\frac{n-1}{2}-a\right)\mathcal{P}_{a+n,b}\hspace{2.5mm};\hspace{2.5mm}\left[\bar{\mathcal{L}}_n\hspace{1mm},\hspace{1mm}\mathcal{P}_{a,b}\right]=i\left(\frac{n-1}{2}-b\right)\mathcal{P}_{a,b+n}\hspace{2.5mm};\hspace{2.5mm}\left[\mathcal{P}_{a,b}\hspace{1mm},\hspace{1mm}\mathcal{P}_{c,d}\right]=0\nonumber
\end{align}

\medskip

From the explicit forms \eqref{26}-\eqref{28}, it is evident that the super-rotation and super-translation generators are singular at $z=0=\bar{z}$ for $n<-1$ and $a,b<-1$ respectively. By performing the following global $1+2$D CC transformation:
\begin{align*}
z\rightarrow z^\prime=-\frac{1}{z}\hspace{2.5mm},\hspace{2.5mm}\bar{z}\rightarrow \bar{z}^\prime=-\frac{1}{\bar{z}}\hspace{2.5mm},\hspace{2.5mm}t\rightarrow t^\prime=\frac{t}{z\bar{z}}
\end{align*}
one concludes, on the other hand, that the super-rotation generators are singular at $z=\infty=\bar{z}$ for $n>1$ and the super-translation generators are singular for $a,b>0$. Thus, we recover the fact that $\mathcal{L}_n$, $\bar{\mathcal{L}}_n$ and $\mathcal{P}_{a,b}$ with $n\in\{0,\pm1\}$ and $a,b\in\{0,-1\}$ generate the $1+2$D global CC transformations of functions defined on $\mathbb{R}\times S^2$.

\medskip

We now discuss on the transformation properties of the $1+2$D classical Carrollian fields.

\medskip 

\subsection{Carrollian spin-boost multiplet}\label{s2.1} 
The Carroll algebra in $1+2$D is given by \cite{LL}:
\begin{align}
&\left[\mathbf{P}_i\hspace{1mm},\hspace{1mm}\mathbf{H}\right]=0\hspace{2.5mm};\hspace{2.5mm}\left[\mathbf{H}\hspace{1mm},\hspace{1mm}\mathbf{B}_i\right]=0\hspace{2.5mm};\hspace{2.5mm}\left[\mathbf{B}_i\hspace{1mm},\hspace{1mm}\mathbf{P}_j\right]=i\delta_{ij}\mathbf{H}\hspace{2.5mm};\hspace{2.5mm}\left[\mathbf{B}_i\hspace{1mm},\hspace{1mm}\mathbf{B}_j\right]=0\hspace{2.5mm};\hspace{2.5mm}\left[\mathbf{P}_i\hspace{1mm},\hspace{1mm}\mathbf{P}_j\right]=0\nonumber\\
&\left[\mathbf{J}\hspace{1mm},\hspace{1mm}\mathbf{H}\right]=0\hspace{2.5mm};\hspace{2.5mm}\left[\mathbf{B}_i\hspace{1mm},\hspace{1mm}\mathbf{P}_j\right]=i\delta_{ij}\mathbf{H}\hspace{2.5mm};\hspace{2.5mm}\left[\mathbf{J}\hspace{1mm},\hspace{1mm}\mathbf{P}_i\right]=i\epsilon_{ij}\mathbf{P}_j\hspace{2.5mm};\hspace{2.5mm}\left[\mathbf{J}\hspace{1mm},\hspace{1mm}\mathbf{B}_i\right]=i\epsilon_{ij}\mathbf{B}_j
\end{align}
where $\mathbf{P}_i$, $\mathbf{H}$, $\mathbf{J}$ and $\mathbf{B}_i$ respectively are the generators of the space-translations, time-translation, spatial rotation and Carrollian boosts. Only the Carrollian boosts and spatial rotation leave the origin $(t,\vec{x})=(0,\vec{0})$ invariant, out of these six generators. So, we look into the structure of the Carrollian boosts and rotation which will lead to the construction of the Carrollian tensors, in the same way that one uses the Lorentz generators to define the tensorial transformation laws in the relativistic case. 

\medskip

With the complexification $\mathbf{B}=\mathbf{B}_x+i\mathbf{B}_y$ and $\bar{\mathbf{B}}=\mathbf{B}_x-i\mathbf{B}_y$, the spin-boost sub-algebra in $1+2$D reads: 
\begin{align}
[\mathbf{J},\mathbf{B}]=\mathbf{B}\hspace{2.5mm};\hspace{2.5mm}[\mathbf{J},\bar{\mathbf{B}}]=-\bar{\mathbf{B}}\hspace{2.5mm};\hspace{2.5mm}[\mathbf{B},\bar{\mathbf{B}}]=0\label{eq:7}
\end{align}
In 1+2 dimensions, the Carrollian space-time boost transformation (CB) is defined as: $(x,y,t)\rightarrow (x^\prime,y^\prime,t^\prime)=(x\text{ },\text{ }y\text{ },\text{ }t+v_xx+v_yy)\text{ }$; or equivalently, as:
\begin{align*}
\begin{pmatrix}
x\\
y\\
t
\end{pmatrix}\longrightarrow \begin{pmatrix}
x^\prime\\
y^\prime\\
t^\prime
\end{pmatrix}= \left[\exp{\begin{pmatrix}
0 & 0 & 0\\
0 & 0 & 0\\
v_x & v_y & 0
\end{pmatrix}}\right]\begin{pmatrix}
x\\
y\\
t
\end{pmatrix}\text{ }\Longleftrightarrow\text{ } x^\mu\rightarrow {x^{\prime}}^\mu={\left[e^{v_x{\mathbf{B}^{(3)}_x}+v_y{\mathbf{B}^{(3)}_y}}\right]}^{\mu}_{\hspace{2mm}\nu}\text{ }x^\nu
\end{align*}
where 
\begin{align}
{\mathbf{B}^{(3)}_x}:=\begin{pmatrix}
0 & 0 & 0\\
0 & 0 & 0\\
1 & 0 & 0
\end{pmatrix} \text{\hspace{2mm} and \hspace{2mm}} {\mathbf{B}^{(3)}_y}:=\begin{pmatrix}
0 & 0 & 0\\
0 & 0 & 0\\
0 & 1 & 0
\end{pmatrix}\label{8}
\end{align}
are three-dimensional representations of the CB generators while the spatial rotation by angle $\theta$ is defined as usual:
\begin{align*}
\begin{pmatrix}
x\\
y\\
t
\end{pmatrix}\longrightarrow \begin{pmatrix}
x^\prime\\
y^\prime\\
t^\prime
\end{pmatrix}= \left[\exp{\begin{pmatrix}
0 & -\theta & 0\\
\theta & 0 & 0\\
0 & 0 & 0
\end{pmatrix}}\right]\begin{pmatrix}
x\\
y\\
t
\end{pmatrix}\text{ }\Longleftrightarrow\text{ } x^\mu\rightarrow {x^{\prime}}^\mu={\left[e^{-i\theta \mathbf{J}^{(3)}}\right]}^{\mu}_{\hspace{2mm}\nu}\text{ }x^\nu
\end{align*}
with
\begin{align*}
\mathbf{J}^{(3)}:=\begin{pmatrix}
0 & -i & 0\\
i & 0 & 0\\
0 & 0 & 0
\end{pmatrix}
\end{align*}

\medskip

Since \eqref{eq:7} is a solvable Lie algebra, it can have only 1-dimensional irreducible representations (irreps); its only irrep is generated by: $J=1, B=0=\bar{B}$.

\medskip

But it can also have reducible but indecomposable representations, the construction of which we begin by observing that the classical boost generators \eqref{8} are not diagonalizable as their only generalized eigenvalue 0 has geometric multiplicity 2; only the spatial rotation generator $\mathbf{J}$ (which is hermittian) can be diagonalized. Consequently, we work in the diagonal basis of $\mathbf{J}$ and motivated by the knowledge of the quantum mechanics of angular momentum, we assume that the matrix representation of $\mathbf{J}$ has no degenerate eigenvalue. 

\medskip

From \eqref{eq:7}, it can then be shown that in the indecomposable but reducible matrix representation of this algebra, the difference between any two consecutive (real) eigenvalues of $\mathbf{J}$ is 1. We organize the basis such that the largest eigenvalue of $\mathbf{J}$ is at the first row/column; then $\mathbf{B}$ is a super-diagonal matrix and $\bar{\mathbf{B}}$ is a sub-diagonal matrix such that: 
\begin{align*}
[\mathbf{B}]_{i,i+1}[\bar{\mathbf{B}}]_{i+1,i}=0\text{\hspace{10mm}(no sum over $i$)}
\end{align*}
E.g. a reducible but indecomposable three dimensional representation of the algebra \eqref{eq:7} is given by:
\begin{align}
\mathbf{J}^\prime=\begin{pmatrix}
l+1 & 0 & 0\\
0 & l & 0\\
0 & 0 & l-1
\end{pmatrix} \hspace{2.5mm},\hspace{2.5mm} \mathbf{B}^\prime=\begin{pmatrix}
0 & 0 & 0\\
0 & 0 & a\\
0 & 0 & 0
\end{pmatrix} \hspace{2.5mm},\hspace{2.5mm} \bar{\mathbf{B}}^\prime=\begin{pmatrix}
0 & 0 & 0\\
b & 0 & 0\\
0 & 0 & 0
\end{pmatrix}\label{71}
\end{align}
An explicit example of a multiplet transforming under this matrix representation will arise in section \ref{s5.3}. More general representations were discussed in \cite{Chen:2021xkw}.

\medskip

A general $1+2$D Carrollian Cartesian tensor field transform under a decomposible representation of the spin-boost sub-algebra \eqref{eq:7}, just like in the relativistic case. A multi-component field transforming under a $d$-dimensional reducible but indecomposible representation of the spin-boost sub-algebra will be called a Carrollian multiplet of rank $d$. Under a spatial rotation by angle $\theta$ and Carrolean-boost by parameter $\vec{v}$, a rank $d$ Carrollian multiplet transforms as:
\begin{align}
\mathbf{\Phi}_{(d)}(t,\vec{x})\longrightarrow\tilde{\mathbf{\Phi}}_{(d)}(t^\prime,\vec{x}^\prime)={\left[e^{-i\theta \mathbf{J}_{(d)}-\vec{v}\cdot\vec{\mathbf{B}}_{(d)}}\right]}\cdot\mathbf{{\Phi}}_{(d)}(t,\vec{x})\label{32}
\end{align}
Clearly, a Carrollian scalar field transforms under the trivial irrep. The rotation eigenvalue $l$ is taken to be integers.

\medskip

\subsection{Carrollian conformal transformations of classical fields}\label{s2.2}
From \eqref{25}, we see that the subgroup of the $1+2$D Carrollian conformal group that keeps the space-time origin invariant is generated by the Carrollian boosts, rotation, dilation, TSCT and SSCTs. The corresponding sub-algebra is isomorphic to $1+2$D Carrollian algebra augmented by dilation because of the similar roles played by space-time translation generators and special CC transformation generators in the algebra. So, we just note the commutators involving dilation generator $\mathbf{D}$:
\begin{align*}
\left[\mathbf{D}\hspace{1mm},\hspace{1mm}\mathbf{J}\right]=\left[\mathbf{D}\hspace{1mm},\hspace{1mm}\mathbf{B}_i\right]=0\hspace{2.5mm};\hspace{2.5mm}
\left[\mathbf{D}\hspace{1mm},\hspace{1mm}\mathbf{K}_t\right]=-i\mathbf{K}_t\hspace{2.5mm};\hspace{2.5mm}\left[\mathbf{D}\hspace{1mm},\hspace{1mm}\mathbf{K}_i\right]=-i\mathbf{K}_i
\end{align*}
where $\mathbf{K}_t$ and $\mathbf{K}_i$ are the generators of TSCT and SSCTs respectively.

\medskip

Given that the spin-boost sub-algebra has finite-dimensional indecomposable matrix representation, we now find the matrix representation of the algebra corresponding to the full invariant subgroup of the origin, generated by\footnote{To clarify the notation, e.g. $\bm{\mathcal{L}}_n$ is the full generator whose space-time differential part is ${\mathcal{L}}_n$.} $\bm{\mathcal{L}}_n$, $\bar{\bm{\mathcal{L}}}_n$ for $n\in\{0,1\}$ and $\bm{\mathcal{P}}_{0,-1}$, $\bm{\mathcal{P}}_{-1,0}$, $\bm{\mathcal{P}}_{0,0}$. The dilation generator of the level 2 Carrollian conformal algebra in 1+2D acts uniformly on space and time via the transformation rule: $(x,y,t)\rightarrow (x^\prime,y^\prime,t^\prime)=(\lambda x,\lambda y,\lambda t)$. Since the dilation generator commutes with the spatial rotation and the boost generators, in the finite-dimensional indecomposable spin-boost representation (and trivially, in the 1D irrep), $\mathbf{D}(=\bm{\mathcal{L}}_0+\bar{\bm{\mathcal{L}}}_0)$ must be proportional to the Identity matrix $\mathbf{I}$. Though this result seems to have followed from the Schur's lemma, that is not the case since we are working with reducible representations. Moreover, none of the SSCT and TSCT generators commutes with the dilation generator, so  $\mathbf{K}_t$ and $\mathbf{K}_i$ (and thus ${\bm{\mathcal{P}}}_{0,0}$, ${\bm{\mathcal{L}}}_1$ and $\bar{\bm{\mathcal{L}}}_1$) can only be $\mathbf{0}$-matrices in these representations.

\medskip

Since, $\mathbf{{D}}$ is proportional to $\mathbf{I}$, the proportionality constant will be related to the scaling dimension of the field. A Carrollian multiplet with scaling dimension $\Delta$ transforms under the dilation as:
\begin{align}
\mathbf{\Phi}_{(d)}(t,\vec{x})\longrightarrow\mathbf{\tilde{\Phi}}_{(d)}(t^\prime,\vec{x}^\prime)=\lambda^{-\Delta}\mathbf{\Phi}_{(d)}(t,\vec{x})\label{30}
\end{align}

\medskip

Now looking at the infinitesimal versions of the transformations \eqref{32} and \eqref{30}, we find the action of the Carrollian boost, rotation and dilation generators on a classical Carrollian multiplet by applying the definition \eqref{31}:
\begin{align}
&\bm{\mathcal{P}}_{0,-1}(t,\vec{x})\mathbf{\Phi}_{(d)}(t,\vec{x})=\mathbf{{B}}(t,\vec{x})\mathbf{\Phi}_{(d)}(t,\vec{x})=-i\left[\mathbf{I}z\partial_t+\mathbf{B}_{(d)}\right]\cdot\mathbf{\Phi}_{(d)}(t,\vec{x})\\
&\bm{\mathcal{P}}_{-1,0}(t,\vec{x})\mathbf{\Phi}_{(d)}(t,\vec{x})=\bar{\mathbf{{B}}}(t,\vec{x})\mathbf{\Phi}_{(d)}(t,\vec{x})=-i\left[\mathbf{I}\bar{z}\partial_t+\bar{\mathbf{B}}_{(d)}\right]\cdot\mathbf{\Phi}_{(d)}(t,\vec{x})\\
&\bm{\mathcal{L}}_{0}(t,\vec{x})\mathbf{\Phi}_{(d)}(t,\vec{x})=\frac{\mathbf{D}-i\mathbf{{J}}}{2}(t,\vec{x})\mathbf{\Phi}_{(d)}(t,\vec{x})=-i\left[\mathbf{I}\left(z\partial_z+\frac{1}{2}t\partial_t\right)+\frac{\Delta\mathbf{I}+\mathbf{J}_{(d)}}{2}\right]\cdot\mathbf{\Phi}_{(d)}(t,\vec{x})\\
&\bar{\bm{\mathcal{L}}}_{0}(t,\vec{x})\mathbf{\Phi}_{(d)}(t,\vec{x})=\frac{\mathbf{D}+i\mathbf{{J}}}{2}(t,\vec{x})\mathbf{\Phi}_{(d)}(t,\vec{x})=-i\left[\mathbf{I}\left(\bar{z}\partial_{\bar{z}}+\frac{1}{2}t\partial_t\right)+\frac{\Delta\mathbf{I}-\mathbf{J}_{(d)}}{2}\right]\cdot\mathbf{\Phi}_{(d)}(t,\vec{x})\label{56}
\end{align}
We note that these four generators act as constant matrix-multiplications on a multiplet situated at the origin.

\medskip

It remains to find the actions of the generators $\bm{\mathcal{P}}_{0,0}$, $\bm{\mathcal{L}}_1$ and $\bar{\bm{\mathcal{L}}}_1$ on the Carrollian multiplets. Since these generators have no non-trivial finite dimensional matrix representation and they keep the space-time origin invariant, we infer that:
\begin{align}
\bm{\mathcal{P}}_{0,0}(\mathbf{0})=\mathbf{0}\hspace{2.5mm};\hspace{2.5mm}\bm{\mathcal{L}}_1(\mathbf{0})=\mathbf{0}\hspace{2.5mm};\hspace{2.5mm}\bar{\bm{\mathcal{L}}}_1(\mathbf{0})=\mathbf{0}\label{33}
\end{align}
To find the corresponding actions on a multiplet situated at an arbitrary location in space-time, we use the finite actions of the space-time translation generators via the BCH lemma, as illustrated below for the TSCT generator:
\begin{align}
\bm{\mathcal{P}}_{0,0}(t,\vec{x})&=e^{iz\mathcal{L}_{-1}+i\bar{z}\bar{\mathcal{L}}_{-1}+it\bm{\mathcal{P}}_{-1,-1}}\text{ }\bm{\mathcal{P}}_{0,0}(\mathbf{0})\text{ }e^{-iz\mathcal{L}_{-1}-i\bar{z}\bar{\mathcal{L}}_{-1}-it\bm{\mathcal{P}}_{-1,-1}}\nonumber\\
&=\bm{\mathcal{P}}_{0,0}(\mathbf{0})+\left[z\bm{\mathcal{P}}_{-1,0}+\bar{z}\bm{\mathcal{P}}_{0,-1}\right](\mathbf{0})+z\bar{z}\bm{\mathcal{P}}_{-1,-1}\nonumber\\
\Longrightarrow\text{ }\bm{\mathcal{P}}_{0,0}(t,\vec{x})&\mathbf{\Phi}_{(d)}(t,\vec{x})=-i\left[\mathbf{I}z\bar{z}\partial_t+z\bar{\mathbf{B}}_{(d)}+\bar{z}\mathbf{B}_{(d)}\right]\cdot\mathbf{\Phi}_{(d)}(t,\vec{x})
\end{align}
Similarly, one obtains the following actions of the SSCT generators on Carrollian multiplets:
\begin{align}
&\bm{\mathcal{L}}_{1}(t,\vec{x})\mathbf{\Phi}_{(d)}(t,\vec{x})=-i\left[\mathbf{I}\left(z^2\partial_z+zt\partial_t\right)+2z\text{ }\frac{\Delta\mathbf{I}+\mathbf{J}_{(d)}}{2}+t\mathbf{B}_{(d)}\right]\cdot\mathbf{\Phi}_{(d)}(t,\vec{x})\\
&\bar{\bm{\mathcal{L}}}_{1}(t,\vec{x})\mathbf{\Phi}_{(d)}(t,\vec{x})=-i\left[\mathbf{I}\left(\bar{z}^2\partial_{\bar{z}}+\bar{z}t\partial_t\right)+2\bar{z}\text{ }\frac{\Delta\mathbf{I}-\mathbf{J}_{(d)}}{2}+t\bar{\mathbf{B}}_{(d)}\right]\cdot\mathbf{\Phi}_{(d)}(t,\vec{x})\label{57}
\end{align}

\medskip

If a Carrollian multiplet infinitesimally transforms under all the super-translations and super-rotations (i.e. for any $n\geq-1$ and $a,b\geq-1$) as following\footnote{These transformation laws are derived in exactly the same way as illustrated, with the conditions \eqref{33} extended to $n\geq1$ and $a,b\geq0$.} (with $\mathbf{h}=\frac{\Delta\mathbf{I}+\mathbf{J}}{2}$):
\begin{align}
&\bm{\mathcal{P}}_{a,b}(t,\vec{x})\mathbf{\Phi}(t,\vec{x})=-i\left[z^{a+1}\bar{z}^{b+1}\partial_t+(a+1)z^a\bar{z}^{b+1}{\mathbf{B}}+(b+1)z^{a+1}\bar{z}^b\bar{\mathbf{B}}\right]\cdot\mathbf{\Phi}(t,\vec{x})\label{36}\\
&\bm{\mathcal{L}}_{n}(t,\vec{x})\mathbf{\Phi}(t,\vec{x})=-i\left[z^{n+1}\partial_z+\frac{n+1}{2}z^nt\partial_t+(n+1)z^n\mathbf{h}+(n+1)nz^{n-1}\frac{t}{2}\mathbf{B}\right]\cdot\mathbf{\Phi}(t,\vec{x})\label{37}\\
&\bar{\bm{\mathcal{L}}}_{n}(t,\vec{x})\mathbf{\Phi}(t,\vec{x})=-i\left[\bar{z}^{n+1}\partial_{\bar{z}}+\frac{n+1}{2}\bar{z}^nt\partial_t+(n+1)\bar{z}^n\bar{\mathbf{h}}+(n+1)n\bar{z}^{n-1}\frac{t}{2}\bar{\mathbf{B}}\right]\cdot\mathbf{\Phi}(t,\vec{x})\label{38}
\end{align}
it is called a $1+2$D Carrollian conformal primary multiplet field. These are the infinitesimal forms of the $1+2$D CC indecomposible tensorial transformation laws. If a multiplet transforms tensorially (i.e. like above) only for $n\in\{0,\pm1\}$ and $a,b\in\{0,-1\}$ i.e. only under the $1+2$D CC group, it is called a quasi-primary multiplet.

\medskip

We end this section by mentioning the finite transformation rules of $1+2$D CC primary multiplet fields. Under a general finite $1+2$D CC transformation \eqref{34}, a rank $d$ primary multiplet with scaling dimension $\Delta$ transforms as:
\begin{align}
\mathbf{\Phi}_{(d)}(t,\vec{x})\longrightarrow\mathbf{\tilde{\Phi}}_{(d)}(t^\prime,\vec{x}^\prime)={\left(\frac{df_0}{dz}\cdot\frac{d\bar{f}_0}{d\bar{z}}\right)}^{-\frac{\Delta}{2}}e^{-\frac{\mathbf{J}_{(d)}}{2}\text{Ln}\left(\frac{\partial_zf_0}{\partial_{\bar{z}}\bar{f}_0}\right)-{\mathbf{B}}_{(d)}{\frac{\partial_z t^\prime}{\partial_zf_0}}-\bar{{\mathbf{B}}}_{(d)}\frac{\partial_{\bar{z}} t^\prime}{\partial_{\bar{z}}\bar{f}_0}}\cdot\mathbf{{\Phi}}_{(d)}(t,\vec{x})\label{82}
\end{align}
If a multiplet transforms like this only under the global transformations \eqref{25}, it is called a quasi-primary field. It is worth noting that a primary field that transforms under spin-boost irrep has a 2D CFT-primary like transformation property under \eqref{34}:
\begin{align}
{\Phi}(t,\vec{x})\longrightarrow{\tilde{\Phi}}(t^\prime,\vec{x}^\prime)={\left(\frac{df_0}{dz}\right)}^{-h}{\left(\frac{d\bar{f}_0}{d\bar{z}}\right)}^{-\bar{h}}{{\Phi}}(t,\vec{x})\label{69}
\end{align}
if it has scaling dimension $\Delta=h+\bar{h}$ and spin $J=h-\bar{h}$.

\medskip
 
\subsection{Classical EM tensor}\label{s2.3}
We want to investigate on the consequences for a $1+2$D field theory (on flat (Carrollian) background with topology $\mathbb{R}\times S^2$) possessing the original BMS$_4$ algebra \cite{Bondi:1962px,Sachs:1962wk,Sachs:1962zza} as the kinematical (global space-time) symmetry of the action. In classical field theory, Noether's theorem states that each of the continuous symmetries of the action must have an associated on-shell conserved current. The (finite) conserved charges constructed out of these currents generate the continuous symmetry transformations on the classical phase-space, thereby providing a classical dynamical realization of the continuous symmetry. The conserved Noether currents corresponding to the kinematical space-time symmetries are related to the EM tensor of the theory. Hence, we now look at the classical properties of the EM tensor of a BMS$_{4}$ invariant theory.

\medskip

Following \cite{Bagchi:2019clu}, let the action describing a classical theory of fields in $1+2$D flat Carrollian space-time (with topology $\mathbb{R}\times S^2$) be:
\begin{align*}
S[\bm{\Phi}]=\int dt\int\limits_{S^2} d^2\vec{x}\text{ }\mathcal{L}(\bm{\Phi},\partial_\mu\bm{\Phi})
\end{align*} 
where, assuming Weyl invariance, we use the (mostly) flat-metric on $S^2$. Under an infinitesimal space-time transformation \eqref{35} when the field transforms as \eqref{79}, this action transforms on-shell as \cite{DiFrancesco:1997nk}:
\begin{align*}
S[\bm{\Phi}]\rightarrow S^\prime[\bm{\tilde{\Phi}}]=S[\bm{\Phi}]+\int dt\int\limits_{S^2} d^2\vec{x}\text{ }\epsilon^a\partial_\mu j^\mu_{\hspace{1.5mm}a}(\mathbf{x})
\end{align*}
where $j^\mu_{\hspace{1.5mm}a}(\mathbf{x})$ is the corresponding Noether current:
\begin{align*}
j^\mu_{\hspace{1.5mm}a}(\mathbf{x})= {T_{(c)}}^\mu_{\hspace{1.5mm}\nu} f^\nu_{\hspace{1.5mm}a}-{(\mathcal{F}_a\cdot\Phi)}^i\frac{\partial\mathcal{L}}{\partial(\partial_\mu\Phi^i)}
\end{align*}
with ${T_{(c)}}^\mu_{\hspace{1.5mm}\nu}$ being the canonical (on-shell conserved) EM tensor which is the Noether current corresponding to the global Carrollian translation-invariance.

\medskip

In \cite{Baiguera:2022lsw}, the special properties of the classical Carrollian EM tensor in any space-time dimension, that arise as the consequences of local boost invariance and Weyl invariance of the classical action on an arbitrary Carrollian manifold, were extracted. For the flat Carrollian manifold, the local boosts are nothing other than the super-translations. Classical super-translation symmetry of the flat-Carrollian action implies that the canonical EM tensor ${T_{(c)}}^\mu_{\hspace{1.5mm}\nu}$ can be Belinfante-improved in any dimension to have vanishing energy flux densities off-shell:
\begin{align}
{T_{(B)}}^i_{\hspace{1.5mm}t}=0\text{ }\text{ }(\text{with }\partial_\mu{T_{(B)}}^\mu_{\hspace{1.5mm}t}=0)\text{ }\text{ }\Longrightarrow\text{ }\text{ } \partial_t{T_{(B)}}^t_{\hspace{1.5mm}t}=0\label{77}
\end{align} 

\medskip

Classical Weyl invariance implies that the EM tensor can additionally be improved to become off-shell traceless: ${T_{(B)}}^\mu_{\hspace{1.5mm}\mu}=0$ in any dimension. Moreover, due to the global spatial-rotation invariance of the flat-Carrollian theory, the EM tensor can further be made off-shell symmetric in spatial indices: ${T_{(B)}}^{ij}={T_{(B)}}^{ji}$. Together they lead to the following classical constraints in $1+2$D:
\begin{align}
{T}^\mu_{\hspace{1.5mm}\mu}=0\hspace{2.5mm}\text{ and }\hspace{2.5mm}{T}^z_{\hspace{1.5mm}z}={T}^{\bar{z}}_{\hspace{1.5mm}\bar{z}}\hspace{2.5mm}\Longrightarrow\hspace{2.5mm} {T}^z_{\hspace{1.5mm}z}={T}^{\bar{z}}_{\hspace{1.5mm}\bar{z}}=-\frac{1}{2}{T}^t_{\hspace{1.5mm}t}\hspace{5mm}\text{(off-shell)}\label{78}
\end{align}

\medskip

The properties \eqref{77} and \eqref{78} enable us to off-shell express the (on-shell) conserved Noether current associated to an arbitrary symmetry transformation \eqref{eq:1} or \eqref{35}, compactly expressed as $x^{\mu}\rightarrow x^{\mu}+\epsilon^af^{\mu}_{\hspace{1.5mm}a}(\mathbf{x})$, as below:
\begin{equation}
{j}^\mu_{\hspace{1.5mm}a}={T}^\mu_{\hspace{1.5mm}\nu} f^\nu_{\hspace{1.5mm}a}\hspace{5mm}\text{(off-shell)}\label{80}
\end{equation}
Our derivation of the corresponding Ward identities will heavily rely on this simple form of the Noether currents.

\medskip

\section{Ward identities in $1+2$D Carrollian CFTs}\label{s3}
We now turn our attention to the quantum theory (QFT) to derive the (source-less) $1+2$D Carrollian conformal Ward identities that are the QFT analogue of Noether's theorem. We closely follow the treatment presented in \cite{DiFrancesco:1997nk} for relativistic CFTs. 

\medskip

We shall show how the memory effects \cite{Strominger:2014pwa,Pasterski:2015tva,Strominger:2017zoo} corresponding to the leading \cite{Weinberg:1965nx,Strominger:2013jfa,He:2014laa} and sub-leading \cite{Cachazo:2014fwa,Kapec:2014opa} soft graviton theorems arise directly from these Ward identities. Along the way, we shall construct the leading and sub-leading soft graviton fields as well as the Virasoro EM tensor of the Celestial CFT \cite{Kapec:2016jld,Cheung:2016iub} purely in terms of the Carrollian EM tensor components. From such a construction, one can actually show that the Celestial (anti-)holomorphic Virasoro EM tensor is the 2D shadow transformation \cite{Kapec:2016jld} of the (positive)negative helicity sub-leading soft graviton field. 

\medskip

Finally, we reach the soft-factorization property of the bulk AFS mass-less $S$-matrices involving an outgoing soft graviton simply by performing a temporal Fourier transformation of the appropriate Carrollian Ward identities. This soft-factorization property along with the null four-momentum conservation for mass-less scattering will hint towards an identification of the temporal Fourier transformation of a special subset of the position-space Carrollian conformal primary correlators with the bulk AFS momentum-space correlators \cite{Donnay:2022wvx}. 

\medskip

In the path-integral formalism of QFT, correlation functions are the main objects of interest. A general (covariant time-ordered\footnote{As defined in section $6.1.4.$ of \cite{Itzykson:1980rh}, the covariant time-ordering commutes with space-time differentiation and integration.}) $n$-point correlator is defined as (suppressing the field tensor indices):
\begin{align}
\langle X\rangle\equiv\langle\hat{\mathcal{T}}\Phi_1(\mathbf{x_1})\Phi_2(\mathbf{x_2})...\Phi_n(\mathbf{x_n})\rangle:=\frac{\int[\mathcal{D}\bm{\Phi}]\text{ }\Phi_1(\mathbf{x_1})\Phi_2(\mathbf{x_2})...\Phi_n(\mathbf{x_n})\text{ }e^{iS[\bm{\Phi}]}}{\int[\mathcal{D}\bm{{\Phi}}]\text{ }e^{iS[\bm{{\Phi}}]}}
\end{align}
A field transformation, e.g. \eqref{eq:1}, will be a (local) symmetry of the QFT if, for any $X$:
\begin{align}
\langle X\rangle={\langle X\rangle}^\prime\equiv\langle\hat{\mathcal{T}}\tilde{\Phi}_1(\mathbf{x_1})\tilde{\Phi}_2(\mathbf{x_2})...\tilde{\Phi}_n(\mathbf{x_n})\rangle:=\frac{\int[\mathcal{D}\bm{\tilde{\Phi}}]\text{ }\tilde{\Phi}_1(\mathbf{x_1})\tilde{\Phi}_2(\mathbf{x_2})...\tilde{\Phi}_n(\mathbf{x_n})\text{ }e^{iS^\prime[\bm{\tilde{\Phi}}]}}{\int[\mathcal{D}\bm{\tilde{\Phi}}]\text{ }e^{iS^\prime[\bm{\tilde{\Phi}}]}}
\end{align} 
With the assumption that the path-integral measure is invariant: $[\mathcal{D}\bm{\Phi}]=[\mathcal{D}\bm{\tilde{\Phi}}]$, this symmetry condition leads to the Ward identity (at the 1st order in $\bm\epsilon$):
\begin{align}
-\delta_{\bm{\epsilon}}\langle X\rangle\equiv\sum_{i=1}^n\langle\hat{\mathcal{T}}\Phi_1(\mathbf{x_1}))...(i\epsilon^aG_a\Phi_i(\mathbf{x_i}))...\Phi_n(\mathbf{x_n})\rangle=i\int dt\text{ } d^{2}\vec{x}\text{ }\epsilon^a\langle\hat{\mathcal{T}}\partial_\mu j^\mu_{\hspace{1.5mm}a}(\mathbf{x})\left(X-\langle X\rangle\right)\rangle\label{9}
\end{align}

\medskip

Carrollian path-integral formalism was discussed in \cite{Chen:2023pqf}. For the $1+2$D Carrollian CFTs, the Ward identity takes the following differential form when $X$ is a string of Carrollian conformal primary fields, with (covariant) time-ordering implicit from now on:
\begin{align}
\partial_\mu\langle j^\mu_{\hspace{1.5mm}a}(\mathbf{x})(X-\langle X\rangle)\rangle=\sum\limits_{i=1}^n\text{ }\delta(t-t_i)\delta^2(\vec{x}-\vec{x_i}){{(G_a)}_i\langle X\rangle}
\end{align}
We emphasize that had $X$ been a string of arbitrary Carrollian conformal (non-primary) fields, spatial derivatives of arbitrary orders of the spatial delta-function would have also appeared on the R.H.S. of this Ward identity, just as in 2D CFT or in $1+1$D Carrollian CFT \cite{Saha:2022gjw}.

\medskip

Now, the space-time transformation currents must satisfy: $\partial_\mu\langle{j}^\mu_{\hspace{1.5mm}a}(\mathbf{x})\rangle=0$ since in the Carrollian CFT, these currents are expressed as: $\hat{j}_{\hspace{1.5mm}a}^\mu=\hat{T}_{\hspace{1.5mm}\nu}^\mu f^\nu_{\hspace{1.5mm}a}$. The generator for global translation is: $iG_\nu\Phi^a(\mathbf{x})=\partial_\nu\Phi^a(\mathbf{x})$, giving the translation Ward identity:
\begin{align}
\partial_\mu\langle T^\mu_{\hspace{1.5mm}\nu}(\mathbf{x})X\rangle=-i\sum\limits_{p=1}^n\text{ }{{\partial_{\nu_p}}\langle X\rangle}\text{ }\delta(t-t_p)\delta^2(\vec{x}-\vec{x}_p)\label{eq:2}
\end{align}
The Carrollian boost generators being $iG_i\Phi^a(\mathbf{x})=\left(x_i\partial_t+\bm{\xi_i}\right)\Phi^a(\mathbf{x})$ with $\bm{\xi}$ being the classical boost-matrix\footnote{The finite-dimensional boost-matrix $\mathbf{B}$($\bar{\mathbf{B}}$) from section \ref{s2.1} is denoted as $\bm{\xi}$($\bar{\bm{\xi}}$) in the rest of this work.} for the Carrollian multiplet field $\Phi^a(\mathbf{x})$, give rise to the boost Ward identity:
\begin{align}
\langle T^i_{\hspace{1.5mm}t}(\mathbf{x})X\rangle=-i\sum\limits_{p=1}^n\text{ }{(\bm{\xi_i})}_p{\langle X\rangle}\text{ }\delta(t-t_p)\delta^2(\vec{x}-\vec{x}_p)\label{eq:3}
\end{align}
Due to the dilation generator $iG\Phi^a(\mathbf{x})=(\Delta_a+x^\mu\partial_\mu)\Phi^a(\mathbf{x})$ where $\Delta_a$ is the scaling dimension of the field $\Phi^a(\mathbf{x})$, the dilation Ward identity takes the following form:
\begin{align}
\langle T^\mu_{\hspace{1.5mm}\mu}(\mathbf{x})X\rangle=-i\sum\limits_{p=1}^n\text{ }{\Delta}_p{\langle X\rangle}\text{ }\delta(t-t_p)\delta^2(\vec{x}-\vec{x}_p)\label{eq:4}
\end{align}
Finally, the spatial rotation generator $iG\Phi^a(\mathbf{x})=\left(x\partial_y-y\partial_x+i\bm{J}\right)\Phi^a(\mathbf{x})$ dictates the form of the rotation Ward identity:
\begin{align}
i\langle (T_{yx}-T_{xy})(\mathbf{x})X\rangle=-\langle T^{\bar{z}}_{\hspace{1.5mm}\bar{z}}(\mathbf{x})X\rangle+\langle T^z_{\hspace{1.5mm}z}(\mathbf{x})X\rangle=-i\sum\limits_{p=1}^n\text{ }m_p{\langle X\rangle}\text{ }\delta(t-t_p)\delta^2(\vec{x}-\vec{x}_p)\label{eq:5}
\end{align}
where $\{m\}$ are eigenvalues of the $\bm{J}$ operator. No new Ward identity is obtained for any of the three special Carrollian conformal transformations or for other super-translations (or super-rotations). These Ward identities can be readily generalized to higher dimensions by thinking of $X$ as a string of primary fields that transform covariantly under that higher-dimensional analog of the original BMS$_4$ group.

\medskip

Except the Carrollian boost Ward identity, all the others exactly match to those derived in \cite{Donnay:2022aba,Donnay:2022wvx} for the source-less case. Crucially in our case, the energy flux (boost) Ward identities $\langle T^i_{\hspace{1.5mm}t}X\rangle$ are not identically zero.    

\medskip

We emphasize that though the above Ward identities completely follow just from the $1+2$D Carrollian symmetry augmented by dilation, we needed the full BMS$_4$ invariance as well as the Weyl invariance to be able to write the Noether currents as \eqref{80} in the first place that facilitated this calculation. 

\medskip

We shall now show how, by manipulating these (source-less) Ward identities, the super-translation \cite{Strominger:2014pwa} and super-rotation \cite{Pasterski:2015tva} memory-effects, manifested by the presence of a temporal step-function \cite{Strominger:2017zoo}, can be obtained. The two corresponding soft graviton theorems \cite{Weinberg:1965nx,Cachazo:2014fwa} can then be reached simply via a temporal Fourier transformation \cite{Banerjee:2018fgd,Strominger:2014pwa,Pasterski:2015tva}. As a very significant byproduct, we shall get to build the soft graviton operators from the Carrollian EM tensor components in a completely boundary-theoretic set-up.  

\medskip

\subsection{Super-translation Ward identity}\label{s3.1}
Subtraction of the spatial divergence of \eqref{eq:3} from \eqref{eq:2}$_{\nu=t}$ leads to:
\begin{align}
\partial_t\langle T^t_{\hspace{1.5mm}t}(t,\vec{x})X\rangle=-i\sum\limits_{p=1}^n\text{ }\delta(t-t_p)\left[\delta^2(\vec{x}-\vec{x}_p)\partial_{t_p}-\left(\vec{\bm{\xi}}_p\cdot\bm{\nabla}\right)\delta^2(\vec{x}-\vec{x}_p)\right]\langle X\rangle\label{eq:6}
\end{align}
To solve this differential equation, an initial condition is needed. Upon choosing the following initial condition:
\begin{align}
\lim\limits_{t\rightarrow-\infty}\langle\hat{\mathcal{T}} T^t_{\hspace{1.5mm}t}(t,\vec{x})X\rangle=0\label{a11}
\end{align}
the solution turns out to be:
\begin{align}
\langle T^t_{\hspace{1.5mm}t}(t,\vec{x})X\rangle=-i\sum\limits_{p=1}^n\text{ }\theta(t-t_p)\left[\delta^2(\vec{x}-\vec{x}_p)\partial_{t_p}-\left(\vec{\bm{\xi}}_p\cdot\bm{\nabla}\right)\delta^2(\vec{x}-\vec{x}_p)\right]\langle X\rangle\label{7}
\end{align}
This temporal step-function captures the essence of the super-translation memory effect \cite{Strominger:2017zoo} because it reveals a DC shift between the (sufficiently) late and early time values of the energy-density correlator $\langle T^t_{\hspace{1.5mm}t}(t,\vec{x})X\rangle$.

\medskip

To convert this correlator into a non-contact expression (on $S^2$), we note the following useful representations of the 2D delta-function:
\begin{align}
\delta^2(\vec{x}-\vec{x}_p)=\frac{1}{\pi}\partial_z\frac{1}{\bar{z}-\bar{z}_p}=\frac{1}{\pi}\partial_{\bar{z}}\frac{1}{z-z_p}\label{a12}
\end{align}
Using these, we can express the R.H.S. of \eqref{7} as a $\partial$-derivative or a $\bar{\partial}$-derivative (with $\bm{\xi}=\bm{\xi}_x+i\bm{\xi}_y$ and $\bar{\bm{\xi}}=\bm{\xi}_x-i\bm{\xi}_y$):
\begin{align}
&\langle T^t_{\hspace{1.5mm}t}(t,\vec{x})X\rangle=-\frac{i}{\pi}\sum\limits_{p=1}^n\text{ }\theta(t-t_p)\text{ }\bar{\partial}\left[\frac{\partial_{t_p}}{z-z_p}+\frac{\bm{\xi}_p}{{(z-z_p)}^2}-\pi\bar{{\bm{\xi}}}_p\delta^2(\vec{x}-\vec{x}_p)\right]\langle X\rangle\label{a8}\\
&\langle T^t_{\hspace{1.5mm}t}(t,\vec{x})X\rangle=-\frac{i}{\pi}\sum\limits_{p=1}^n\text{ }\theta(t-t_p)\text{ }\partial\left[\frac{\partial_{t_p}}{\bar{z}-\bar{z}_p}+\frac{\bm{\bar{\xi}}_p}{{(\bar{z}-\bar{z}_p)}^2}-\pi\bm{\xi}_p\delta^2(\vec{x}-\vec{x}_p)\right]\langle X\rangle\label{a9}
\end{align}
This is the onset of the different treatments to time and space in the $1+2$D Carrollian CFT. The `DBAR' problem posed in \eqref{a8} has the following solution \cite{Ab}:
\begin{align}
\int\limits_{S^2} d^2{r}^\prime\frac{\langle T^t_{\hspace{1.5mm}t}(t,\vec{x}^\prime)X\rangle}{z-z^\prime}=&-i\sum\limits_{p=1}^n\text{ }\theta(t-t_p)\left[\left\{\frac{\partial_{t_p}}{z-z_p}+\frac{\bm{\xi}_p}{{(z-z_p)}^2}-\pi\bar{{\bm{\xi}}}_p\delta^2(\vec{x}-\vec{x}_p)\right\}\langle X\rangle\right.\nonumber\\
&\left.+(\text{a holomorphic function in $z$})\right]\label{a10}
\end{align}
Now, since the covariant time-ordering inside the correlator and spatial integration commute, taking inspiration from \cite{Strominger:2013jfa,He:2014laa}, we define the following operator:
\begin{align}
P(t,z,\bar{z}):=\int\limits_{S^2} d^2{r}^\prime\frac{T^t_{\hspace{1.5mm}t}(t,\vec{x}^\prime)}{z-z^\prime}\hspace{2.5mm}\Longrightarrow\hspace{2.5mm}{\partial_t}P=0\hspace{2.5mm},\hspace{2.5mm}\bar{\partial}P=\pi T^t_{\hspace{1.5mm}t}\label{21}
\end{align}
If we also demand that the $\langle P(t,\vec{x})X\rangle$ must be finite except at the positions of other operator insertions, then, by Liouville's theorem, the undetermined holomorphic function in \eqref{a10} must be a constant (that will be later shown to be zero). Thus we are led to the following Ward identity for the $P$ operator:
\begin{align}
&\langle P(t,z,\bar{z})X\rangle=-i\sum\limits_{p=1}^n\text{ }\theta(t-t_p)\left[\frac{\partial_{t_p}}{z-z_p}+\frac{\bm{\xi}_p}{{(z-z_p)}^2}-\pi\bar{{\bm{\xi}}}_p\delta^2(\vec{x}-\vec{x}_p)\right]\langle X\rangle\label{19}\\
\text{with\hspace{2.5mm}}&\langle \bar{\partial}P(t,z,\bar{z})X\rangle=-i\sum\limits_{p=1}^n\text{ }\theta(t-t_p)\left[\text{contact terms on $S^2$}\right]\nonumber
\end{align} 
Comparing the behavior of both sides under global dilation and rotation, we see that the field $P$ has scaling dimension $\Delta=2$ and spin $m=1$. This Ward identity reduces to the form \cite{Banerjee:2018fgd,Strominger:2013jfa} equivalent to Weinberg (positive-helicity) soft graviton theorem \cite{Weinberg:1965nx} when all the fields in $X$ have $\bm{\xi}=0=\bar{\bm{\xi}}$ i.e., they transform under the irrep of the spin-boost subgroup \eqref{eq:7}. 

\medskip

Since there is still a contact term in \eqref{a8}, we try to make it finite below by noticing that:
\begin{align}
\langle T^t_{\hspace{1.5mm}t}(t,\vec{x})X\rangle=-\frac{i}{\pi}\sum\limits_{p=1}^n\text{ }\theta(t-t_p)\text{ }{\bar{\partial}}^2\left[\frac{\bar{z}-\bar{z}_p}{z-z_p}\partial_{t_p}+\frac{\bar{z}-\bar{z}_p}{{(z-z_p)}^2}\bm{\xi}_p-\frac{\bar{{\bm{\xi}}}_p}{z-z_p}\right]\langle X\rangle
\end{align}
Inverting the $\bar{\partial}^2$ , we obtain:
\begin{align}
\int\limits_{S^2} d^2{r}^\prime\text{ }\frac{\bar{z}-\bar{z}^\prime}{z-z^\prime}\text{ }\langle T^t_{\hspace{1.5mm}t}(t,\vec{x}^\prime)X\rangle=&-i\sum\limits_{p=1}^n\text{ }\theta(t-t_p)\text{ }\left[\left\{\frac{\bar{z}-\bar{z}_p}{z-z_p}\partial_{t_p}+\frac{\bar{z}-\bar{z}_p}{{(z-z_p)}^2}\bm{\xi}_p-\frac{\bar{{\bm{\xi}}}_p}{z-z_p}\right\}\langle X\rangle\right.\nonumber\\
&\left.+\bar{z}A(z)-B(z)\right]
\end{align}
where $A(z)$ and $B(z)$ are two holomorphic functions (independent of time). So, we define a field $S_0^+$ as below:
\begin{align}
S_0^+(t,z,\bar{z}):=\int\limits_{S^2} d^2{r}^\prime\text{ }\frac{\bar{z}-\bar{z}^\prime}{z-z^\prime}\text{ }T^t_{\hspace{1.5mm}t}(t,\vec{x}^\prime)\hspace{2.5mm}\Longrightarrow\hspace{2.5mm}{\partial_t}S^+_0=0\hspace{2.5mm},\hspace{2.5mm}\bar{\partial}^2S_0^+=\bar{\partial}P=\pi T^t_{\hspace{1.5mm}t}\label{81}
\end{align}
whose Ward identity is expressed below in a suggestive form:
\begin{align}
\langle S_0^+(t,z,\bar{z})X\rangle=-i\sum\limits_{p=1}^n\text{ }\theta(t-t_p)&\left[\bar{z}\left\{\left(\frac{\partial_{t_p}}{z-z_p}+\frac{\bm{\xi}_p}{{(z-z_p)}^2}\right)\langle X\rangle+A(z)\right\}\right.\nonumber\\
&\left.\text{ }-\left(\frac{\bar{z}_p\partial_{t_p}+\bar{{\bm{\xi}}}_p}{z-z_p}+\frac{\bar{z}_p\bm{\xi}_p}{{(z-z_p)}^2}\right)\langle X\rangle-B(z)\right]\label{16}\\
\text{with\hspace{2.5mm}}\langle \bar{\partial}^2S^+_0(t,z,\bar{z})X\rangle=-i\sum\limits_{p=1}^n\text{ }&\theta(t-t_p)\left[\text{contact terms on $S^2$}\right]\nonumber
\end{align}
Thus, the field $P$ is an $\overline{SL(2,\mathbb{R})}$ descendant of the field $S_0^+$ with scaling dimension $\Delta=1$ and spin $m=2$.

\medskip

The term proportional to $\frac{\bar{z}-\bar{z}_p}{z-z_p}$ in \eqref{16} is singular only if $z$ and $\bar{z}$ are treated as independent variables \cite{Fotopoulos:2019tpe}; otherwise, it represents just a phase ambiguity. So, considering the form of the above Ward identity, inside the correlator we now re-express the field $S^+_0$, following \cite{Banerjee:2020zlg}, as:
\begin{align}
S_0^+(t,z,\bar{z})=\bar{z}P_{-1}(t,z,\bar{z})-P_0(t,z,\bar{z})
\end{align}
where the field $P_0$ has dimensions same as those of $S^+_0$; it satisfies the following Ward identity:
\begin{align*}
&\langle P_0(t,z,\bar{z})X\rangle=-i\sum\limits_{p=1}^n\text{ }\theta(t-t_p)\left[\left(\frac{\bar{z}_p\partial_{t_p}+\bar{{\bm{\xi}}}_p}{z-z_p}+\frac{\bar{z}_p\bm{\xi}_p}{{(z-z_p)}^2}\right)\langle X\rangle+B(z)\right]\\
\text{with\hspace{2.5mm}}&\langle \bar{\partial}P_0(t,z,\bar{z})X\rangle=-i\sum\limits_{p=1}^n\text{ }\theta(t-t_p)\left[\text{contact terms on $S^2$}\right]
\end{align*}
It is noteworthy that $\langle P_{-1}(t,z,\bar{z})X\rangle$ differs only by contact terms on $S^2$ from $\langle P(t,z,\bar{z})X\rangle$:
\begin{align*}
&\langle P_{-1}(t,z,\bar{z})X\rangle=-i\sum\limits_{p=1}^n\text{ }\theta(t-t_p)\left[\left(\frac{\partial_{t_p}}{z-z_p}+\frac{\bm{\xi}_p}{{(z-z_p)}^2}\right)\langle X\rangle+A(z)\right]\\
\text{with\hspace{2.5mm}}&\langle \bar{\partial}P_{-1}(t,z,\bar{z})X\rangle=-i\sum\limits_{p=1}^n\text{ }\theta(t-t_p)\left[\text{contact terms on $S^2$}\right]
\end{align*}
The dimensions of the fields $P_{-1}$ and $P$ are same. But we recall that, unlike $P$, the fields $P_{-1}$ and $P_0$ are defined only inside a correlator; we do not know about their classical counterparts.

\medskip

Again, we demand that the correlator $\langle S^+_0(t,z,\bar{z})X\rangle$ be finite everywhere (as a function of only $z$ since $\bar{z}$ is now an independent variable) except at the positions of insertions of other fields in $X$. Thus, Liouville's theorem restricts the holomorphic functions $A(z)$ and $B(z)$ to be merely constants. Moreover, since $S^+_0$ have positive holomorphic weight $h=\frac{3}{2}$, by demanding the finite-ness of the $\langle S^+_0(t,z,\bar{z})X\rangle$ correlator both\footnote{Being a primary field (as will be independently shown) $S^+_0$ transforms under $(z,\bar{z},t)\rightarrow\left(-\frac{1}{z},\bar{z},\frac{t}{z}\right)$ as \eqref{69}:\begin{align*}
\lim\limits_{z\rightarrow\infty} S^+_0(t,z,\bar{z})=\lim\limits_{z\rightarrow\infty}z^{-3} S^+_0(\frac{t}{z},-\frac{1}{z},\bar{z})
\end{align*}} at $z=\infty$ and $z=0$ keeping $\bar{z}$ fixed, we conclude that both of those constants must vanish. Thus, the super-translation Ward identities are given by (for $z\neq\{z_p\}$):
\begin{align}
&\langle P(t,z,\bar{z})X\rangle=\langle P_{-1}(t,z,\bar{z})X\rangle=-i\sum\limits_{p=1}^n\text{ }\theta(t-t_p)\left(\frac{\partial_{t_p}}{z-z_p}+\frac{\bm{\xi}_p}{{(z-z_p)}^2}\right)\langle X\rangle\\
&\langle P_0(t,z,\bar{z})X\rangle=-i\sum\limits_{p=1}^n\text{ }\theta(t-t_p)\left(\frac{\bar{z}_p\partial_{t_p}+\bar{{\bm{\xi}}}_p}{z-z_p}+\frac{\bar{z}_p\bm{\xi}_p}{{(z-z_p)}^2}\right)\langle X\rangle\\
&\langle S^+_0(t,z,\bar{z})X\rangle=-i\sum\limits_{p=1}^n\text{ }\theta(t-t_p)\text{ }\left\{\frac{\bar{z}-\bar{z}_p}{z-z_p}\partial_{t_p}+\frac{\bar{z}-\bar{z}_p}{{(z-z_p)}^2}\bm{\xi}_p-\frac{\bar{{\bm{\xi}}}_p}{z-z_p}\right\}\langle X\rangle\label{17}
\end{align}
From these Ward identities, we see that both $\langle \partial_t S^+_0(t,z,\bar{z})X\rangle$ and $\langle \partial_t P(t,z,\bar{z})X\rangle$ vanish upto temporal contact terms that also holds classically. 

\medskip

We recognize the Ward identity \eqref{17} to be describing the super-translation memory-effect \cite{Strominger:2014pwa} manifested by the presence of the temporal step-function \cite{Strominger:2017zoo}. With all the primaries having $\bm{\xi}=\bar{\bm{\xi}}=0$ i.e. transforming under the spin-boost irrep, it is the temporal Fourier transformed version \cite{Banerjee:2018fgd,He:2014laa} of Weinberg (positive-helicity) soft graviton theorem \cite{Weinberg:1965nx}. Thus, \eqref{81} provides a purely boundary-theoretic construction of the $1+3$D positive-helicity soft graviton operator in terms of the energy-density component of the $1+2$D Carrollian EM tensor, free of any metric components of the bulk AFS. The Carrollian construction of the super-translation generator of \cite{Strominger:2013jfa} is then given as \eqref{21}.    

\medskip

Similarly as \eqref{21} and \eqref{81}, the following fields associated with the negative-helicity soft graviton theorem can be constructed:
\begin{align}
{\partial}^2S_0^-={\partial}\bar{P}=\pi T^t_{\hspace{1.5mm}t}
\end{align}
where $\bar{P}$ has $(\Delta,m)=(2,-1)$ and $S_0^-$ has $(\Delta,m)=(1,-2)$ that satisfy the `complex conjugate' Ward identities.

\medskip

However, the two fields $S_0^\pm$ are the 2D shadow transformations of each other \cite{Pasterski:2017kqt} since:
\begin{align}
\bar{\partial}^2S_0^+=\pi T^t_{\hspace{1.5mm}t}={\partial}^2S_0^-\hspace{2.5mm}&\Longrightarrow\hspace{2.5mm}S_0^-(t,z,\bar{z})=\frac{2}{\pi}\int\limits_{S^2} d^2{r}^\prime\frac{z-z^\prime}{(\bar{z}-\bar{z}^\prime)^3}\text{ }S_0^+(t,z^\prime,\bar{z}^\prime)\\
&\Longrightarrow\hspace{2.5mm}S_0^+(t,z,\bar{z})=\frac{2}{\pi}\int\limits_{S^2} d^2{r}^\prime\frac{\bar{z}-\bar{z}^\prime}{({z}-{z}^\prime)^3}\text{ }S_0^-(t,z^\prime,\bar{z}^\prime)\nonumber
\end{align}
Since, a field and its shadow transformation can not both be local operators in a theory \cite{Banerjee:2022wht}, we shall later choose to treat $S^+_0$ as the local field in this work.

\medskip
  
Before moving on to the super-rotation Ward identities, a comment on the nature of the Carrollian energy-density $T^t_{\hspace{1.5mm}t}$ is in order. We saw that by construction it is a Lorentz descendant of both $S^\pm_0$. Moreover, its Ward identity \eqref{7} vanishes whenever the position of insertion of $T^t_{\hspace{1.5mm}t}$ does not coincide with that of any field in $X$ i.e. the correlator vanishes in the OPE limit. Besides, in section \ref{s5.3}, it will be evident that both $S_0^+$ and $T^t_{\hspace{1.5mm}t}$ are $1+2$D Carrollian conformal primary fields. All these properties together point to $T^t_{\hspace{1.5mm}t}$ being a primary-descendant or null-field (in the usual 2D CFT language), in agreement with \cite{Banerjee:2019aoy,Banerjee:2019tam}.

\medskip

\subsection{Super-rotation Ward identity}\label{s3.2}
We start by combining \eqref{eq:4}, \eqref{eq:5} and \eqref{7} into the following form (with $h=\frac{\Delta+m}{2}$ and $\bar{h}=\frac{\Delta-m}{2}$):
\begin{align*}
&\langle T^z_{\hspace{1.5mm}z}(\mathbf{x})X\rangle+\frac{1}{2}\langle T^t_{\hspace{1.5mm}t}(\mathbf{x})X\rangle=-i\sum\limits_{p=1}^n\text{ }{h}_p{\langle X\rangle}\text{ }\delta(t-t_p)\delta^2(\vec{x}-\vec{x}_p)\\
\Rightarrow\text{ }&\langle T^z_{\hspace{1.5mm}z}(\mathbf{x})X\rangle=-i\sum\limits_{p=1}^n\left[\delta(t-t_p)\delta^2(\vec{x}-\vec{x}_p)h_p-\frac{\theta(t-t_p)}{2}\left\{\delta^2(\vec{x}-\vec{x}_p)\partial_{t_p}-\left(\vec{\bm{\xi}}_p\cdot\bm{\nabla}\right)\delta^2(\vec{x}-\vec{x}_p)\right\}\right]{\langle X\rangle}
\end{align*}
Thus, subtraction of $\partial_z\langle T^z_{\hspace{1.5mm}z}(\mathbf{x})X\rangle$ from \eqref{eq:2}$_{\nu=z}$ results into:
\begin{align}
\partial_{\bar{z}}\langle T^{\bar{z}}_{\hspace{1.5mm}z}(\mathbf{x})X\rangle+\partial_t\langle T^t_{\hspace{1.5mm}z}(\mathbf{x})X\rangle=&-i\sum\limits_{p=1}^n\left[\delta(t-t_p)\left\{\delta^2(\vec{x}-\vec{x}_p)\partial_{z_p}-h_p\partial_{z}\delta^2(\vec{x}-\vec{x}_p)\right\}\right.\nonumber\\
&\left.+\frac{1}{2}\theta(t-t_p)\partial_z\left\{\delta^2(\vec{x}-\vec{x}_p)\partial_{t_p}-\left(\vec{\bm{\xi}}_p\cdot\bm{\nabla}\right)\delta^2(\vec{x}-\vec{x}_p)\right\}\right]{\langle X\rangle}
\end{align}
Solving the temporal part with an initial condition similar to \eqref{a11}, we get a hint of the super-rotation memory effect \cite{Pasterski:2015tva}:
\begin{align}
\langle T^t_{\hspace{1.5mm}z}(\mathbf{x})X\rangle+\int\limits_{t_0}^tdt^\prime\partial_{\bar{z}}\langle T^{\bar{z}}_{\hspace{1.5mm}z}(t^\prime,\vec{x})X\rangle=&-i\sum\limits_{p=1}^n\theta(t-t_p)\left[\delta^2(\vec{x}-\vec{x}_p)\partial_{z_p}-h_p\partial_{z}\delta^2(\vec{x}-\vec{x}_p)\right.\nonumber\\
&\left.+\frac{t-t_p}{2}\partial_z\left\{\delta^2(\vec{x}-\vec{x}_p)\partial_{t_p}-\left(\vec{\bm{\xi}}_p\cdot\bm{\nabla}\right)\delta^2(\vec{x}-\vec{x}_p)\right\}\right]{\langle X\rangle}\label{a13}
\end{align}
(where $t_0$ is an arbitrary reference time) since it shows the DC shift between the late and the early time values of the Carrollian momentum-density correlator $\langle T^t_{\hspace{1.5mm}z}X\rangle$ and Carrollian space-translations are actually rotations on the celestial sphere. 

\medskip

Choosing appropriate representations from \eqref{a12}, we re-express the above correlator in the following form involving derivatives of non-contact terms:
\begin{align}
\langle T^t_{\hspace{1.5mm}z}(\mathbf{x})X\rangle+\int\limits_{t_0}^tdt^\prime\partial_{\bar{z}}\langle T^{\bar{z}}_{\hspace{1.5mm}z}(t^\prime,\vec{x})X\rangle=&-\frac{i}{\pi}\sum\limits_{p=1}^n\text{ }\theta(t-t_p)\text{ }{\bar{\partial}}\left[\frac{h_p}{(z-z_p)^2}+\frac{\partial_{z_p}}{z-z_p}\right.\nonumber\\
&\left.-\frac{t-t_p}{2}\left\{\frac{\partial_{t_p}}{(z-z_p)^2}+\frac{2\bm{\xi}_p}{{(z-z_p)}^3}+\pi\bar{\bm{\xi}}_p\partial_z\delta^2(\vec{x}-\vec{x}_p)\right\}\right]\langle X\rangle\label{14}
\end{align}
The solution of this `DBAR' problem is given as \cite{Ab}:
\begin{align}
&i\int\limits_{S^2} d^2{r}^\prime\frac{\langle T^t_{\hspace{1.5mm}z}(t,\vec{x}^\prime)X\rangle}{z-z^\prime}+i\pi\int\limits_{t_0}^tdt^\prime\langle T^{\bar{z}}_{\hspace{1.5mm}z}(t^\prime,\vec{x})X\rangle\nonumber\\
=&\sum\limits_{p=1}^n\theta(t-t_p)\left[\frac{h_p}{(z-z_p)^2}+\frac{\partial_{z_p}}{z-z_p}-\frac{t-t_p}{2}\left\{\frac{\partial_{t_p}}{(z-z_p)^2}+\frac{2\bm{\xi}_p}{{(z-z_p)}^3}+\pi\bar{\bm{\xi}}_p\partial_z\delta^2(\vec{x}-\vec{x}_p)\right\}\right]\langle X\rangle\nonumber\\
&+(\text{a function holomorphic in $z$ and linear in $t$})\label{83}
\end{align}

\medskip

Then let us define an operator $T(t,z,\bar{z})$ as below:
\begin{align}
T(t,z,\bar{z}):=\int\limits_{S^2} d^2{r}^\prime\frac{T^t_{\hspace{1.5mm}z}(t,\vec{x}^\prime)}{z-z^\prime}+\pi\int\limits_{t_0}^tdt^\prime T^{\bar{z}}_{\hspace{1.5mm}z}(t^\prime,\vec{x})\hspace{2.5mm}\Longrightarrow\hspace{2.5mm}\bar{\partial}T=\pi T^t_{\hspace{1.5mm}z}+\pi\int\limits_{t_0}^tdt^\prime \partial_{\bar{z}} T^{\bar{z}}_{\hspace{1.5mm}z}\label{20}
\end{align}
that gives rise to the following Ward identity that is non-zero finite in the OPE limit (the undetermined function vanishes\footnote{It will be shown that the $T$ field is a Lorentz quasi-primary field and under the global $1+2$D CC transformations except the TSCT, it transforms as \eqref{82}.} as in the case of the super-translation Ward identity):
\begin{align}
&i\langle T(t,z,\bar{z})X\rangle\nonumber\\
=&\sum\limits_{p=1}^n\theta(t-t_p)\left[\frac{h_p}{(z-z_p)^2}+\frac{\partial_{z_p}}{z-z_p}-\frac{t-t_p}{2}\left\{\frac{\partial_{t_p}}{(z-z_p)^2}+\frac{2\bm{\xi}_p}{{(z-z_p)}^3}+\text{contact}\right\}\right]\langle X\rangle\label{18}\\
&\text{with\hspace{2.5mm}}\langle \bar{\partial} T(t,z,\bar{z})X\rangle=-i\sum\limits_{p=1}^n\theta(t-t_p)\left[\text{contact terms on $S^2$}\right]\nonumber
\end{align}
As will be shown, from this Ward identity with all $\bm{\xi}_p=0=\bar{\bm{\xi}}_p$, we can obtain the Virasoro Ward identity of the Celestial holography \cite{Kapec:2016jld,Cheung:2016iub} that is equivalent to the Cachazo-Strominger subleading soft-graviton theorem \cite{Cachazo:2014fwa}. The field $T(t,z,\bar{z})$ has $(\Delta,m)=(2,2)$.

\medskip

Comparing \eqref{18} with the super-translation Ward identity \eqref{19}, one finds that:
\begin{align}
\langle \partial_t T(t,z,\bar{z})X\rangle-\frac{1}{2}\langle \partial_z P(t,z,\bar{z})X\rangle=0+\text{temporal contact terms}
\end{align}
It needs to be emphasized that the relation $\partial_t T= \frac{1}{2}\partial_z P$ does not automatically hold classically, but only within a correlator. Classically, we can only say that:
\begin{align*}
\partial_{\bar{z}}\left(\partial_t T-\frac{1}{2}\partial_z P\right)=0
\end{align*}
following from the defining relations \eqref{21} and \eqref{20}. So, we define a quantum field $T_e(t,z,\bar{z})$ (with $h=2,\bar{h}=0$) as:
\begin{align}
T(t,z,\bar{z})=\frac{t}{2}\partial P(z,\bar{z})+T_e(t,z,\bar{z})\text{\hspace{2.5mm}with\hspace{2.5mm}}\langle{\partial}_t T_{e}(t,z,\bar{z}) X\rangle=\text{temporal contact terms}\label{24}
\end{align}
Its Ward identity is extracted to be:
\begin{align}
&\langle T_e(t,z,\bar{z})X\rangle=-i\sum\limits_{p=1}^n\theta(t-t_p)\left[\frac{h_p+\frac{t_p}{2}\partial_{t_p}}{(z-z_p)^2}+\frac{\partial_{z_p}}{z-z_p}+\frac{t_p\bm{\xi}_p}{{(z-z_p)}^3}\right]\langle X\rangle\label{39}\\
&\text{with\hspace{2.5mm}}\langle\bar{\partial} T_e(t,z,\bar{z}) X\rangle=-\frac{i}{2}\sum\limits_{p=1}^n\theta(t-t_p)\left[\text{contact terms on $S^2$}\right]\text{\hspace{7.5mm}(Classically, $\bar{\partial} T_e\neq0$)}\nonumber
\end{align}
which has exactly the same form to that of the 2D CFT holomorphic EM tensor (Virasoro) Ward identity if we define $h_p^\prime:=h_p+\frac{t_p}{2}\partial_{t_p}$, following \cite{Kapec:2014opa,Kapec:2016jld,Cheung:2016iub,Banerjee:2020kaa}. Hence, this $T_e$ field is the Virasoro stress-tensor of the Celestial CFT.

\medskip

It is to be noted that the $\langle T_e(t,z,\bar{z})X\rangle$ correlator does not have any contact term on $S^2$. Clearly, only those Carrollian conformal primary fields having $\bm{\xi}=0$ can be the primary fields of this Virasoro-like symmetry.

\medskip

Thus \eqref{20}, together with \eqref{21}, provides a purely boundary-theoretic construction of the 2D Celestial EM tensor \cite{Kapec:2016jld,Cheung:2016iub} in terms of the Carrollian EM tensor components that does not involve a shadow transformation of the sub-leading (energetically) soft negative helicity graviton \cite{Kapec:2014opa}. Below, we shall build the sub-leading (conformally \cite{Donnay:2018neh,Puhm:2019zbl}) soft graviton operator $S^-_1$ from the same Carrollian EM tensor components and show that the fields $T$ and $S^-_1$ are indeed 2D shadow transformations of each other.   

\medskip

Since we have already figured out that the $S^2$ contact-term in \eqref{83} is coming from the $\partial$-derivative of the contact term in \eqref{19}, we do not further need\footnote{We may define a $(\Delta=1,m=3)$ field $\tilde{S}^+$ such that $\bar{\partial}\tilde{S}^+=T$ ; but, following the arguments in section \ref{s5.3}, it can be shown that such a field $\tilde{S}^+$ can not be a non-descendant local Lorentz quasi-primary besides preventing $S^+_0$ from being one.} to try to make it non-contact. Instead, let us extract $\partial$-derivatives from \eqref{a13} as:
\begin{align}
&\langle T^t_{\hspace{1.5mm}z}(\mathbf{x})X\rangle+\int\limits_{t_0}^tdt^\prime\partial_{\bar{z}}\langle T^{\bar{z}}_{\hspace{1.5mm}z}(t^\prime,\vec{x})X\rangle\nonumber\\
=&-\frac{i}{\pi}\sum\limits_{p=1}^n\theta(t-t_p)\partial^2\left[\frac{z-z_p}{\bar{z}-\bar{z}_p}\partial_{z_p}-\frac{h_p}{\bar{z}-\bar{z}_p}+\frac{t-t_p}{2}\left\{\frac{\partial_{t_p}}{\bar{z}-\bar{z}_p}+\frac{\bar{\bm{\xi}}_p}{(\bar{z}-\bar{z}_p)^2}-\pi{\bm{\xi}}_p\delta^2(\vec{x}-\vec{x}_p)\right\}\right]{\langle X\rangle}\nonumber\\
=&-\frac{i}{2\pi}\sum\limits_{p=1}^n\theta(t-t_p)\partial^3\left[\frac{(z-z_p)^2}{\bar{z}-\bar{z}_p}\partial_{z_p}-2h_p\frac{(z-z_p)}{\bar{z}-\bar{z}_p}\right.\nonumber\\
&\hspace{53.5mm}\left.+(t-t_p)\left\{\frac{z-z_p}{\bar{z}-\bar{z}_p}\partial_{t_p}+\frac{z-z_p}{(\bar{z}-\bar{z}_p)^2}\bar{\bm{\xi}}_p-\frac{{\bm{\xi}}_p}{\bar{z}-\bar{z}_p}\right\}\right]{\langle X\rangle}
\end{align}
We could have stopped at the second line recognizing that the contact term has originated from the $\bar{P}$ Ward identity. But doing so will prevent us from reaching the sub-leading soft graviton theorem \cite{Cachazo:2014fwa}; the reason is as follows.

\medskip

The soft-factor of the bulk AFS scattering amplitude is actually a (Laurent) series expansion starting at the simple pole order in energy $\omega$ near the soft limit $\omega\rightarrow0$ \cite{Cachazo:2014fwa}; it is the simple pole term which is Weinberg (leading) soft graviton theorem \cite{Weinberg:1965nx} and the $\omega^0$ term is the Cachazo-Strominger sub-leading soft graviton theorem \cite{Cachazo:2014fwa}. Here, we have similar results if we note that the temporal Fourier transformations of $\theta(t-t_p)$ and $(t-t_p)\theta(t-t_p)$ give rise to $\frac{1}{\omega}$ and $\frac{1}{\omega^2}$ poles respectively. On both sides, performing the temporal Fourier transformation, multiplying with $\omega$ and taking the soft limit successively do not give rise to Weinberg soft graviton theorem \cite{Weinberg:1965nx,He:2014laa} properly if we stop at the second line. But extracting a $\partial^3$-derivative leads to all the desired features because the coefficient of $(t-t_p)\theta(t-t_p)$ in the third line is the leading soft-graviton Ward identity \eqref{17}. Moreover, the other terms are then identified with the sub-leading (negative-helicity) soft-graviton theorem as expressed in \cite{Cachazo:2014fwa,Kapec:2014opa}.    

\medskip
  
Now we invert the $\partial^3$ to obtain:
\begin{align}
&i\int\limits_{S^2} d^2{r}^\prime\text{ }\frac{{(z-z^\prime)}^2}{\bar{z}-\bar{z}^\prime}\left[\langle T^t_{\hspace{1.5mm}z}(t,\vec{x}^\prime)X\rangle+\int\limits_{t_0}^tdt^\prime\partial_{\bar{z}^\prime}\langle T^{\bar{z}}_{\hspace{1.5mm}z}(t^\prime,\vec{x}^\prime)X\rangle\right]\nonumber\\
=&\sum\limits_{p=1}^n\theta(t-t_p)\left[\left\{\frac{(z-z_p)^2}{\bar{z}-\bar{z}_p}\partial_{z_p}-2h_p\frac{z-z_p}{\bar{z}-\bar{z}_p}+(t-t_p)\left(\frac{z-z_p}{\bar{z}-\bar{z}_p}\partial_{t_p}+\frac{z-z_p}{(\bar{z}-\bar{z}_p)^2}\bar{\bm{\xi}}_p-\frac{{\bm{\xi}}_p}{\bar{z}-\bar{z}_p}\right)\right\}{\langle X\rangle}\right.\nonumber\\
&\left.\hspace{21mm} z^2C(t,\bar{z})-2zD(t,\bar{z})+E(t,\bar{z})\right]\label{84}
\end{align}
where the functions $C(t,\bar{z})$, $D(t,\bar{z})$ and $E(t,\bar{z})$ are anti-holomorphic in $\bar{z}$ and linear in $t$. So, we define a field $S_1^-(t,z,\bar{z})$ as:
\begin{align}
&S_1^-(t,z,\bar{z})=\frac{1}{2}\int\limits_{S^2} d^2{r}^\prime\text{ }\frac{{(z-z^\prime)}^2}{\bar{z}-\bar{z}^\prime}\left[T^t_{\hspace{1.5mm}z}(t,\vec{x}^\prime)+\int\limits_{t_0}^tdt^\prime\partial_{\bar{z}^\prime}T^{\bar{z}}_{\hspace{1.5mm}z}(t^\prime,\vec{x}^\prime)\right]\label{85}\\
\Longrightarrow\hspace{2.5mm}&\partial^3S_1^-=\pi T^t_{\hspace{1.5mm}z}+\pi\int\limits_{t_0}^tdt^\prime \partial_{\bar{z}} T^{\bar{z}}_{\hspace{1.5mm}z}=\bar{\partial}T\label{15}
\end{align}
whose Ward identity is expressed in the following suggestive form:
\begin{align}
\langle S_1^-(t,z,\bar{z}) X\rangle&=-\frac{i}{2}\sum\limits_{p=1}^n\theta(t-t_p)\left[z^2\left\{\frac{\partial_{z_p}}{\bar{z}-\bar{z}_p}\langle X\rangle+C(t,\bar{z})\right\}\right.\nonumber\\
&\left.-2z\left\{\frac{z_p\partial_{z_p}+h_p}{\bar{z}-\bar{z}_p}\langle X\rangle-\frac{t-t_p}{2}\left(\frac{\partial_{t_p}}{\bar{z}-\bar{z}_p}+\frac{\bar{\bm{\xi}}_p}{(\bar{z}-\bar{z}_p)^2}\right)\langle X\rangle+D(t,\bar{z})\right\}\right.\nonumber\\
&\left.+\left\{\frac{z_p^2\partial_{z_p}+2z_ph_p}{\bar{z}-\bar{z}_p}\langle X\rangle-(t-t_p)\left(\frac{z_p\partial_{t_p}+{\bm{\xi}}_p}{\bar{z}-\bar{z}_p}+\frac{z_p\bar{\bm{\xi}}_p}{(\bar{z}-\bar{z}_p)^2}\right)\langle X\rangle+E(t,\bar{z})\right\}\right]\nonumber\\
\text{with\hspace{2.5mm}}&\langle \partial^3 S_1^-(t,z,\bar{z}) X\rangle=-i\sum\limits_{p=1}^n\theta(t-t_p)\left[\text{contact terms on $S^2$}\right]\nonumber
\end{align}
The dimensions of the $S_1^-$ field are $(\Delta,m)=(0,-2)$. 

\medskip

The relation \eqref{15} implies that the $T$ field is automatically the 2D shadow transformation of $S_1^-$ since:
\begin{align}
\bar{\partial}T=\partial^3S_1^-\hspace{2.5mm}\Longrightarrow\hspace{2.5mm}T(t,z,\bar{z})=-\frac{3!}{\pi}\int\limits_{S^2} d^2{r}^\prime\frac{S_1^-(t,z^\prime,\bar{z}^\prime)}{(z-z^\prime)^4}\label{49}
\end{align}
So, only one of $T$ or $S^-_1$ can be treated as a local field in a theory \cite{Banerjee:2022wht}.

\medskip

Obviously, there is an anti-holomorphic version of the super-rotation Ward identities derived above. We can similarly construct a field $\bar{T}(t,z,\bar{z})$ as:
\begin{align}
\bar{T}(t,z,\bar{z}):=\int\limits_{S^2} d^2{r}^\prime\frac{T^t_{\hspace{1.5mm}\bar{z}}(t,\vec{x}^\prime)}{\bar{z}-\bar{z}^\prime}+\pi\int\limits_{t_0}^tdt^\prime T^{{z}}_{\hspace{1.5mm}\bar{z}}(t^\prime,\vec{x})\hspace{2.5mm}\Longrightarrow\hspace{2.5mm}{\partial}\bar{T}=\pi T^t_{\hspace{1.5mm}\bar{z}}+\pi\int\limits_{t_0}^tdt^\prime \partial_{{z}} T^{{z}}_{\hspace{1.5mm}\bar{z}}\label{86}
\end{align}
Its Ward identity is given below:
\begin{align}
&i\langle \bar{T}(t,z,\bar{z})X\rangle\nonumber\\
=&\sum\limits_{p=1}^n\theta(t-t_p)\left[\frac{\bar{h}_p}{(\bar{z}-\bar{z}_p)^2}+\frac{\partial_{\bar{z}_p}}{\bar{z}-\bar{z}_p}-\frac{t-t_p}{2}\left\{\frac{\partial_{t_p}}{(\bar{z}-\bar{z}_p)^2}+\frac{2\bar{\bm{{\xi}}}_p}{{(\bar{z}-\bar{z}_p)}^3}+\pi{\bm{\xi}}_p\partial_{\bar{z}}\delta^2(\vec{x}-\vec{x}_p)\right\}\right]\langle X\rangle\nonumber\\
&\text{with\hspace{2.5mm}}\langle {\partial}\bar{T}(t,z,\bar{z})X\rangle=-i\sum\limits_{p=1}^n\theta(t-t_p)\left[\text{contact terms on $S^2$}\right]
\end{align}
which has the exactly opposite pole-structure to that of \eqref{18}. The dimensions of the $\bar{T}$ field are given by $(\Delta,m)=(2,-2)$.

\medskip

Similarly, the anti-holomorphic counterpart $S^+_1$ of the field $S^-_1$ is defined as:
\begin{align}
&S_1^+(t,z,\bar{z})=\frac{1}{2}\int\limits_{S^2} d^2{r}^\prime\text{ }\frac{{(\bar{z}-\bar{z}^\prime)}^2}{{z}-{z}^\prime}\left[T^t_{\hspace{1.5mm}\bar{z}}(t,\vec{x}^\prime)+\int\limits_{t_0}^tdt^\prime\partial_{{z}^\prime} T^{{z}}_{\hspace{1.5mm}\bar{z}}(t^\prime,\vec{x}^\prime)\right]\label{22}\\
\Longrightarrow\hspace{2.5mm}&\bar{\partial}^3S_1^+=\pi T^t_{\hspace{1.5mm}\bar{z}}+\pi\int\limits_{t_0}^tdt^\prime \partial_{{z}} T^{{z}}_{\hspace{1.5mm}\bar{z}}={\partial}\bar{T}\label{87}
\end{align}
The last relation says that $S^+_1$ is the 2D shadow-transformation of $\bar{T}$. Its Ward identity is similarly derived to be:
\begin{align}
\langle S_1^+(t,z,\bar{z}) X\rangle&=-\frac{i}{2}\sum\limits_{p=1}^n\theta(t-t_p)\left[\bar{z}^2\left\{\frac{\partial_{\bar{z}_p}}{{z}-{z}_p}\langle X\rangle+C_1(t,{z})\right\}\right.\nonumber\\
&\left.-2\bar{z}\left\{\frac{\bar{z}_p\partial_{\bar{z}_p}+\bar{h}_p}{{z}-{z}_p}\langle X\rangle-\frac{t-t_p}{2}\left(\frac{\partial_{t_p}}{{z}-{z}_p}+\frac{{\bm{\xi}}_p}{({z}-{z}_p)^2}\right)\langle X\rangle+D_1(t,{z})\right\}\right.\nonumber\\
&\left.+\left\{\frac{\bar{z}_p^2\partial_{\bar{z}_p}+2\bar{z}_p\bar{h}_p}{{z}-{z}_p}\langle X\rangle-(t-t_p)\left(\frac{\bar{z}_p\partial_{t_p}+\bar{{\bm{\xi}}}_p}{{z}-{z}_p}+\frac{\bar{z}_p{\bm{\xi}}_p}{({z}-{z}_p)^2}\right)\langle X\rangle+E_1(t,{z})\right\}\right]\nonumber\\
\text{with\hspace{2.5mm}}&\langle \bar{\partial}^3 S_1^+(t,z,\bar{z}) X\rangle=-i\sum\limits_{p=1}^n\theta(t-t_p)\left[\text{contact terms on $S^2$}\right]\label{23}
\end{align}
with $C_1(t,z)$, $D_1(t,z)$ and $E_1(t,z)$ being holomorphic functions of $z$ and linear in $t$. The dimensions of the $S^+_1$ field are $(\Delta,m)=(0,2)$.

\medskip

We note that all the holomorphic poles in the $S^+_1$ Ward identity \eqref{23} are true singularities (in the coincident-position limit) only if $z$ and $\bar{z}$ are treated as independent variables since this Ward identity can be expressed analogously to \eqref{84}. So, following \cite{Banerjee:2020zlg,Polyakov:1987zb}, we re-express the field $S_1^+$ inside a correlator as below:
\begin{align}
S_1^+(t,z,\bar{z})=j^{(+)}(t,z,\bar{z})-2\bar{z}j^{(0)}(t,z,\bar{z})+\bar{z}^2j^{(-)}(t,z,\bar{z})
\end{align}
As we shall see, this will make the interpretation of the $S^+_1$ Ward identity clearer. The Ward identities satisfied by the $j^a$ fields are obvious from \eqref{23}. Like $S^+_1$, each of these fields has holomorphic weight $h=1$.

\medskip

As usual, we demand that the correlators $\langle S^+_1(t,z,\bar{z}) X\rangle$ be finite everywhere except at $z=\{z_p\}$. Finite-ness of these correlators at $z=\infty$ renders the functions $C_1,D_1,E_1$ independent of $z$, by Liouville's theorem. Moreover, since the $S^+_1$ field has positive holomorphic weight and, as will be shown, it transforms as \eqref{69} under the $\text{ISL}(2,\mathbb{C})$ group, by considering the finite-ness of these correlators both at $z=\infty$ and $z=0$ keeping $\bar{z}$ fixed, it can be shown that each of $C_1,D_1,E_1$ vanishes. Thus, the Ward identities for the $j^a$ fields are finally reduced to:
\begin{align}
&\langle j^{(-)}(t,z,\bar{z}) X\rangle=-\frac{i}{2}\sum\limits_{p=1}^n\theta(t-t_p)\frac{\partial_{\bar{z}_p}}{{z}-{z}_p}\langle X\rangle\\
&\langle j^{(0)}(t,z,\bar{z}) X\rangle=-\frac{i}{2}\sum\limits_{p=1}^n\theta(t-t_p)\left\{\frac{\bar{z}_p\partial_{\bar{z}_p}+\bar{h}_p}{{z}-{z}_p}-\frac{t-t_p}{2}\left(\frac{\partial_{t_p}}{{z}-{z}_p}+\frac{{\bm{\xi}}_p}{({z}-{z}_p)^2}\right)\right\}\langle X\rangle\\
&\langle j^{(+)}(t,z,\bar{z}) X\rangle=-\frac{i}{2}\sum\limits_{p=1}^n\theta(t-t_p)\left\{\frac{\bar{z}_p^2\partial_{\bar{z}_p}+2\bar{z}_p\bar{h}_p}{{z}-{z}_p}-(t-t_p)\left(\frac{\bar{z}_p\partial_{t_p}+\bar{{\bm{\xi}}}_p}{{z}-{z}_p}+\frac{\bar{z}_p{\bm{\xi}}_p}{({z}-{z}_p)^2}\right)\right\}\langle X\rangle\\
&\text{with\hspace{2.5mm}}\langle\bar{\partial} j^{a}(t,z,\bar{z}) X\rangle=-\frac{i}{2}\sum\limits_{p=1}^n\theta(t-t_p)\left[\text{contact terms on $S^2$}\right]\nonumber
\end{align}
We emphasize that the last relation holds not classically but only as a correlator statement. 

\medskip

The Ward identity for $S^+_1$ follows obviously from the above that is readily identified with the sub-leading conformally soft graviton theorem \cite{Banerjee:2020zlg} when all the fields in $X$ transform under the spin-boost irrep. Thus, we provide the purely boundary-theoretic construction of the sub-leading conformally soft \cite{Donnay:2018neh,Puhm:2019zbl} graviton operators in terms of the Carrollian EM tensor components by \eqref{22} for positive-helicity and by \eqref{85} for negative-helicity. The anti-holomorphic counterpart $\bar{T}_e$ of the 2D Celestial CFT stress-tensor $T_e$ can be obtained similarly as \eqref{24} from \eqref{86}, providing an entirely boundary-theoretic construction of the same.

\medskip

We now observe that:
\begin{align}
\langle\partial_t S_1^+(t,z,\bar{z}) X\rangle-\frac{1}{2}\langle S_0^+(t,z,\bar{z}) X\rangle=0+(\text{temporal contact terms})
\end{align} 
This is valid only at the level of correlators, since classically, we could only conclude that:
\begin{align*}
\bar{\partial}^3\left(\partial_tS_1^+-\frac{1}{2}S_0^+\right)=0
\end{align*}
So, we define a quantum field $S_{1e}^+$ such that:
\begin{align*}
S^+_1(t,z,\bar{z})=\frac{t}{2}S^+_0(z,\bar{z})+S^+_{1e}(t,z,\bar{z})\text{\hspace{2.5mm}with\hspace{2.5mm}}\langle{\partial}_t S_{1e}^+(t,z,\bar{z}) X\rangle=\text{temporal contact terms}
\end{align*}
The corresponding $j^a_e$ fields are then obviously defined as:
\begin{align}
j^{(-)}=j^{(-)}_e\hspace{2.5mm};\hspace{2.5mm}j^{(0)}=-\frac{t}{4}P_{-1}+j^{(0)}_e\hspace{2.5mm};\hspace{2.5mm}j^{(+)}=-\frac{t}{2}P_{0}+j^{(+)}_e
\end{align}
The Ward identities of the $j^a_e$ fields (with holomorphic dimension $h=1$) then take exactly same forms with those generated by 2D CFT holomorphic Kac-Moody currents:
\begin{align}
&\langle j^{(-)}_e(t,z,\bar{z}) X\rangle=-\frac{i}{2}\sum\limits_{p=1}^n\theta(t-t_p)\frac{\partial_{\bar{z}_p}}{{z}-{z}_p}\langle X\rangle\nonumber\\
&\langle j^{(0)}_e(t,z,\bar{z}) X\rangle=-\frac{i}{2}\sum\limits_{p=1}^n\theta(t-t_p)\left\{\frac{\bar{z}_p\partial_{\bar{z}_p}+\frac{t_p}{2}\partial_{t_p}+\bar{h}_p}{{z}-{z}_p}+\frac{\frac{1}{2}t_p{\bm{\xi}}_p}{({z}-{z}_p)^2}\right\}\langle X\rangle\\
&\langle j^{(+)}_e(t,z,\bar{z}) X\rangle=-\frac{i}{2}\sum\limits_{p=1}^n\theta(t-t_p)\left\{\frac{\bar{z}_p^2\partial_{\bar{z}_p}+\bar{z}_pt_p\partial_{t_p}+2\bar{z}_p\bar{h}_p+t_p\bar{{\bm{\xi}}}_p}{{z}-{z}_p}+\frac{\bar{z}_pt_p{\bm{\xi}}_p}{({z}-{z}_p)^2}\right\}\langle X\rangle\\
&\text{with\hspace{2.5mm}}\langle\bar{\partial} j^{a}_e(t,z,\bar{z}) X\rangle=-\frac{i}{2}\sum\limits_{p=1}^n\theta(t-t_p)\left[\text{contact terms on $S^2$}\right]\nonumber
\end{align}
Clearly, only those fields that have $\bm{\xi}=0$ can be the primary fields of this Kac-Moody-like symmetry. It will be confirmed in section \ref{s5.3} that the $j^a_e$ fields indeed generate (and not merely resemble) an $\overline{\text{sl}(2,\mathbb{R})}$ Kac-Moody symmetry. 

\medskip

The Ward identity of the $S^+_{1e}$ field is now readily obtained that resembles the temporal Fourier transformed version of the Cachazo-Strominger sub-leading (positive-helicity, energetically) soft graviton theorem \cite{Cachazo:2014fwa,Kapec:2014opa} upon defining $h_p^\prime:=h_p+\frac{t_p}{2}\partial_{t_p}$, when all the primaries in $X$ have $\bm{\xi}=\bar{\bm{\xi}}=0$. The $S^\pm_{1e}$ Ward identities then obviously describe the super-rotation memory effects \cite{Pasterski:2015tva} due to the presence of the temporal step-function \cite{Strominger:2017zoo}. Thus $S^\pm_{1e}$ are the Carrollian constructions of the sub-leading energetically soft graviton fields.

\medskip

As an aside, we note that the Ward identities of the fields $P_0$ and $P_{-1}$ also resemble Kac-Moody Ward identities. But, since both of these fields have $h=\frac{3}{2}$, they can not be interpreted as possible Kac-Moody generators. 

\medskip

We emphasize that all the fields in $X$ were already primary fields of Carrollian CFT; so not all of the Carrollian conformal primary fields are the primaries of the above-mentioned Virasoro or Kac-Moody symmetries.

\medskip

Similar to a $T^t_{\hspace{1.5mm}t}$ insertion in a Carrollian conformal correlator, any insertion of the following two operators:
\begin{align*}
T^t_{\hspace{1.5mm}z}+\int\limits_{t_0}^tdt^\prime \partial_{\bar{z}} T^{\bar{z}}_{\hspace{1.5mm}z}\hspace{7.5mm}\text{ and }\hspace{7.5mm}T^t_{\hspace{1.5mm}\bar{z}}+\int\limits_{t_0}^tdt^\prime \partial_{{z}} T^{{z}}_{\hspace{1.5mm}\bar{z}}
\end{align*}
vanishes in the OPE limit, as seen from the Ward identity \eqref{a13} and its conjugate. Both of these are also global descendants of Lorentz quasi-primaries, as seen respectively from \eqref{15} and \eqref{87}. Moreover, from the discussion in section \ref{s5.3}, it will be clear that both of these are actually $1+2$D Carrollian conformal primary fields. Hence, both of these fields are primary-descendants or null-fields \cite{Banerjee:2019aoy,Banerjee:2019tam}, just like $T^t_{\hspace{1.5mm}t}$.

\medskip

Now we provide an explicit derivation of the leading and sub-leading energetically soft graviton theorems \cite{Weinberg:1965nx,Cachazo:2014fwa} simply by performing a temporal Fourier transformation \cite{Strominger:2014pwa,Pasterski:2015tva,Strominger:2017zoo} of the conformally soft graviton $S^\pm_1$ Ward identities \cite{Banerjee:2020zlg}.

\medskip

\subsection{The soft graviton theorems}\label{s3.3}
We begin by recalling two facts: that soft(graviton)-factorization of a bulk AFS mass-less scattering amplitude occurs when an outgoing (external) graviton goes soft and that in the relativistic LSZ formula which connects the (bulk) $S$-matrix elements with (bulk, time-ordered) position-space correlators, opposite Fourier transformation (to go from the position- to the momentum-space) convention is used. 

\medskip

Now, as stated in the introduction, the Carrollian CFT lives only on one of $\mathcal{I}^\pm$ and not on the other, following \cite{Banerjee:2018fgd,Banerjee:2018gce}. Moreover, in this work, there is no notion of Carrollian incoming/outgoing particles because we are not interested in Carrollian momentum-space physics of Carrollian scattering. Instead, it will be evident below that the temporal Fourier transformation of a Carrollian position-space primary field $\Phi(t,z,\bar{z})$ with $\Delta=1$ that transforms under a spin-boost irrep can describe an external bulk AFS mass-less particle with either initial or final (bulk) null four-momentum completely characterized by $(\omega,z,\bar{z})$. Whether it describes an incoming or an outgoing bulk AFS mass-less particle is decided by the temporal Fourier transformation factor to go from the (Carrollian time) $t$-space to the (initial/final energy of the bulk mass-less particle) $\omega$-space \cite{Donnay:2022wvx}. This is similar to the LSZ formula scenario (see also \cite{Bekaert:2022ipg,Liu:2022mne}). Thus, a single position-space Carrollian field can correspond to either an incoming or the same outgoing bulk AFS mass-less particle. Below we fix the convention (with $\omega\geq0$):
\begin{align}
\tilde{\Phi}_{\text{out}}(\omega,z,\bar{z})=\frac{1}{2\pi}\int\limits_{-\infty}^{\infty}dt \text{ }e^{-i\omega t}{\Phi}(t,z,\bar{z})\hspace{5mm}\text{ and }\hspace{5mm}\tilde{\Phi}_{\text{in}}(\omega,z,\bar{z})=\frac{1}{2\pi}\int\limits_{-\infty}^{\infty}dt \text{ }e^{i\omega t}{\Phi}(t,z,\bar{z})\label{88}
\end{align}

\medskip

We first check below that this prescription reproduces the correct global energy conservation law for bulk mass-less scattering from the Carrollian time-translation invariance. Due to the global Carrollian time-translation symmetry of the action, any Carrollian correlator should be invariant under this transformation, i.e.:
\begin{align*}
-{\delta}\langle X\rangle:=\sum_{p=1}^n\partial_{t_p}\left\langle\hat{\mathcal{T}}\Phi_1(t_1,z_1,\bar{z}_1)\Phi_2(t_2,z_2,\bar{z}_2)\ldots\Phi_n(t_n,z_n,\bar{z}_n)\right\rangle=0
\end{align*} 
Now, let $k$ out of these $n$ Carrollian fields represent bulk outgoing mass-less particles and the rest represent bulk incoming mass-less particles. So, using the inverse of the prescription \eqref{88}, we readily discover the desired conservation law:
\begin{align}
&\sum_{p=1}^n\partial_{t_p}\left\langle\hat{\mathcal{T}}\int\limits_{0}^{\infty}d\omega_1 \text{ }e^{i\omega_1 t_1}\tilde{\Phi}_{\text{(out)}1}(\omega_1,z_1,\bar{z}_1)\ldots\int\limits_{0}^{\infty}d\omega_k\text{ }e^{i\omega_k t_k}\tilde{\Phi}_{\text{(out)}k}(\omega_k,z_k,\bar{z}_k)\right.\nonumber\\
&\hspace{10mm}\left.\int\limits_{0}^{\infty}d\omega_{k+1} \text{ }e^{-i\omega_{k+1} t_{k+1}}\tilde{\Phi}_{\text{(in)}k+1}(\omega_{k+1},z_{k+1},\bar{z}_{k+1})\ldots\int\limits_{0}^{\infty}d\omega_n \text{ }e^{-i\omega_n t_n}\tilde{\Phi}_{\text{(in)}n}(\omega_n,z_n,\bar{z}_n)\right\rangle=0\nonumber\\
\Rightarrow&\text{ }\left(\omega^{\text{total}}_{\text{out}}-\omega^{\text{total}}_{\text{in}}\right)\left\langle\tilde{\Phi}_{\text{(out)}1}(\omega_1,z_1,\bar{z}_1)\ldots\tilde{\Phi}_{\text{(in)}n}(\omega_n,z_n,\bar{z}_n)\right\rangle=0
\end{align}

\medskip

From this conservation law, along with those for the other components of the bulk null four-momentum, it seems that $\tilde{\Phi}_{\text{out}}(\omega,z,\bar{z})$ is related to the Fourier transformation of a $1+3$D bulk AFS (position-space) field describing an outgoing mass-less particle and $\tilde{\Phi}_{\text{in}}(\omega,z,\bar{z})$ is related to the Fourier transformation of a $1+3$D bulk AFS field describing that same but incoming particle\footnote{The Carrollian field $\Phi^\prime$ that describes an anti-particle of a bulk particle described by the Carrollian field $\Phi$ should be in general different from $\Phi$.} with null four-momentum completely specified by $(\omega,z,\bar{z})$. The temporal Fourier transformation of the (appropriate) Carrollian correlator then corresponds to the Fourier transformation of the bulk position-space correlator to the null four-momentum space.

\medskip

We now proceed to find the soft-graviton theorems remembering that the Carrollian field $S^-_1$ should correspond to an outgoing graviton which will be taken energetically soft. We first write the $S^-_1$ Ward identity for Carrollian conformal primaries transforming under a Carrollian spin-boost irrep, i.e. with $\bm{\xi}=\bar{\bm{\xi}}=0$, from \eqref{84} as:
\begin{align*}
\langle S_1^-(t,z,\bar{z}) X\rangle=-\frac{i}{2}\sum\limits_{p=1}^n\theta(t-t_p)\left[\frac{(z-z_p)^2}{\bar{z}-\bar{z}_p}\partial_{z_p}-2h_p\frac{z-z_p}{\bar{z}-\bar{z}_p}+(t-t_p)\frac{z-z_p}{\bar{z}-\bar{z}_p}\partial_{t_p}\right]{\langle X\rangle}
\end{align*}
Taking the temporal Fourier transformation of this Ward identity according to the convention \eqref{88} and using the fact that:
\begin{align*}
\int\limits_{-\infty}^{\infty}dt \text{ }e^{-i\omega t}\theta(t-t_p)=\frac{e^{-i\omega t_p}}{i\omega}-\lim\limits_{R\rightarrow\infty}\frac{e^{-i\omega R}}{i\omega}=\frac{e^{-i\omega t_p}}{i\left(\omega-i0^+\right)}
\end{align*}
we obtain, after taking the soft limit (and performing a `soft' Taylor expansion of the leading order pole's residue), that:
\begin{align}
\lim\limits_{\omega\rightarrow0}\omega\left\langle S_1^-(\omega,z,\bar{z}) \tilde{X}_{\text{out}}\tilde{X}_{\text{in}}\right\rangle=&\lim\limits_{\omega\rightarrow0}-\frac{1}{2}\left[\mathcal{O}(\omega)+\sum\limits_{p\in\text{all}}\left\{\frac{(z-z_p)^2}{\bar{z}-\bar{z}_p}\partial_{z_p}-\left(2h_p-1-\omega_p\partial_{\omega_p}\right)\frac{z-z_p}{\bar{z}-\bar{z}_p}\right\}\right.\nonumber\\
&\hspace{13mm}\left.+\frac{1}{\omega}\left(\sum\limits_{i\in\text{out}}\frac{z-z_i}{\bar{z}-\bar{z}_i}\omega_i-\sum\limits_{i\in\text{in}}\frac{z-z_i}{\bar{z}-\bar{z}_i}\omega_i\right)\right]\left\langle  \tilde{X}_{\text{out}}\tilde{X}_{\text{in}}\right\rangle
\end{align}
which reveals the universal soft-factorization and soft-expansion of the soft-factor of a temporal Fourier transformed Carrollian conformal correlator involving the field $S^-_1$ which in Celestial CFT is described as the conformally soft sub-leading graviton operator \cite{Donnay:2018neh,Puhm:2019zbl}. Had $\bm{\xi}$ and $\bar{\bm{\xi}}$ not been set to zero, we would have got modified coefficients in the above soft-expansion.

\medskip

Similarly, from the $S^+_1$ Ward identity, we reach the `conjugate' universal soft-factorization property:
\begin{align}
\lim\limits_{\omega\rightarrow0}\omega\left\langle S_1^+(\omega,z,\bar{z}) \tilde{X}_{\text{out}}\tilde{X}_{\text{in}}\right\rangle=&\lim\limits_{\omega\rightarrow0}-\frac{1}{2}\left[\mathcal{O}(\omega)+\sum\limits_{p\in\text{all}}\left\{\frac{(\bar{z}-\bar{z}_p)^2}{{z}-{z}_p}\partial_{\bar{z}_p}-\left(2\bar{h}_p-1-\omega_p\partial_{\omega_p}\right)\frac{\bar{z}-\bar{z}_p}{z-z_p}\right\}\right.\nonumber\\
&\hspace{13mm}\left.+\frac{1}{\omega}\left(\sum\limits_{i\in\text{out}}\frac{\bar{z}-\bar{z}_i}{z-z_i}\omega_i-\sum\limits_{i\in\text{in}}\frac{\bar{z}-\bar{z}_i}{z-z_i}\omega_i\right)\right]\left\langle  \tilde{X}_{\text{out}}\tilde{X}_{\text{in}}\right\rangle
\end{align} 

\medskip
  
We immediately recognize from the above, following the convention of \cite{Banerjee:2018fgd}, that the residues of the $\frac{1}{\omega}$ poles ($\omega$ is now complexified) are the Weinberg soft graviton theorems \cite{Weinberg:1965nx,He:2014laa}, respectively of the negative and the positive helicity. Moreover, the $\mathcal{O}(1)$ terms reproduce the Cachazo-Strominger sub-leading soft graviton theorems \cite{Cachazo:2014fwa,Kapec:2014opa,Kapec:2016jld} if all of the Carrollian conformal primaries in $X$ have $\Delta_p=1$. The $1+2$D Carrollian spin $m_p$ of a primary field in $X$ is then identified with the helicity of the corresponding $1+3$D bulk AFS mass-less external particle. Hence, we conclude that:
\begin{itemize}
\item a $1+2$D position-space Carrollian conformal field whose temporal Fourier transformation corresponds to a $1+3$D bulk null momentum-space field describing a mass-less external (hard) scattering particle, must be a primary field with scaling dimension $\Delta=1$ and transforming under a Carrollian spin-boost irrep; besides, the Carrollian spin $m$ of such a primary is the same as the helicity of the corresponding bulk particle,
\end{itemize}
in perfect agreement with \cite{Donnay:2022aba,Donnay:2022wvx} where the same conclusions were reached by analyzing the gravitational radiative fall-off conditions. 

\medskip

We clarify that there definitely are other Carrollian conformal primaries with $\Delta\neq1$ (as we will see in section \ref{s5.3}, $S^\pm_1$ and $T^t_{\hspace{1.5mm}t}$ are some important examples) and/or transforming under a reducible but indecomposable representation of the Carrollian spin-boost sub-algebra. But, as deduced above, they can not correspond to any bulk AFS mass-less external scattering particles.

\medskip

Thus, we have given a Carrollian derivation, completely independent of the physics of the bulk mass-less scattering, of the universal soft-factorization property \cite{Cachazo:2014fwa} of the bulk $S$-matrices involving an external soft graviton, using only general Carrollian symmetry arguments. This identification reinforces the interpretation of the temporal Fourier transformation of a $1+2$D position-space Carrollian conformal correlator as the null momentum-space $S$-matrix of the corresponding $1+3$D bulk AFS mass-less particles \cite{Donnay:2022wvx}.

\medskip

Before concluding the discussion on the $1+2$D Carrollian conformal Ward identities, a comment on the prefix `soft' in the Carrollian context is in order. As we have noticed, to reach the universal soft(graviton)-factorization from the Carrollian conformal Ward identities, only performing the temporal Fourier transformations of the latter is not sufficient; doing so only gives rise to the pole structure at $\omega=0$ originating from the temporal step-function which is the Carrollian manifestation of the corresponding memory effects \cite{Strominger:2014pwa,Pasterski:2015tva,Strominger:2017zoo}. The crucial next step is to impose an explicit $\omega\rightarrow0$ (soft) limit that actually leads to the soft-factorization.  Accordingly, it is more appropriate to not name the Carrollian fields $S^\pm_0$ and $S^\pm_1$ as the `soft' graviton fields since their insertions in a position-space Carrollian correlator do not automatically lead to soft-factorization upon only a temporal Fourier transformation. A prime example, in view of the above discussion, will be the Carrollian fields $S^\pm_0$ that can describe the bulk external hard graviton particles as $S^\pm_0$ have $\Delta=1$ and, as will be shown in section \ref{s5.3}, $\bm{\xi}\cdot S^\pm_0=\bar{\bm{\xi}}\cdot S^\pm_0=0$. Nevertheless, as stated before, we keep the nomenclature same as in the Celestial holography literature, e.g. \cite{Banerjee:2020zlg}, to avoid confusion. This will not create any problems as the rest of this work is in the Carrollian position-space(-time) representation.  

\medskip

\section{Transformation of Quantum Fields}\label{s4}
In this section, our goal is to find out the quantum generators that implement the extended BMS$_4$ transformations \cite{Barnich:2009se,Barnich:2010ojg,Barnich:2010eb,Barnich:2011mi} on the Carrollian conformal quantum fields. For this purpose, we shall now study the changes suffered by the correlators of primaries under the infinitesimal transformations \eqref{35}. We shall see that the quantum charges generating the extended BMS$_4$ transformations on the Hilbert space differ from the expected Noether charges built out of the conserved Noether currents of the form \eqref{80}. This signifies the non-conservation of the  Carrollian conformal Noether charges even in a source-less $1+2$D Carrollian CFT, due to the presence of the gravitational radiation in the $1+3$D bulk AFS.

\medskip

Before starting, we recall that in the operator formalism of QFT, the conserved charge $Q_a$ is the generator of an infinitesimal symmetry transformation on the quantum fields:
\begin{align*}
Q_a=\int\limits_{\Sigma^d} d^d\vec{x}\text{ }j^t_{\hspace{1.5mm}a}(t,\vec{x})\hspace{5mm};\hspace{5mm}\delta_{\bm{\epsilon}}{\Phi}(t,\vec{x})=-i\epsilon^a[Q_a\text{ },\text{ }{\Phi}(t,\vec{x})]
\end{align*}
where $\Sigma^d$ is a space-like hypersurface. The L.H.S. of an OPE must be time-ordered if it is to have an operator meaning. In view of this, the above generator relation is interpreted as (in the limit $t^\pm=t+0^\pm$) \cite{DiFrancesco:1997nk}:
\begin{align*}
[Q_a\text{ },{\Phi}(t,\vec{x})]=\int\limits_{\Sigma^d} d^d\vec{x}^\prime\text{ }j^t_{\hspace{1.5mm}a}(t^+,\vec{x}^\prime){\Phi}(t,\vec{x})-\int\limits_{\Sigma^d} d^d\vec{x}^\prime\text{ }{\Phi}(t,\vec{x})j^t_{\hspace{1.5mm}a}(t^-,\vec{x}^\prime)\text{\hspace{7.5mm}(as an OPE)}
\end{align*}
i.e. in the R.H.S., the OPE between the density $j^t_{\hspace{1.5mm}a}$ and the field ${\Phi}$ is to be used.

\medskip

We now closely follow the treatment presented in \cite{Saha:2022gjw}.

\medskip

\subsection{The super-rotation generators}\label{s4.1}
We begin by considering the holomorphic super-rotation $z^{m+1}$ ($m>-2$). From \eqref{9}, we see that a primary correlator undergoes the following change due to the primary transformation property \eqref{37}:
\begin{align}
-&\delta_\epsilon\langle X\rangle=\sum_{i=1}^n\langle\hat{\mathcal{T}}\Phi_1(\mathbf{x_1})\ldots(i\epsilon \bm{\mathcal{L}}_m\Phi_i(\mathbf{x_i}))\ldots\Phi_n(\mathbf{x_n})\rangle\nonumber\\
=&\epsilon\sum_{i=1}^n\langle\hat{\mathcal{T}}\Phi_1(\mathbf{x_1})\ldots\left[z_i^{m+1}\partial_{z_i}+\frac{m+1}{2}z_i^mt_i\partial_{t_i}+(m+1)z_i^m{h}_i+(m+1)mz_i^{m-1}\frac{t_i}{2}\bm{\xi}_i\right]\Phi_i(\mathbf{x_i})\ldots\Phi_n(\mathbf{x_n})\rangle\nonumber\\
=&\frac{\epsilon}{2\pi i}\oint\limits_{C}dz\text{ }z^{m+1}\sum\limits_{p=1}^n\left[\frac{h_p+\frac{t_p}{2}\partial_{t_p}}{(z-z_p)^2}+\frac{\partial_{z_p}}{z-z_p}+\frac{t_p\bm{\xi}_p}{{(z-z_p)}^3}\right]\langle X\rangle\nonumber\\
=&\frac{\epsilon}{2\pi}\oint\limits_{C}dz\text{ }z^{m+1}\langle T_e(t,z,\bar{z})X\rangle\big|_{t>\{t_p\}}\text{\hspace{10mm}[from \eqref{39}]}\label{43}
\end{align}
where the counter-clockwise contour $C$ in the complex $z$-plane encloses all the positions of insertion $\{z_p\}$ and the condition $t>\{t_p\}$ arises clearly from the presence of the temporal step-function in \eqref{39}. This expression can be brought into the following suggestive forms using the property of the step-function (with $t^\pm:=t+0^\pm$) and the deformation of the contour $C$ respectively:
\begin{align}
-\delta_\epsilon\langle X\rangle&=\frac{\epsilon}{2\pi}\sum_{p=1}^n\oint\limits_{C}dz\text{ }\left[z^{m+1}\langle\hat{\mathcal{T}} T_e(t_p^+,z,\bar{z})X\rangle-z^{m+1}\langle\hat{\mathcal{T}} T_e(t_p^-,z,\bar{z})X\rangle\right]\label{40}\\
&=\frac{\epsilon}{2\pi}\sum_{p=1}^n\oint\limits_{z_p}dz\text{ }\left[z^{m+1}\langle\hat{\mathcal{T}} T_e(t_p^+,z,\bar{z})X\rangle-z^{m+1}\langle\hat{\mathcal{T}} T_e(t_p^-,z,\bar{z})X\rangle\right]\label{41}
\end{align}

\medskip

Now, we recall from \eqref{24} that $\langle\partial_t T_e(t,z,\bar{z})X\rangle=0$ if $t$ does not coincide with any other time of insertion; thus (for non-coincident time insertion and recalling that the covariant time-ordering and time-derivative commute):
\begin{align*}
\oint\limits_{C_e}dz\text{ }\langle\hat{\mathcal{T}}\partial_t T_e(t,z,\bar{z})X\rangle=0\hspace{2.5mm}\Longrightarrow\hspace{2.5mm}\partial_t\langle\hat{\mathcal{T}}(\oint\limits_{C_e}dz\text{ } T_e(t,z,\bar{z})\text{ })X\rangle=0
\end{align*}
which is valid only as a correlator statement. This implies that the operator $L_m$ defined below can be thought of as a conserved quantity within a correlator:
\begin{align}
L_m:=\frac{1}{2\pi i}\oint\limits_{C_e}dz\text{ }z^{m+1}\text{ }T_e(t,z,\bar{z})\label{42}
\end{align}
where the contour $C_e$ can be effectively taken to be enclosing the entire complex $z$-plane\footnote{The significance of the adverb `effectively' will be elaborated in section \ref{s4.3}.}. \eqref{40} can now be written as \cite{DiFrancesco:1997nk}:
\begin{align}
\delta_\epsilon\langle X\rangle=-i\epsilon\sum_{p=1}^n\langle\hat{\mathcal{T}}\Phi_1(\mathbf{x_1})\ldots\left[L_m\text{ },\text{ }\Phi_p(\mathbf{x_p})\right]\ldots\Phi_n(\mathbf{x_n})X\rangle=-i\epsilon\langle\hat{\mathcal{T}}\left[L_m\text{ },\text{ }X\right]\rangle\label{46}
\end{align}
showing that $L_m$ is the conserved quantum charge implementing the $z^{m+1}$ super-rotation on the quantum fields. 

\medskip

Comparing this with \eqref{41}, we see that:
\begin{align}
\left[L_m\text{ },\text{ }\Phi(\mathbf{x}_p)\right]=\frac{1}{2\pi i}\oint\limits_{z_p}dz\text{ }z^{m+1}\hat{\mathcal{T}} T_e(t_p^+,z,\bar{z})\Phi(\mathbf{x}_p)-\frac{1}{2\pi i}\oint\limits_{z_p}dz\text{ }z^{m+1}\hat{\mathcal{T}} T_e(t_p^-,z,\bar{z})\Phi(\mathbf{x}_p)\label{48}
\end{align}
Due to the presence of the temporal step-function factor in the OPE, the second term vanishes giving rise to the following relation between operator commutation relation and (time-ordered) OPE without performing any radial quantization, similarly as in $1+1$D Carrollian CFT \cite{Saha:2022gjw}:
\begin{align}
\left[L_m\text{ },\text{ }\Phi(\mathbf{x}_p)\right]=\frac{1}{2\pi i}\oint\limits_{z_p}dz\text{ }z^{m+1}\hat{\mathcal{T}} T_e(t_p^+,z,\bar{z})\Phi(\mathbf{x}_p)\label{53}
\end{align}

\medskip

Till now, we have the restriction: $m\geq-1$; this can be lifted in the definition \eqref{42} of the conserved charge $L_m$ to include all $m\in\mathbb{Z}$. The contour $C_e$ then encloses the singularities of the vector field $z^{m+1}$; $C$ encloses only the positions of insertion $z_p$ but not the singularities of the vector field so that the contour deformation leading to \eqref{41} remains valid.

\medskip

Similarly, in the anti-holomorphic sector, the conserved charge $\bar{L}_m$ can be defined as below that generates the $\bar{z}^{m+1}$ super-rotation on the space of the quantum fields (for $m\in\mathbb{Z}$):
\begin{align}
\bar{L}_m:=\frac{1}{2\pi i}\oint\limits_{\bar{C}_e}d\bar{z}\text{ }\bar{z}^{m+1}\text{ }\bar{T}_e(t,z,\bar{z})
\end{align}
with $\bar{C}_e$ being a clock-wise contour enclosing the entire complex $\bar{z}$-plane.

\medskip

\subsection{The problem with super-translation generators}\label{s4.2}
To discuss on the super-translations, let us first note down the generalized Cauchy integral formula \cite{Ab}:
\begin{align}
&f(z,\bar{z})=\frac{1}{2\pi i}\oint\limits_Cdz^\prime\text{ }\frac{f(z^\prime,\bar{z}^\prime)}{z^\prime-z}-\frac{1}{\pi}\int\limits_A d^2\vec{r}^\prime\text{ }\frac{\bar{\partial}^\prime f(z^\prime,\bar{z}^\prime)}{z^\prime-z}\\
\Longrightarrow\hspace{2.5mm}&\partial^nf(z,\bar{z})=\frac{n!}{2\pi i}\oint\limits_Cdz^\prime\text{ }\frac{f(z^\prime,\bar{z}^\prime)}{{(z^\prime-z)}^{n+1}}-\frac{n!}{\pi}\int\limits_A d^2\vec{r}^\prime\text{ }\frac{\bar{\partial}^\prime f(z^\prime,\bar{z}^\prime)}{{(z^\prime-z)}^{n+1}}
\end{align}
where the counter-clockwise contour $C$ encloses a region $A$ containing the point $(z,\bar{z})$ and $n\geq0$. 

\medskip

Now we consider the infinitesimal change of a primary correlator due to the transformation property \eqref{36} under the super-translation $z^{a+1}\bar{z}^{b+1}$ (with $a,b\geq-1$):
\begin{align}
-&\delta_\epsilon\langle X\rangle=\sum_{i=1}^n\langle\hat{\mathcal{T}}\Phi_1(\mathbf{x_1})\ldots(i\epsilon \bm{\mathcal{P}}_{a,b}\Phi_i(\mathbf{x_i}))\ldots\Phi_n(\mathbf{x_n})\rangle\nonumber\\
=&\epsilon\sum_{i=1}^n\langle\hat{\mathcal{T}}\Phi_1(\mathbf{x_1})\ldots\left[z_i^{a+1}\bar{z}_i^{b+1}\partial_{t_i}+(a+1)z_i^a\bar{z}_i^{b+1}{\bm{\xi}}_i+(b+1)z_i^{a+1}\bar{z}_i^b\bar{\bm{\xi}}_i\right]\Phi_i(\mathbf{x_i})\ldots\Phi_n(\mathbf{x_n})\rangle\nonumber\\
=&\frac{\epsilon}{2\pi}\oint\limits_{C}dz\text{ }z^{a+1}\bar{z}^{b+1}\langle P(t,z,\bar{z})X\rangle\big|_{t>\{t_p\}}-\frac{i\epsilon}{\pi}\int\limits_A d^2\vec{r}\text{ }\bar{\partial}\left(z^{a+1}\bar{z}^{b+1}\right)\langle P(t,z,\bar{z})X\rangle\big|_{t>\{t_p\}}\label{44}
\end{align}
The last line follows from \eqref{19} and the generalized Cauchy integral formula with $n=1,2$. The contour $C$ encloses the region $A$ containing all the positions of insertion but whether it encloses the singularities of the vector field $z^{a+1}\bar{z}^{b+1}$ does not make any difference; so, we inflate $C$ to $C_e$ to enclose the entire complex plane. The restriction $a,b\geq-1$ is then readily lifted to include all $a,b\in\mathbb{Z}$.

\medskip

Interestingly, the effects \eqref{43} of the super-rotations with any $m\in\mathbb{Z}$ on the correlator can be expressed analogously, as:
\begin{align}
-\delta_\epsilon\langle X\rangle=\frac{\epsilon}{2\pi}\oint\limits_{C^\prime}dz\text{ }z^{m+1}\langle T_e(t,z,\bar{z})X\rangle\big|_{t>\{t_p\}}-\frac{i\epsilon}{\pi}\int\limits_{A^\prime} d^2\vec{r}\text{ }\bar{\partial}\left(z^{m+1}\right)\langle T_e(t,z,\bar{z})X\rangle\big|_{t>\{t_p\}}\label{45}
\end{align}
where the contour $C^\prime$ encloses the region $A^\prime$ containing all the positions of insertion; it does not matter if $C^\prime$ encloses the singularities of the vector field $z^{m+1}$ or not. Thus, $C^\prime$ can be taken to be $C_e$. 

\medskip

After obtaining the analogues of \eqref{46} from \eqref{44} and \eqref{45}, it is tempting to conclude that:
\begin{align*}
\tilde{P}_{a,b}&:=\frac{1}{2\pi i}\oint\limits_{C_e}dz\text{ }z^{a+1}\bar{z}^{b+1} \text{ }P(t,z,\bar{z})-\frac{1}{\pi}\int\limits_{S^2} d^2\vec{r}\text{ }\bar{\partial}\left(z^{a+1}\bar{z}^{b+1}\right) P(t,z,\bar{z})\\
&=\frac{1}{\pi}\int\limits_{S^2} d^2\vec{r}\text{ }z^{a+1}\bar{z}^{b+1}\text{ }\bar{\partial}{P}(t,z,\bar{z})=\int\limits_{S^2} d^2\vec{r}\text{ }z^{a+1}\bar{z}^{b+1}\text{ }T^t_{\hspace{1.5mm}t}(t,z,\bar{z})\\
\tilde{L}_m&:=\frac{1}{2\pi i}\oint\limits_{C_e}dz\text{ }z^{m+1} \text{ }T_e(t,z,\bar{z})-\frac{1}{\pi}\int\limits_{S^2} d^2\vec{r}\text{ }\bar{\partial}\left(z^{m+1}\right) T_e(t,z,\bar{z})=\frac{1}{\pi}\int\limits_{S^2} d^2\vec{r}\text{ }z^{m+1}\text{ }\bar{\partial}{T}_e(t,z,\bar{z})\nonumber
\end{align*}
are the respective conserved charges generating these transformations. $\tilde{P}_{a,b}$ and $\tilde{L}_m$ even look exactly like the expected (conserved) Noether charges for super-translations and super-rotations (at least for $T^z_{\hspace{1.5mm}\bar{z}}=0=T^{\bar{z}}_{\hspace{1.5mm}z}$); but these are actually divergent\footnote{The integral of the Noether charge density over $S^2$ can not be finite for all $a,b\in\mathbb{Z}$ even if we use the `round sphere' metric, unless the Carrollian EM tensor components themselves are linear combinations of the spatial delta-function and its derivatives.}. They can generate the infinitesimal BMS$_4$ transformations properly only because the two Ward identities \eqref{7} and \eqref{a13} contain only $S^2$ contact terms. But, since these Ward identities vanish in the OPE limit, it is not possible to form consistent mode-expansions of these EM tensor component fields in terms of the quantum generators $\tilde{P}_{a,b}$, $\tilde{L}_m$ and $\tilde{\bar{L}}_m$. So, instead of the Noether charges $\tilde{L}_m$, the holomorphic super-rotation generators are given by the 2D CFT like form \eqref{42}.

\medskip

Unfortunately, the following candidates $P^\prime_{a,b}$ for the super-translation generators, defined analogously to the super-rotation generators \eqref{42}:
\begin{align*}
P^\prime_{a,b}:=\frac{1}{2\pi i}\oint\limits_{C_e}dz\text{ }z^{a+1}\bar{z}^{b+1} \text{ }P(t,z,\bar{z})
\end{align*}
clearly do not generate super-translations on the space of quantum fields for $b\neq-1$, as seen from \eqref{19}. Thus, only the holomorphic (and analogously, the anti-holomorphic) super-translation generators can be defined in this way. 

\medskip

Another possible extension of the definition \eqref{42} to the super-translation case would be:
\begin{align*}
P_{a,b}:=\frac{1}{4\pi^2}\oint\limits_{C_e}dz\oint\limits_{\bar{C}_e}d\bar{z}\text{ }z^{a+1}\bar{z}^{b+1} \text{ }\mathcal{P}(t,z,\bar{z})
\end{align*}
where both $C_e$ and $\bar{C}_e$ are counter-clockwise contours enclosing the entire $z$- and $\bar{z}$-planes respectively (i.e. $z$ and $\bar{z}$ are now treated as completely independent variables) and $\mathcal{P}$ is a (local) field \cite{Barnich:2017ubf} with weights $(\Delta,m)=(3,0)$ that embeds the field $P$ as \cite{Fotopoulos:2019vac}:
\begin{align*}
{P}(t,z,\bar{z})=-\frac{1}{2\pi i}\oint\limits_{\bar{C}_e}d\bar{z}^\prime\text{ }\mathcal{P}(t,z,\bar{z}^\prime)
\end{align*}
This relation should be thought to be valid only inside a correlator so that $\langle\bar{\partial}P\ldots\rangle=0$ holds. Its Ward identity (involving only CC primaries) then reads:
\begin{align*}
\langle \mathcal{P}(t,z,\bar{z})X\rangle=-i\sum\limits_{p=1}^n\text{ }\theta(t-t_p)\left[\frac{\partial_{t_p}}{\left(z-z_p\right)\left(\bar{z}-\bar{z}_p\right)}+\frac{\bm{\xi}_p}{{(z-z_p)}^2\left(\bar{z}-\bar{z}_p\right)}+\frac{\bar{\bm{\xi}}_p}{{(z-z_p)}{\left(\bar{z}-\bar{z}_p\right)}^2}\right]\langle X\rangle
\end{align*}
The price to pay now is to similarly embed the fields $T_e$ and $\bar{T}_e$ respectively into the (local) fields $\mathcal{J}_e$ with $(\Delta,m)=(3,1)$ and $\bar{\mathcal{J}}_e$ with $(\Delta,m)=(3,-1)$ as \cite{Donnay:2021wrk}:
\begin{align*}
T_e(t,z,\bar{z})=-\frac{1}{2\pi i}\oint\limits_{\bar{C}_e}d\bar{z}^\prime\text{ }\mathcal{J}_e(t,z,\bar{z}^\prime)\hspace{5mm};\hspace{5mm}\bar{T}_e(t,z,\bar{z})=\frac{1}{2\pi i}\oint\limits_{{C}_e}d{z}^\prime\text{ }\bar{\mathcal{J}}_e(t,z^\prime,\bar{z})
\end{align*}
so that $\mathcal{J}_e$ and $\bar{\mathcal{J}}_e$ OPEs have similar pole structures to that of $\mathcal{P}$. This way we have a chance to obtain a closed algebra from the mutual OPEs of $\mathcal{J}_e$, $\bar{\mathcal{J}}_e$ and $\mathcal{P}$.

\medskip

But in \cite{Schwarz:2022dqf}, it was shown that, to form a consistent algebra respecting the Jacobi identity from the OPEs between $\mathcal{J}_e\bar{\mathcal{J}}_e$, $\bar{\mathcal{J}}_e{\mathcal{J}}_e$, $\mathcal{J}_e{\mathcal{J}}_e$ and $\bar{\mathcal{J}}_e\bar{\mathcal{J}}_e$, another field with $(\Delta,m)=(2,0)$ must be introduced. Since we do not have any natural candidate for such a field, we refrain from introducing the $\mathcal{P}$ field at all in this work.

\medskip

Thus, till now, we do not have the quantum generators of the mixed super-translations.  

\medskip

\subsection{The $j\epsilon$-prescription}\label{s4.3}
Now we establish the effective definition \eqref{42} of the super-rotation generator by introducing a $j\epsilon$-prescription, following \cite{Saha:2022gjw}. Its origin is the temporal step-function factor appearing in the previously discussed Ward identities. 

\medskip

First, we hyper-complexify the stereographic coordinates $z,\bar{z}$ as:
\begin{align}
\hat{z}:=z+j t\hspace{5mm};\hspace{5mm}\hat{\bar{z}}:=\bar{z}+j t
\end{align}
where $j$ is a second complex unit. This can be alternatively viewed as a complexification of only the $x$-coordinate as: $\hat{x}:=x+j t$ keeping the $y$-coordinate real. On any $y=ax+b$ plane (with complex coordinate $z+jt$ ; $z$ is a real quantity with respect to the complex unit $j$) of the $1+2$D Carrollian space-time, all the positions of insertion $z_p$ are projected onto the $t=0$ line; $t>0$ denotes the upper half of this plane.

\medskip

Next, motivated by \eqref{40}, we introduce the $j\epsilon$-form of the holomorphic super-rotation Ward identity \eqref{39}, with $\Delta\tilde{z}_p:=\hat{z}-{z}_p-j\epsilon(t-t_p)$ :
\begin{align}
\langle T_e(t,\hat{z},\hat{\bar{z}})X\rangle=\lim\limits_{\epsilon\rightarrow0^+}-i\sum\limits_{p=1}^n\left[\frac{h_p+\frac{t_p}{2}\partial_{t_p}}{(\Delta\tilde{z}_p)^2}+\frac{\partial_{z_p}}{\Delta\tilde{z}_p}+\frac{t_p\bm{\xi}_p}{{(\Delta\tilde{z}_p)}^3}\right]\langle X\rangle\label{47}
\end{align} 
Thus, the poles at $\left\{{z}_p+j\epsilon(t-t_p)\right\}$ in this Laurent series (with $\hat{z}=z+jt$ being the complex variable) are projected onto the upper half plane for $t>t_p$ and onto the lower half plane for $t<t_p$.

\medskip
       
We shall now consider complex contour integral on a $y=ax+b$ plane, with the hyper-complexified $\hat{z}$ being the integration variable. Looking at \eqref{47}, we infer that the relation \eqref{48} can be expressed in the $j\epsilon$-prescription, for $m\in\mathbb{Z}$, as:
\begin{align}
\left[L_m\text{ },\text{ }\Phi(\mathbf{x}_p)\right]&=\frac{1}{2\pi j}\oint\limits_{C_u}d\hat{z}\text{ }\hat{z}^{m+1}\hat{\mathcal{T}} T_e(t_p^+,\hat{z},\hat{\bar{z}})\Phi(\mathbf{x}_p)-\frac{1}{2\pi j}\oint\limits_{C_u}d\hat{z}\text{ }\hat{z}^{m+1}\hat{\mathcal{T}} T_e(t_p^-,\hat{z},\hat{\bar{z}})\Phi(\mathbf{x}_p)\nonumber\\
&=\frac{1}{2\pi j}\oint\limits_{C^\prime_u}d\hat{z}\text{ }\hat{z}^{m+1}\hat{\mathcal{T}} T_e(t_p^+,\hat{z},\hat{\bar{z}})\Phi(\mathbf{x}_p)-\frac{1}{2\pi j}\oint\limits_{C^\prime_u}d\hat{z}\text{ }\hat{z}^{m+1}\hat{\mathcal{T}} T_e(t_p^-,\hat{z},\hat{\bar{z}})\Phi(\mathbf{x}_p)\nonumber\\
&=[\frac{1}{2\pi j}\oint\limits_{C^\prime_u}d\hat{z}\text{ }\hat{z}^{m+1}\text{ } T_e(t,\hat{z},\hat{\bar{z}})\text{ },\text{ }\Phi(\mathbf{x}_p)]\label{89}
\end{align}
where the contours $C_u$ and $C_u^\prime$ are depicted in Figure 1 (for a $y=b$ plane). In the last line, we have replaced $t_p$ by $t$ as we recall that $\partial_tT_e\sim0$ as an OPE statement.
\begin{figure}[h]
\begin{center}
\begin{tikzpicture}[decoration={markings,
mark=at position 1.1cm with {\arrow[line width=1pt]{>}}}]
\draw[help lines,->] (-1.1,0) -- (1.1,0) coordinate (xaxis);
\draw[help lines,->] (0,-0.3) -- (0,1.1) coordinate (yaxis);
\path[draw,line width=0.8pt,postaction=decorate] (-.9,0)  (.9,0) arc (0:180:.9);
\path[draw,line width=0.8pt,postaction=decorate] (-0.6,0)  (-.3,0) arc (0:180:.15);
\path[draw,line width=0.8pt,postaction=decorate] (0.3,0)  (.6,0) arc (0:180:.15);
\path[draw,line width=0.8pt,postaction=decorate] (-.9,0)--(-.6,0);
\path[draw,line width=0.8pt,postaction=decorate] (-.3,0)--(.3,0);
\path[draw,line width=0.8pt,postaction=decorate] (.6,0)--(.9,0);
\node[above] at (xaxis) {$x$};
\node[left] at (yaxis) {$t$};
\draw (-.45,0) node{$\times$};
\draw (.45,0) node{$\times$};
\draw (-.2,0.2) node{$\bullet$};
\node at (-.2,0.5) {$z^\prime_p$};
\node at (-.9,.8) {$C_{u}$};
\draw[help lines,->] (2.4,0) -- (4.6,0) coordinate (xaxis);
\draw[help lines,->] (3.5,-0.3) -- (3.5,1.1) coordinate (yaxis);
\path[draw,line width=0.8pt,postaction=decorate] (2.6,0)  (4.4,0) arc (0:180:.9);
\path[draw,line width=0.8pt,postaction=decorate] (2.9,0)  (3.2,0) arc (0:180:.15);
\path[draw,line width=0.8pt,postaction=decorate] (3.8,0)  (4.1,0) arc (0:180:.15);
\path[draw,line width=0.8pt,postaction=decorate] (2.6,0)--(2.9,0);
\path[draw,line width=0.8pt,postaction=decorate] (3.2,0)--(3.8,0);
\path[draw,line width=0.8pt,postaction=decorate] (4.1,0)--(4.4,0);
\draw (3.05,0) node{$\times$};
\draw (3.95,0) node{$\times$};
\draw (3.3,-0.2) node{$\bullet$};
\node at (3.3,-0.45) {$z^\prime_p$};
\node at (2.6,.8) {$C_{u}$};
\node at (1.75,.4) {-};
\draw[help lines,->] (6.4,0) -- (8.6,0) coordinate (xaxis);
\draw[help lines,->] (7.5,-0.4) -- (7.5,1) coordinate (yaxis);
\path[draw,line width=0.8pt,postaction=decorate] (6.6,-0.15)--(8.4,-.15) arc (0:180:.9);
\node[above] at (xaxis) {$x$};
\node[left] at (yaxis) {$t$};
\draw (7.05,0) node{$\times$};
\draw (7.95,0) node{$\times$};
\draw (7.3,0.2) node{$\bullet$};
\node at (7.3,0.5) {${z}^\prime_p$};
\node at (6.7,.8) {$C^\prime_{u}$};
\node at (5.5,.4) {=};
\draw[help lines,->] (9.9,0) -- (12.1,0) coordinate (xaxis);
\draw[help lines,->] (11,-0.4) -- (11,1) coordinate (yaxis);
\path[draw,line width=0.8pt,postaction=decorate] (10.1,-0.15)--(11.9,-.15) arc (0:180:.9);
\draw (10.55,0) node{$\times$};
\draw (11.45,0) node{$\times$};
\draw (10.8,-0.3) node{$\bullet$};
\node at (10.8,-0.55) {${z}^\prime_p$};
\node at (10.2,.8) {$C^\prime_{u}$};
\node at (9.25,.4) {-};
\end{tikzpicture}
\caption{Equality of subtractions of contours in $j\epsilon$ prescription for $1+2$D CCFT OPEs : $z^\prime_p\equiv z_p+j\epsilon(t-t_p)$ where $t$ is the time of insertion of the quantum charge density operator; $\times$ are the singularities of the vector field. $C_u$ does not enclose the singularities of the vector field but $C_u^\prime$ does so by enclosing the $t=0$ line. Both $C_u$ and $C_u^\prime$ enclose the upper half plane.} 
\end{center}
\end{figure}
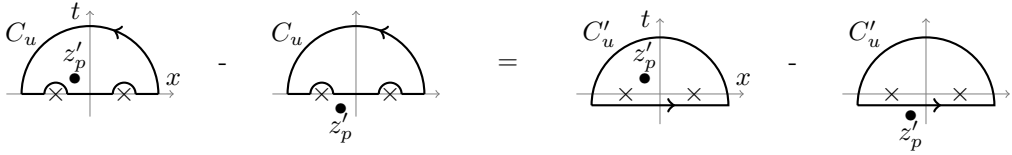 

\medskip

Since, the contour $C_u^\prime$ is in a sense universal i.e. independent of the field $\Phi(\mathbf{x}_p)$ as well as the vector fields, the holomorphic super-rotation generators can be defined as:
\begin{align}
L_m:=\frac{1}{2\pi j}\oint\limits_{C_u^\prime}d\hat{z}\text{ }\hat{z}^{m+1}\text{ }T_e(t,\hat{z},\hat{\bar{z}})
\end{align}
where the counter-clockwise contour $C_u^\prime$ encloses the entire upper half of a $y=ax+b$ plane as well as the line $t=0$; thus it encloses the projections of the singularities of the vector field $\hat{z}^{m+1}$. 

\medskip

Because the singularities of the vector fields $\hat{z}^{m+1}$ are independent of $t$, this definition of $L_m$ under the $j\epsilon$-prescription is practically equivalent to the definition \eqref{42} under the $\theta$-prescription where the contour $C_e$ (on $x$\textendash$y$ plane) must enclose the entire complex $z$-plane (hence, the singularities of the vector field). As is evident from the construction, both of these definitions should be thought to be valid only inside correlators.

\medskip

Thus the $j\epsilon$-prescription plays a very crucial rule to fix the definitions of the (anti-)holomorphic quantum conserved charge operators but not those of the mixed super-translation generators. As a very significant byproduct, we get to establish the relation between the OPEs and the operator commutation relations, e.g. \eqref{89}, via complex contour integrals (on $S^2$).

\medskip

The initial conditions like \eqref{a11} have clearer technical meaning in the $j\epsilon$-prescription. E.g. if we put $t\rightarrow-\infty$ in \eqref{47}, all the poles are pushed into the lower half of a $y=ax+b$ plane; these poles are not then enclosed by the contours $C_u$ or $C_u^\prime$, hence contributing nothing to the integral.

\medskip

It may seem that all the above conclusions remain intact even if we replace the $j\epsilon(t-t_p)$ part of $\Delta\tilde{z}_p$ by $j\lambda{\left(t-t_p\right)}^{2n+1}$ with $\lambda=\lambda_1+j\lambda_2$ such that $\lambda_1>0$. But we wanted the dimension of $\lambda$ to be that of speed, so we chose $n=0$. Moreover, remembering that $\partial_t\langle T_e(t,\hat{z},\hat{\bar{z}})X\rangle=0$ in the OPE limit, we are forced to take $\lambda_1\rightarrow0^+$ and $\lambda_2=0$ that reduces to the $j\epsilon$-prescription.    

\medskip

We shall always write the OPEs after analytically continuing the expressions involving $\hat{z},\hat{\bar{z}}$ back to `real' (wrt complex unit $j$) $z,\bar{z}$. 

\medskip

\section{Symmetry Algebra from the OPE}\label{s5}
We now proceed to find the symmetry algebra that is manifest at the level of the $1+2$D Carrollian conformal OPEs. From the discussion in section \ref{s3}, it already appears that all of the $S^\pm_0$, $S^\pm_1$, $T(\bar{T})$ fields can not be simultaneously taken to form consistent mutual OPEs. In this section, we pinpoint the reasons and choose an appropriate subset from these six generator-fields that allows the formation of consistent OPEs. Under two crucial assumptions (one is inspired from the usual 2D CFT \cite{Zamolodchikov:1989mz} and the another from the Celestial CFT \cite{Fotopoulos:2019tpe}), we are then able to fix the pole singularities of these mutual OPEs using only the OPE-commutativity property, just as in \cite{Zamolodchikov:1985wn,Zamolodchikov:1989mz,Saha:2022gjw,Bagchi:2023dzx}. The ansatz for these OPEs are made from the corresponding Carrollian conformal Ward identities (themselves derived from general symmetry principles) in the first place. Our results for the singular parts of these OPEs match with those obtained in the Celestial CFT \cite{Guevara:2019ypd,Fotopoulos:2019vac,Guevara:2021abz,Banerjee:2022wht,Puhm:2019zbl} where the starting point was the (linearized) Einstein theory in the bulk AFS. Finally, we translate these OPEs into the language of the algebra of the modes, using the complex contour integral prescription developed in section \ref{s4}.

\medskip
  
\subsection{The OPEs}\label{s5.1}
We begin by writing down the OPEs directly in the $j\epsilon$ form from the corresponding Ward identities derived in section \ref{s3}. The Bosonic or Fermionic exchange properties of the two composite (or local) operators involved can be readily implemented in this form. On the contrary, in the form involving the temporal step-function, it is not straight-forward to use these exchange properties in the OPE because the discontinuous initial conditions like \eqref{a11} needs to be simultaneously altered in the process. 

\medskip

As discussed in \cite{Saha:2022gjw}, in a Carrollian field theory, two fields inside a correlator are to be treated as a single composite operator if they are inserted at the same spatial location (but possibly at different times) causing the Carrollian-invariant norm to vanish\footnote{The flat Carrollian-invariant norm is the Euclidean (spatial) distance: $\big|\vec{x}_1-\vec{x}_2\big|$ .}. This is the essence of the spatial absoluteness in Carrollian field theory. So, the two local fields whose operator-product is considered are always inserted at different spatial locations (thus, contact terms can not appear as OPE coefficients). Moreover, to avoid the time-ordering ambiguity, they are also inserted at different times. 

\medskip

We now note down various OPEs involving a primary Carrollian conformal multiplet $\Phi(t_p,\vec{x}_p)$, with $\Delta\tilde{z}_p:=z-z_p-j\epsilon(t-t_p)$, :
\begin{align}
&P(t,z,\bar{z})\Phi(t_p,z_p,\bar{z}_p)\sim\lim\limits_{\epsilon\rightarrow0^+}-i\left(\frac{\partial_{t_p}}{(\Delta\tilde{z}_p)}+\frac{\bm{\xi}_p}{{(\Delta\tilde{z}_p)}^2}\right)\Phi(t_p,z_p,\bar{z}_p)\label{51}\\
&S^+_0(t,z,\bar{z})\Phi(t_p,z_p,\bar{z}_p)\sim\lim\limits_{\epsilon\rightarrow0^+}-i\left\{\frac{\bar{z}-\bar{z}_p}{(\Delta\tilde{z}_p)}\partial_{t_p}+\frac{\bar{z}-\bar{z}_p}{{(\Delta\tilde{z}_p)}^2}\bm{\xi}_p-\frac{\bar{{\bm{\xi}}}_p}{(\Delta\tilde{z}_p)}\right\}\Phi(t_p,z_p,\bar{z}_p)\\
&T(t,z,\bar{z})\Phi(\mathbf{x}_p)\sim\lim\limits_{\epsilon\rightarrow0^+}-i\left[\frac{h_p}{(\Delta\tilde{z}_p)^2}+\frac{\partial_{z_p}}{\Delta\tilde{z}_p}-\frac{t-t_p}{2}\left\{\frac{\partial_{t_p}}{(\Delta\tilde{z}_p)^2}+\frac{2\bm{\xi}_p}{{(\Delta\tilde{z}_p)}^3}\right\}\right]\Phi(\mathbf{x}_p)\label{50}\\
&S_1^+(t,z,\bar{z})\Phi(\mathbf{x}_p)\sim\lim\limits_{\epsilon\rightarrow0^+}-\frac{i}{2}\left[\frac{(\bar{z}-\bar{z}_p)^2}{\Delta\tilde{z}_p}\partial_{\bar{z}_p}-2\bar{h}_p\frac{\bar{z}-\bar{z}_p}{\Delta\tilde{z}_p}+\frac{0}{\Delta\tilde{z}_p}\right.\nonumber\\
&\hspace{43mm}\left.+(t-t_p)\left(\frac{\bar{z}-\bar{z}_p}{(\Delta\tilde{z}_p)}\partial_{t_p}+\frac{\bar{z}-\bar{z}_p}{{(\Delta\tilde{z}_p)}^2}\bm{\xi}_p-\frac{\bar{{\bm{\xi}}}_p}{(\Delta\tilde{z}_p)}\right)\right]\Phi(\mathbf{x}_p)\label{52}
\end{align}
where $\sim$ denotes `modulo terms holomorphic (regular) in $\Delta\tilde{z}_p$ '. That there are regular terms in these OPEs can be easily checked by expanding the corresponding Ward identities in $j\epsilon$-prescription as power series simultaneously in e.g. ${z}-{z}_1-j\epsilon(t-t_1)$ and $\bar{z}-\bar{z}_1$ remembering that $\big\vert z-z_1\big\vert<\min\limits_{p\neq1}\big\vert z_p-z_1\big\vert$, following \cite{Belavin:1984vu}. Here, we have collected only those OPEs that are expressed as Laurent series in the holomorphic variable $z$ (or $\tilde{z}=z-j\epsilon t$) but are (anti-)holomorphic in $\bar{z}$.

\medskip

The OPEs with the opposite pole-structures are noted below, with $\Delta\tilde{\bar{z}}_p:=\bar{z}-\bar{z}_p-j\epsilon(t-t_p)$, :
\begin{align}
&\bar{P}(t,z,\bar{z})\Phi(\mathbf{x}_p)\sim\lim\limits_{\epsilon\rightarrow0^+}-i\left(\frac{\partial_{t_p}}{\Delta\tilde{\bar{z}}_p}+\frac{\bar{\bm{\xi}}_p}{(\Delta\tilde{\bar{z}}_p)^2}\right)\Phi(\mathbf{x}_p)\\
&S_0^-(t,z,\bar{z})\Phi(\mathbf{x}_p)\sim\lim\limits_{\epsilon\rightarrow0^+}-i\left(\frac{z-z_p}{\Delta\tilde{\bar{z}}_p}\partial_{t_p}+\frac{z-z_p}{(\Delta\tilde{\bar{z}}_p)^2}\bar{\bm{\xi}}_p-\frac{{\bm{\xi}}_p}{\Delta\tilde{\bar{z}}_p}\right)\Phi(\mathbf{x}_p)\\
&\bar{T}(t,z,\bar{z})\Phi(\mathbf{x}_p)\sim\lim\limits_{\epsilon\rightarrow0^+}-i\left[\frac{\bar{h}_p}{(\Delta\tilde{\bar{z}}_p)^2}+\frac{\partial_{\bar{z}_p}}{\Delta\tilde{\bar{z}}_p}-\frac{t-t_p}{2}\left\{\frac{\partial_{t_p}}{(\Delta\tilde{\bar{z}}_p)^2}+\frac{2\bar{\bm{\xi}}_p}{{(\Delta\tilde{\bar{z}}_p)}^3}\right\}\right]\Phi(\mathbf{x}_p)\\
&S_1^-(t,z,\bar{z})\Phi(\mathbf{x}_p)\sim\lim\limits_{\epsilon\rightarrow0^+}-\frac{i}{2}\left[\frac{(z-z_p)^2}{\Delta\tilde{\bar{z}}_p}\partial_{z_p}-2h_p\frac{z-z_p}{\Delta\tilde{\bar{z}}_p}+\frac{0}{\Delta\tilde{\bar{z}}_p}\right.\nonumber\\
&\hspace{43mm}\left.+(t-t_p)\left(\frac{z-z_p}{\Delta\tilde{\bar{z}}_p}\partial_{t_p}+\frac{z-z_p}{(\Delta\tilde{\bar{z}}_p)^2}\bar{\bm{\xi}}_p-\frac{{\bm{\xi}}_p}{\Delta\tilde{\bar{z}}_p}\right)\right]\Phi(\mathbf{x}_p)
\end{align}
where $\sim$ denotes `modulo terms (anti-)holomorphic (regular) in $\Delta\tilde{\bar{z}}_p$ '.

\medskip

The Carrollian conformal primary fields can be alternatively defined as the fields that, in the above described OPEs, have no higher order poles than those shown explicitly and also have vanishing coefficients for the $\frac{{(\bar{z}-\bar{z}_p)}^0}{\Delta\tilde{z}_p}$ and $\frac{{(z-z_p)}^0}{\Delta\tilde{\bar{z}}_p}$ terms in the OPEs with $S_1^+$ and $S_1^-$ respectively\footnote{When we perform the 2D shadow transformation e.g. \eqref{49} to reach the $T\Phi$ OPE from the $S_1^-\Phi$ OPE, the $\frac{{(z-z_p)}^0}{\Delta\tilde{\bar{z}}_p}$ singularity gives rise to a cubic pole in the $T_e\Phi$ OPE. Thus, to respect the primary OPE \eqref{50}, its coefficient must vanish.}. These coefficients will be non-zero in addition to the appearance of higher order poles in the above OPEs corresponding to a general Carrollian conformal field. 

\medskip

Recalling that an OPE in a translation-invariant theory has the following general form:
\begin{align*}
{\Phi}_1(t_1,\vec{x}_1){\Phi}_2(t_2,\vec{x}_2)=\sum_kC^k_{12}(t_{12},\vec{x}_{12})\text{ }{\Phi}_k(t_2,\vec{x}_2)
\end{align*} 
with $\Phi_1$, $\Phi_2$ and $\Phi_k$ being local fields, we now extend the above OPEs to the case of a general (non-primary) Carrollian conformal field:
\begin{align}
&P(t,z,\bar{z})\Phi(\mathbf{x}_p)\sim\lim\limits_{\epsilon\rightarrow0^+}-i\left[\sum\limits_{n\geq1}\frac{\left(P_{n,-1}\Phi\right)}{{(\Delta\tilde{z}_p)}^{n+2}}+\frac{\partial_{t_p}\Phi}{(\Delta\tilde{z}_p)}+\frac{\bm{\xi}_p\cdot\Phi}{{(\Delta\tilde{z}_p)}^2}\right](\mathbf{x}_p)\label{62}\\
&S^+_0(t,z,\bar{z})\Phi(\mathbf{x}_p)\sim\lim\limits_{\epsilon\rightarrow0^+}-i\left[(\bar{z}-\bar{z}_p)\left(\sum\limits_{n\geq1}\frac{\left(P_{n,-1}\Phi\right)}{{(\Delta\tilde{z}_p)}^{n+2}}+\frac{\partial_{t_p}\Phi}{(\Delta\tilde{z}_p)}+\frac{\bm{\xi}_p\cdot\Phi}{{(\Delta\tilde{z}_p)}^2}\right)\right.\nonumber\\
&\hspace{78mm}\left.-\left(\sum\limits_{n\geq0}\frac{\left(P_{n,0}\Phi\right)}{{(\Delta\tilde{z}_p)}^{n+2}}+\frac{\bar{{\bm{\xi}}}_p\cdot\Phi}{(\Delta\tilde{z}_p)}\right)\right](\mathbf{x}_p)\label{64}\\
&T(t,z,\bar{z})\Phi(\mathbf{x}_p)\sim\lim\limits_{\epsilon\rightarrow0^+}-i\left[\sum\limits_{n\geq1}\frac{\left(L_{n}\Phi\right)}{{(\Delta\tilde{z}_p)}^{n+2}}+\frac{h_p\Phi}{(\Delta\tilde{z}_p)^2}+\frac{\partial_{z_p}\Phi}{\Delta\tilde{z}_p}\right.\nonumber\\
&\hspace{41mm}\left.-\frac{t-t_p}{2}\left\{\sum\limits_{n\geq1}(n+2)\frac{\left(P_{n,-1}\Phi\right)}{{(\Delta\tilde{z}_p)}^{n+3}}+\frac{\partial_{t_p}\Phi}{(\Delta\tilde{z}_p)^2}+\frac{2\bm{\xi}_p\cdot\Phi}{{(\Delta\tilde{z}_p)}^3}\right\}\right](\mathbf{x}_p)\label{70}\\
&\left(S_1^+-\frac{t-t_p}{2}S^+_0\right)(t,z,\bar{z})\Phi(\mathbf{x}_p)\sim\lim\limits_{\epsilon\rightarrow0^+}-\frac{i}{2}\left[(\bar{z}-\bar{z}_p)^2\left(\sum\limits_{n\geq1}\frac{\left(j^{(-)}_{n}\Phi\right)}{{(\Delta\tilde{z}_p)}^{n+1}}+\frac{\partial_{\bar{z}_p}\Phi}{\Delta\tilde{z}_p}\right)+\sum\limits_{n\geq0}\frac{\left(j^{(+)}_{n}\Phi\right)}{{(\Delta\tilde{z}_p)}^{n+1}}\right.\nonumber\\
&\hspace{68mm}\left.-2(\bar{z}-\bar{z}_p)\left(\sum\limits_{n\geq1}\frac{\left(j^{(0)}_{n}\Phi\right)}{{(\Delta\tilde{z}_p)}^{n+1}}+\frac{\bar{h}_p\Phi}{\Delta\tilde{z}_p}\right)\right](\mathbf{x}_p)\label{63}
\end{align} 
The OPE coefficients were labeled in the way that resembles the actions of classical generators at the origin, e.g. $\left(P_{0,-1}\Phi\right)=\bm{\xi}_p\cdot\Phi$ , $\left(P_{-1,0}\Phi\right)=\bar{\bm{\xi}}_p\cdot\Phi$ , $\left(L_{0}\Phi\right)=h_p\Phi$ etc. The coefficients of the $S^+_1$ OPE are labeled keeping in mind that its holomorphic weight is $h=1$. 

\medskip

The OPEs for the fields $T_e$, $j^a_e$, $P_0$ and $P_{-1}$ that are easily derived from above are now listed below:
\begin{align}
&P_{-1}(t,z,\bar{z})\Phi(\mathbf{x}_p)\sim\lim\limits_{\epsilon\rightarrow0^+}-i\left[\sum\limits_{n\geq1}\frac{\left(P_{n,-1}\Phi\right)}{{(\Delta\tilde{z}_p)}^{n+2}}+\frac{\partial_{t_p}\Phi}{(\Delta\tilde{z}_p)}+\frac{\bm{\xi}_p\cdot\Phi}{{(\Delta\tilde{z}_p)}^2}\right](\mathbf{x}_p)\label{59}\\
&P_{0}(t,z,\bar{z})\Phi(\mathbf{x}_p)\sim\lim\limits_{\epsilon\rightarrow0^+}-i\left[\sum\limits_{n\geq-1}\frac{\left(P_{n,0}\Phi\right)+\bar{z}_p\left(P_{n,-1}\Phi\right)}{{(\Delta\tilde{z}_p)}^{n+2}}\right](\mathbf{x}_p)\label{60}\\
&T_e(t,z,\bar{z})\Phi(\mathbf{x}_p)\sim\lim\limits_{\epsilon\rightarrow0^+}-i\left[\sum\limits_{n\geq-1}\frac{\left(L_{n}\Phi\right)+\frac{n+1}{2}t_p\left(P_{n-1,-1}\Phi\right)}{{(\Delta\tilde{z}_p)}^{n+2}}\right](\mathbf{x}_p)\label{58}\\
&j^{(-)}_e(t,z,\bar{z})\Phi(\mathbf{x}_p)\sim\lim\limits_{\epsilon\rightarrow0^+}-\frac{i}{2}\left[\sum\limits_{n\geq1}\frac{\left(j^{(-)}_{n}\Phi\right)}{{(\Delta\tilde{z}_p)}^{n+1}}+\frac{\partial_{\bar{z}_p}\Phi}{\Delta\tilde{z}_p}\right](\mathbf{x}_p)\label{54}\\
&j^{(0)}_e(t,z,\bar{z})\Phi(\mathbf{x}_p)\sim\lim\limits_{\epsilon\rightarrow0^+}-\frac{i}{2}\left[\sum\limits_{n\geq1}\frac{\left(j^{(0)}_{n}\Phi\right)+\bar{z}_p\left(j^{(-)}_{n}\Phi\right)+\frac{t_p}{2}\left(P_{n-1,-1}\Phi\right)}{{(\Delta\tilde{z}_p)}^{n+1}}\right.\nonumber\\
&\hspace{83mm}\left.+\frac{\left(\bar{z}_p\partial_{\bar{z}_p}+\frac{t_p}{2}\partial_{t_p}+\bar{h}_p\right)\Phi}{\Delta\tilde{z}_p}\right](\mathbf{x}_p)\\
&j^{(+)}_e\Phi\sim\lim\limits_{\epsilon\rightarrow0^+}-\frac{i}{2}\left[\sum\limits_{n\geq1}\frac{\left(j^{(+)}_{n}\Phi\right)+2\bar{z}_p\left(j^{(0)}_{n}\Phi\right)+\bar{z}^2_p\left(j^{(-)}_{n}\Phi\right)+t_p\bar{z}_p\left(P_{n-1,-1}\Phi\right)+t_p\left(P_{n-1,0}\Phi\right)}{{(\Delta\tilde{z}_p)}^{n+1}}\right.\nonumber\\
&\hspace{40mm}\left.+\frac{\left(\bar{z}^2_p\partial_{\bar{z}_p}+\bar{z}_pt_p\partial_{t_p}+2\bar{z}_p\bar{h}_p\right)\Phi+t_p\left(\bar{{\bm{\xi}}}_p\cdot\Phi\right)+\left(j^{(+)}_{0}\Phi\right)}{\Delta\tilde{z}_p}\right](\mathbf{x}_p)\label{55}
\end{align}
We recognize that the $j^a_e\Phi$ OPEs looks exactly the same as the holomorphic Kac-Moody current OPEs while the $T_e\Phi$ OPE resembles a holomorphic Virasoro OPE. The interpretation of the $j^a_e$ fields as the generators of a Kac-Moody symmetry \cite{Banerjee:2020zlg,Guevara:2021abz,Polyakov:1987zb}  is further hinted by the fact that each of them has $h=1$. On the other hand, the fact that $T_e$ has $h=2$ hints toward an interpretation of the $T_e$ field as a Virasoro energy-momentum tensor \cite{Kapec:2016jld,Cheung:2016iub}. But these interpretations are not straight-forward, since some of these fields have non-zero $\bar{h}$ unlike in 2D holomorphic CFTs. Nevertheless, below we shall proceed to define the modes of these fields and to find their algebra.

\medskip
 
Similar OPEs can be obtained for the anti-holomorphic sector. 

\medskip

\subsection{The modes and their actions}\label{s5.2}
It is to be noted that these six fields are time-independent and holomorphic when written in an OPE. Thus, they can be mode-expanded (valid only in OPEs) as below:
\begin{align}
&T_e(z)=\sum\limits_{n\in\mathbb{Z}}L_nz^{-n-2}\hspace{2.5mm};\hspace{2.5mm}L_n=\frac{1}{2\pi j}\oint\limits_{C_u^\prime}d\hat{z}\text{ }\hat{z}^{n+1}\text{ }T_e(t,\hat{z},\hat{\bar{z}})=\frac{1}{2\pi i}\oint\limits_{C_e}dz\text{ }z^{n+1}\text{ }T_e(t,z,\bar{z})\nonumber\\
&j^a_e(z)=\frac{1}{2}\sum\limits_{n\in\mathbb{Z}}j^a_nz^{-n-1}\hspace{2.5mm};\hspace{2.5mm}j^a_n=\frac{2}{2\pi j}\oint\limits_{C_u^\prime}d\hat{z}\text{ }\hat{z}^{n}\text{ }j^a_e(t,\hat{z},\hat{\bar{z}})=\frac{2}{2\pi i}\oint\limits_{C_e}dz\text{ }z^{n}\text{ }j^a_e(t,z,\bar{z})\label{75}\\
&P_i(z)=\sum\limits_{n\in\mathbb{Z}}P_{n,i}z^{-n-2}\hspace{2.5mm};\hspace{2.5mm}P_{n,i}=\frac{1}{2\pi j}\oint\limits_{C_u^\prime}d\hat{z}\text{ }\hat{z}^{n+1}\text{ }P_i(t,\hat{z},\hat{\bar{z}})=\frac{1}{2\pi i}\oint\limits_{C_e}dz\text{ }z^{n+1}\text{ }P_{i}(t,z,\bar{z})\label{76}
\end{align}
with $i\in\{0,-1\}$.

\medskip
 
We first note the actions of the zero-modes $j^a_0$ on an arbitrary field; this can be obtained\footnote{Section \ref{s5.3} contains an illustration.} from the analogue of \eqref{53} applied on the OPEs \eqref{54}-\eqref{55}:
\begin{align}
&\left[j_0^{(-)}\text{ },\text{ }\Phi(\mathbf{x}_p)\right]=-i\partial_{\bar{z}_p}\Phi(\mathbf{x}_p)\hspace{5mm};\hspace{5mm}\left[j_0^{(0)}\text{ },\text{ }\Phi(\mathbf{x}_p)\right]=-i\left[\bar{z}_p\partial_{\bar{z}_p}+\frac{t_p}{2}\partial_{t_p}+\bar{h}_p\right]\Phi(\mathbf{x}_p)\nonumber\\
&\left[j_0^{(+)}\text{ },\text{ }\Phi(\mathbf{x}_p)\right]=-i\left[\left(\bar{z}^2_p\partial_{\bar{z}_p}+\bar{z}_pt_p\partial_{t_p}+2\bar{z}_p\bar{h}_p\right)\Phi+t_p\left(\bar{{\bm{\xi}}}_p\cdot\Phi\right)+\left(j^{(+)}_{0}\Phi\right)\right](\mathbf{x}_p)
\end{align}
Comparing these actions with \eqref{56} and \eqref{57} for a general non-multiplet field (i.e. fields with $\bar{\bm{\mathcal{L}}}_1(\mathbf{0})\Phi(\mathbf{0})\neq0$), we see that the zero-modes of the three $j^a_e$ fields generate the three anti-holomorphic Lorentz, i.e. the $\overline{\text{SL}(2,\mathbb{R})}$ transformations on the space of quantum fields. Thus, the possible Kac-Moody algebra generated by the $j^a_e$ fields can be the $\overline{\text{sl}(2,\mathbb{R})}$ current algebra. 

\medskip

On the other hand, from the OPE \eqref{58}, we find that the three modes $L_n$ with $n\in\{0,\pm1\}$ generate the three holomorphic Lorentz, i.e. the $\text{SL}(2,\mathbb{R})$ transformations of the quantum fields: 
\begin{align}
&\left[L_{-1}\text{ },\text{ }\Phi(\mathbf{x}_p)\right]=-i\partial_{z_p}\Phi(\mathbf{x}_p)\hspace{5mm};\hspace{5mm}\left[L_0\text{ },\text{ }\Phi(\mathbf{x}_p)\right]=-i\left[z_p\partial_{z_p}+\frac{t_p}{2}\partial_{t_p}+h_p\right]\Phi(\mathbf{x}_p)\nonumber\\
&\left[L_{1}\text{ },\text{ }\Phi(\mathbf{x}_p)\right]=-i\left[\left(z_p^2\partial_{z_p}+z_pt_p\partial_{t_p}+2z_ph_p\right)\Phi+t_p\left(\bm{\xi}_p\cdot\Phi\right)+\left(L_1\Phi\right)\right](\mathbf{x}_p)
\end{align}
This was expected from the definition \eqref{42} of the quantum holomorphic super-rotation generators.

\medskip

Thus, we see that the six modes $L_n$ with $n\in\{0,\pm1\}$ and $j^a_0$ together generate the Lorentz i.e. $\text{SL}(2,\mathbb{C})$ transformations of the quantum fields in $1+2$D. In addition to that, from the OPEs \eqref{59} and \eqref{60}, we find that the four modes $P_{a,b}$ with $a,b\in\{-1,0\}$ implement the four translations of the Poincare group $\text{ISL}(2,\mathbb{C})$ on the $1+2$D quantum fields:  
\begin{align}
&\left[P_{-1,-1}\text{ },\text{ }\Phi(\mathbf{x}_p)\right]=-i\partial_{t_p}\Phi(\mathbf{x}_p)\nonumber\\
&\left[P_{0,-1}\text{ },\text{ }\Phi(\mathbf{x}_p)\right]=-i\left[z_p\partial_{t_p}+\bm{\xi}_p\right]\Phi(\mathbf{x}_p)\hspace{5mm};\hspace{5mm}\left[P_{-1,0}\text{ },\text{ }\Phi(\mathbf{x}_p)\right]=-i\left[\bar{z}_p\partial_{t_p}+\bar{\bm{\xi}}_p\right]\Phi(\mathbf{x}_p)\\
&\left[P_{0,0}\text{ },\text{ }\Phi(\mathbf{x}_p)\right]=-i\left[z_p\bar{z}_p\partial_{t_p}\Phi+z_p\left(\bar{\bm{\xi}}_p\cdot\Phi\right)+\bar{z}_p\left({\bm{\xi}}_p\cdot\Phi\right)+\left(P_{0,0}\Phi\right)\right](\mathbf{x}_p)\nonumber
\end{align}

\medskip

We shall now list the actions of the modes on a $1+2$D CC primary field that are derived from the primary OPEs (with $n\in\mathbb{Z}$):
\begin{align}
&\left[P_{n,-1}\text{ },\text{ }\Phi(\mathbf{x}_p)\right]=-i\left[z_p^{n+1}\partial_{t_p}+(n+1)z_p^n\bm{\xi}_p\right]\Phi(\mathbf{x}_p)\nonumber\\
&\left[P_{n,0}\text{ },\text{ }\Phi(\mathbf{x}_p)\right]=-i\left[z_p^{n+1}\left(\bar{z}_p\partial_{t_p}+\bar{\bm{\xi}}_p\right)+(n+1)z_p^n\bar{z}_p\bm{\xi}_p\right]\Phi(\mathbf{x}_p)\\
&\left[L_{n}\text{ },\text{ }\Phi(\mathbf{x}_p)\right]=-i\left[z_p^{n+1}\partial_{z_p}+\frac{n+1}{2}z_p^nt_p\partial_{t_p}+(n+1)z_p^nh_p+(n+1)nz_p^{n-1}\frac{t_p}{2}\bm{\xi}_p\right]\Phi(\mathbf{x}_p)\\
&\left[j^{(-)}_{n}\text{ },\text{ }\Phi(\mathbf{x}_p)\right]=-iz_p^{n}\partial_{\bar{z}_p}\Phi(\mathbf{x}_p)\nonumber\\
&\left[j^{(0)}_{n}\text{ },\text{ }\Phi(\mathbf{x}_p)\right]=-i\left[z^n_p\left(\bar{z}_p\partial_{\bar{z}_p}+\frac{t_p}{2}\partial_{t_p}+\bar{h}_p\right)+nz^{n-1}_p\frac{t_p}{2}\bm{\xi}_p\right]\Phi(\mathbf{x}_p)\\
&\left[j^{(+)}_{n}\text{ },\text{ }\Phi(\mathbf{x}_p)\right]=-i\left[z_p^n\left(\bar{z}^2_p\partial_{\bar{z}_p}+\bar{z}_pt_p\partial_{t_p}+2\bar{z}_p\bar{h}_p+t_p\bar{{\bm{\xi}}}_p\right)+nz_p^{n-1}\bar{z}_p{t_p}\bm{\xi}_p\right]\Phi(\mathbf{x}_p)\nonumber
\end{align}
While the first three are recognized respectively to be the $z^{n+1}$ and $z^{n+1}\bar{z}$ super-translations and holomorphic super-rotations from \eqref{36} and \eqref{37}, the classical counterparts of the last three transformations were not discussed. They can be thought of as the infinitesimal versions of the following Carrollian diffeomorphisms, with $q=-1,0,1$ respectively, on the $1+2$D Carrollian space-time:
\begin{align}
z\rightarrow z^\prime=z\hspace{2.5mm};\hspace{2.5mm}\bar{z}\rightarrow \bar{z}^\prime=\bar{z}+\bar{z}^{q+1}f(z)\hspace{2.5mm};\hspace{2.5mm}t\rightarrow t^\prime=t{[1+(q+1)\bar{z}^{q} f(z)]}^{\frac{1}{2}}\label{61}
\end{align}
with $f(z)$ being a meromorphic function. They generalize the $S^2$ diffeomorphisms discussed in \cite{Campiglia:2014yka,Campiglia:2015yka,Compere:2018ylh,Donnay:2020guq,Campiglia:2020qvc}.

\medskip

From the actions of the modes $j^{(+)}_0$ and $L_1$, it is clear that if a field $\Phi$ is to transform covariantly (i.e. like the CC primary fields) under the Lorentz i.e. $\text{SL}(2,\mathbb{C})$ transformations, it must have:
\begin{align}
\left(j^{(+)}_{0}\Phi\right)=0=\left(L_1\Phi\right)\label{67}
\end{align}
These criteria are equivalent to the definition of quasi-primary fields in 2D CFTs \cite{Belavin:1984vu}; these fields will be called `Lorentz quasi-primaries'. If, in addition to these, $\left(P_{0,0}\Phi\right)=0$ holds the field covariantly transforms under the Poincare i.e. $\text{ISL}(2,\mathbb{C})$ group. The quantum field is then called a $1+2$D CC quasi-primary field; thus, it transforms covariantly under the BMS$_4$ transformations globally defined on $\mathbb{R}\times S^2$.

\medskip

Thus from the holomorphic OPEs, we obtain all of the holomorphic super-rotations but only two types of super-translations and the quantum charges (modes) that generate those. This is consistent with our discussion in section \ref{s4.2} where we encountered various difficulties in defining finite charges generating the complete set of the super-translations. Seemingly as a compensation to this, the holomorphic OPEs allowed us to define charges generating a special kind of Carrollian diffeomorphism \eqref{61} that is not a subset of the (extended) BMS$_4$ transformations.

\medskip

Similarly, the anti-holomorphic OPEs give all the anti-holomorphic super-rotations generated by $\bar{L}_n$ but only two types of super-translations generated $P_{-1,n}$ and $P_{0,n}$ that are conjugate to those obtained in the holomorphic case along with the anti-holomorphic version of the diffeomorphism \eqref{61}. Here, the zero-modes of the three Kac-Moody currents generate the $\text{SL}(2,\mathbb{R})$ transformations while $\bar{L}_n$ with $n\in\{0,\pm1\}$ generate the $\overline{\text{SL}(2,\mathbb{R})}$ transformations; thus, together they generate the $\text{SL}(2,\mathbb{C})$ transformations.

\medskip

But, we can not conclude that we have obtained a union of the two sets of quantum transformations coming from the OPEs of the two sectors. This is so because the union of the corresponding quantum generators (modes) does not form an algebra, as we shall now see.

\medskip

\subsection{The symmetry algebra}\label{s5.3}
We begin by recalling that all the field-operators involved in an OPE must be local fields. So, a field and its shadow transformation which is a non-local object can not both be involved in an OPE \cite{Banerjee:2022wht}. Below we shall explicitly see the problems that arise in the case of $1+2$D CCFT when we try to construct an OPE of a field and its shadow. 

\medskip

The seemingly infinite Laurent series in the holomorphic sector OPEs can be truncated at finite order poles by demanding that there be no local field with negative holomorphic weight\footnote{Following \cite{Banerjee:2022wht}, this assumption may be interpreted as to be that the Carrollian conformal fields with $h<0$ are to be treated as the non-local 2D shadow-transformations of the $h>1$ fields which are taken to be the local fields in the holomorphic sector.} i.e. all the local fields have $h\geq0$. In the anti-holomorphic sector, the similar demand would be $\bar{h}\geq0$ for all local fields. One of these will be a very crucial assumption in what follows, similar to the situation in a 2D relativistic CFT \cite{Zamolodchikov:1989mz}. Both of these demands are more relaxing than the demand that scaling dimensions $\Delta\geq0$ for all local fields (in any sector).

\medskip

Following the celestial CFT literature e.g. \cite{Fotopoulos:2019tpe}, we shall also assume that the eight fields $P,\bar{P},T,\bar{T},S^\pm_0$ and $S^\pm_1$ either transform covariantly under the Lorentz i.e. $\text{SL}(2,\mathbb{C})$ transformations (i.e. either they are Lorentz quasi-primaries) or are global descendants of appropriate Lorentz covariant fields. We need not bother at this stage if they are $1+2$D CC quasi-primaries or descendants thereof.

\medskip

As will be demonstrated, these two assumptions stated above together with the bosonic exchange property of these fields (all of them have integer spins) are strong enough to completely specify the pole-singularities of the (allowed) mutual OPEs of those eight fields. From these pole structures, we can readily deduce the algebra of the modes.

\medskip

We now have to choose the sets of the local fields from those eight fields such that members of one set can consistently form mutual OPEs. From the definitions in section \ref{s3}, it is evident that only one member from each of the following sets can be a local field: $\{P,\bar{P}\}$, $\{S^+_0,S^-_0\}$, $\{T,S^-_1\}$ and $\{S^+_1,\bar{T}\}$. We discuss on the following remaining possibilities:
\begin{enumerate}
\item\textbf{Taking $P$ as a local Lorentz quasi-primary field}: This choice immediately tells that $S^+_0$ can not be a Lorentz quasi-primary but still be a local field while $\bar{P}$ and $S^-_0$ can not even be local fields.

\medskip

We recall that, in the derivation of \eqref{44}, the contact term (on $S^2$) in the primary Ward identity \eqref{19} played a very important role; due to this term, \eqref{44} remains valid for arbitrary super-translations. But, as discussed in the beginning of this section, such a contact term does not appear in the OPEs. Hence this choice makes us lose crucial information when we go from Ward identity to the OPE.

\medskip

Besides, we expect from \eqref{62} that any general OPE\footnote{This is due to the fact that the OPE must be associative. Along with the two local fields which are explicitly shown in the L.H.S. of an OPE, there are $n(\geq0)$ other local fields. We can choose any pairs from this $n+2$ local fields and use the OPEs corresponding to those pairs. All of these different pairings (and OPEs) should give the same final result: this is the statement of the crossing symmetry hypothesis.} with $\bar{\partial} P$ on the L.H.S. is regular (up to contact terms). But the $S^+_1\bar{\partial}P$ OPE violates this: since $P$ has non-vanishing $\bar{h}=\frac{1}{2}$ and it is assumed to be a Lorentz quasi-primary, we have a simple pole singularity in the $S^+_1\bar{\partial}P$ OPE coming from \eqref{63} that does not go away even if we assume $P$ to be a primary field. Similarly, the $\bar{T}\bar{\partial}P$ and the $S^-_1\bar{\partial}P$ OPEs contain pole singularities instead of being regular.

\medskip

Thus, if $P$ is to be taken as a Lorentz quasi-primary local field, both $S_1^\pm$ and $\bar{T}$ can not be local fields. But the full $\text{isl}(2,\mathbb{C})$ will then not be a sub-algebra of the resulting mode-algebra because all of the $\overline{\text{sl}(2,\mathbb{R})}$ generators will be absent. 

\medskip

Hence this is not a valid choice i.e. $P$ (or $\bar{P}$) can not be treated as a Lorentz quasi-primary field. 
\item\textbf{Taking $S^+_0$ as a local Lorentz quasi-primary field}: This immediately renders $S^-_0$ and $\bar{P}$ non-local while $P$ is now a local but not Lorentz quasi-primary descendant of $S^+_0$. Also, the $S_0^+$ Ward identities (for primary fields) do not contain any contact terms, so we can directly write the corresponding OPEs without any loss.

\medskip

Now, any general OPE with $\bar{\partial}^2 S^+_0$ on the L.H.S. should be regular (up to contact terms), following from \eqref{64}. But $\bar{T}\bar{\partial}^2 S^+_0$ and $S_1^-\bar{\partial}^2 S^+_0$  OPEs both have pole singularities that persist even if $S^+_0$ is assumed to be a primary field. This happens as both $h,\bar{h}$ are non-zero for $S^+_0$. Thus, under this choice $\bar{T}$ and $S^-_1$ can not be treated as local fields.

\medskip

On the other hand, we can not yet conclude if both ${T}\bar{\partial}^2 S^+_0$ and $S_1^+\bar{\partial}^2 S^+_0$ are regular. So we can proceed assuming for the time being that $T$, $S_1^+$ and $S_0^+$ can be simultaneously treated as local Lorentz quasi-primary fields and we will find that this is indeed the case.
\item\textbf{Taking $S^-_0$ as a local Lorentz quasi-primary field}: Similarly, in this case ${T}$, $S^+_1$, $P$ and $S^+_0$ can not be treated as local fields.
\item\textbf{Taking both $T$ and $\bar{T}$ as local Lorentz quasi-primary fields}: Here both $S^\pm_1$ are to be treated as non-local shadows. But from the discussion similar to the first two cases, it is evident that none of $P$, $\bar{P}$ and $S^\pm_0$ can form consistent OPEs with both of $T$ and $\bar{T}$. From \eqref{70}, any OPE with $\bar{\partial}T$ or $\partial\bar{T}$ on the L.H.S. is expected to be regular; but the OPEs $S^+_0\partial\bar{T}$ (hence $P\partial\bar{T}$) and $S^-_0\bar{\partial}T$ (hence $\bar{P}\bar{\partial}T$) all have pole singularities. Thus, only $T$ and $\bar{T}$ can be treated as local fields in this case.

\medskip

Also, from the definition \eqref{24}, it follows that both $\partial P$ and $\bar{\partial}\bar{P}$ have to be local fields in order to both $T$ and $\bar{T}$ being local. Clearly, $\partial P$ and $\bar{\partial}\bar{P}$ both can not be local simultaneously.

\medskip

Moreover, simultaneously treating both $T_e$ and $\bar{T}_e$ local is also unacceptable since all the super-translation generators then will be absent from the resulting mode-algebra implying that the $\text{isl}(2,\mathbb{C})$ is not even a sub-algebra of the same.
\item\textbf{Taking both $S^\pm_1$ as local Lorentz quasi-primary fields}: This is not possible. A general OPE with $\bar{\partial}^3S_1^+$ (or ${\partial}^3S_1^-$) on the L.H.S. is expected to be regular, as seen from \eqref{63}. This is not what happens to the $S_1^-\bar{\partial}^3S_1^+$ OPE which has anti-meromorphic pole singularities. On the other hand, the OPE $S_1^+{\partial}^3S_1^-$ contains meromorphic poles instead of being regular. Hence, both $S^\pm_1$ can not be treated as local fields simultaneously.
\end{enumerate}  
Therefore we have to pick one among the two sectors to form consistent mutual OPEs \cite{Banerjee:2022wht}. We proceed with the holomorphic one by treating the fields $T$, $S^+_0$ and $S^+_1$ as local Lorentz quasi-primary fields while demoting the others to mere non-local shadows or descendants thereof. This is the holomorphic sector of the $1+2$D Carrollian CFT. Thus, we shall also assume that all the local fields in the theory have holomorphic weights $h\geq0$.

\medskip

Obeying the general form \eqref{64} of the $S^+_0$ OPE and being consistent with the two assumptions, we write down the following ansatz for the $S^+_0S^+_0$ OPE:  
\begin{align}
&S^+_0(t,z,\bar{z})S^+_0(t_p,z_p,\bar{z}_p)\sim\lim\limits_{\epsilon\rightarrow0^+}-i\left[(\bar{z}-\bar{z}_p)\left(\frac{\left(P_{1,-1}S^+_0\right)}{{(\Delta\tilde{z}_p)}^{3}}+\frac{\bm{\xi}\cdot S^+_0}{{(\Delta\tilde{z}_p)}^2}\right)\right.\nonumber\\
&\hspace{60mm}\left.-\left(\frac{\left(P_{1,0}S^+_0\right)}{{(\Delta\tilde{z}_p)}^{3}}+\frac{\left(P_{0,0}S^+_0\right)}{{(\Delta\tilde{z}_p)}^{2}}+\frac{\bar{{\bm{\xi}}}\cdot S^+_0}{(\Delta\tilde{z}_p)}\right)\right](t_p,z_p,\bar{z}_p)\label{65}
\end{align}
Because $\partial_tS^+_0\sim0$ within a correlator (or an OPE), all the local fields appearing on the R.H.S. have this property in common. Also, the condition $S^+_0(t,z,\bar{z})\bar{\partial}^2_p S^+_0(t_p,z_p,\bar{z}_p)\sim0$ gives rise to some constraints between these fields. 

\medskip

Moreover, since $S^+_0$ has integer spin $m=2$, it should obey the bosonic exchange property, i.e. we expect as an OPE statement that:
\begin{align*}
S^+_0(t,z,\bar{z})S^+_0(t_p,z_p,\bar{z}_p)\sim S^+_0(t_p,z_p,\bar{z}_p)S^+_0(t,z,\bar{z})
\end{align*}
We first expand the OPE on the R.H.S. as:
\begin{align}
&S^+_0(t_p,z_p,\bar{z}_p)S^+_0(t,z,\bar{z})\sim\lim\limits_{\epsilon\rightarrow0^+}-i\left[(\bar{z}-\bar{z}_p)\left(\frac{\left(P_{1,-1}S^+_0\right)}{{(\Delta\tilde{z}_p)}^{3}}-\frac{\bm{\xi}\cdot S^+_0}{{(\Delta\tilde{z}_p)}^2}\right)\right.\nonumber\\
&\hspace{60mm}\left.+\left(\frac{\left(P_{1,0}S^+_0\right)}{{(\Delta\tilde{z}_p)}^{3}}-\frac{\left(P_{0,0}S^+_0\right)}{{(\Delta\tilde{z}_p)}^{2}}+\frac{\bar{{\bm{\xi}}}\cdot S^+_0}{(\Delta\tilde{z}_p)}\right)\right](t,z,\bar{z})\label{66}
\end{align}
For comparing the both sides of the bosonic exchange property, we now need to perform a multi-variate Taylor expansion of the fields at $(t,z,\bar{z})$ around $(t_p,z_p,\bar{z}_p)$ on the R.H.S. of \eqref{66} and look for order-by-order matching of the coefficients of $\frac{(\bar{z}-\bar{z}_p)^s}{(z-z_p)^r}$ , where $r\geq1$ and $s\geq0$, with those on the R.H.S. of \eqref{65}.

\medskip

E.g. we can promptly notice that $\left(P_{1,0}S^+_0\right)\sim0$ i.e. $\left(P_{1,0}S^+_0\right)$ is $0$ upto contact terms inside a correlator. The constraints coming from the bosonic exchange property and from the requirement that $S^+_0(t,z,\bar{z})\bar{\partial}^2_p S^+_0(t_p,z_p,\bar{z}_p)\sim0$ have some overlap.

\medskip

Next we make an ansatz for the $S_1^+S^+_0$ OPE consistent with the assumptions\footnote{We recall from \eqref{67} that if $S^+_0$ is to be a Lorentz quasi-primary it must satisfy: $\left(j^{(+)}_{0}S^+_0\right)=0$.} and the fact that $S^+_0$ has $\bar{h}=-\frac{1}{2}$, looking at the general form \eqref{63}:
\begin{align}
&S_1^+(t,z,\bar{z})S^+_0(t_p,z_p,\bar{z}_p)\sim\lim\limits_{\epsilon\rightarrow0^+}-\frac{i}{2}\left[(\bar{z}-\bar{z}_p)^2\left(\frac{\Phi_{\frac{1}{2},\frac{1}{2}}}{{(\Delta\tilde{z}_p)}^{2}}+\frac{\partial_{\bar{z}_p}S^+_0}{\Delta\tilde{z}_p}\right)+(\bar{z}-\bar{z}_p)\left(\frac{\Phi_{\frac{1}{2},-\frac{1}{2}}}{{(\Delta\tilde{z}_p)}^{2}}+\frac{S^+_0}{\Delta\tilde{z}_p}\right)\right.\nonumber\\
&\left.+\frac{\Phi_{\frac{1}{2},-\frac{3}{2}}}{{(\Delta\tilde{z}_p)}^{2}}+(t-t_p)\left\{(\bar{z}-\bar{z}_p)\left(\frac{\left(P_{1,-1}S^+_0\right)}{{(\Delta\tilde{z}_p)}^{3}}+\frac{\bm{\xi}\cdot S^+_0}{{(\Delta\tilde{z}_p)}^2}\right)-\left(\frac{\left(P_{0,0}S^+_0\right)}{{(\Delta\tilde{z}_p)}^{2}}+\frac{\bar{{\bm{\xi}}}\cdot S^+_0}{(\Delta\tilde{z}_p)}\right)\right\}\right](\mathbf{x}_p)\label{68}
\end{align} 
where the $\Phi_{h,\bar{h}}$ are as of yet undetermined local fields.

\medskip

On the other hand, following from \eqref{64}, the ansatz for the $S_0^+S^+_1$ OPE is allowed to be:
\begin{align}
S^+_0(t,z,\bar{z})S^+_1(\mathbf{x}_p)\sim\lim\limits_{\epsilon\rightarrow0^+}-i\left[(\bar{z}-\bar{z}_p)\left(\frac{\bm{\xi}\cdot S^+_1}{{(\Delta\tilde{z}_p)}^2}+\frac{\partial_{t_p} S^+_1}{{(\Delta\tilde{z}_p)}}\right)+\frac{\frac{1}{2}\Phi_{\frac{1}{2},-\frac{3}{2}}}{{(\Delta\tilde{z}_p)}^{2}}-\frac{\bar{{\bm{\xi}}}\cdot S^+_1}{(\Delta\tilde{z}_p)}\right](\mathbf{x}_p)
\end{align} 

\medskip

Now, since both of $S^+_1$ and $S^+_0$ has integer spin $m=2$, they satisfy the following bosonic exchange property:
\begin{align*}
S^+_1(t,z,\bar{z})S^+_0(t_p,z_p,\bar{z}_p)\sim S^+_0(t_p,z_p,\bar{z}_p)S^+_1(t,z,\bar{z})
\end{align*}
Thus, to compare we need to Taylor-expand the R.H.S. of the following OPE around $(t_p,z_p,\bar{z}_p)$ and compare order-by-order with the R.H.S. of \eqref{68}:
\begin{align*}
S^+_0(t_p,z_p,\bar{z}_p)S^+_1(\mathbf{x})\sim\lim\limits_{\epsilon\rightarrow0^+}-i\left[(\bar{z}-\bar{z}_p)\left(-\frac{\bm{\xi}\cdot S^+_1}{{(\Delta\tilde{z}_p)}^2}+\frac{\partial_{t_p} S^+_1}{{(\Delta\tilde{z}_p)}}\right)+\frac{\frac{1}{2}\Phi_{\frac{1}{2},-\frac{3}{2}}}{{(\Delta\tilde{z}_p)}^{2}}+\frac{\bar{{\bm{\xi}}}\cdot S^+_1}{(\Delta\tilde{z}_p)}\right](t,z,\bar{z})
\end{align*}  

\medskip

Upon comparison (and using the fact that $\left(P_{0,0}S^+_0\right)$ has $(h,\bar{h})=(1,-1)$ ) we find:
\begin{align*}
&\left(P_{1,-1}S^+_0\right)\sim{{\bm{\xi}}}\cdot S^+_0\sim\bar{{\bm{\xi}}}\cdot S^+_0\sim0\\
&\partial_t\left(P_{0,0}S^+_0\right)\sim\partial_z\left(P_{0,0}S^+_0\right)\sim\partial_{\bar{z}}\left(P_{0,0}S^+_0\right)\sim0\hspace{2.5mm}\Longrightarrow\hspace{2.5mm}{\left(P_{0,0}S^+_0\right)}\sim0\\
&\partial_t\left({{\bm{\xi}}}\cdot S^+_1\right)\sim\partial_t\left(\bar{{\bm{\xi}}}\cdot S^+_1\right)\sim\partial_t\Phi_{\frac{1}{2},-\frac{3}{2}}\sim\partial_t\Phi_{\frac{1}{2},\frac{1}{2}}\sim\partial_t\Phi_{\frac{1}{2},-\frac{1}{2}}\sim0
\end{align*}
and some other not-so-simple constraints which can also be obtained from the requirements that $S^+_0\bar{\partial}^3 S^+_1\sim0$ ,  $S^+_1\bar{\partial}^2 S^+_0\sim0$ and $\left(\partial_tS^+_1-\frac{1}{2}S^+_0\right)\sim0$. The first two lines ensure that there are no singularity in the $S^+_0S^+_0$ OPE. 

\medskip

Finally, we make the ansatz for the $S^+_1S^+_1$ OPE, keeping in mind that $S^+_1$ has $\bar{h}=-1$ and $\left(j^{(+)}_{0}S^+_1\right)=0$ due to the assumption that $S^+_1$ is Lorentz quasi-primary:
\begin{align}
&S_1^+(t,z,\bar{z})S^+_1(t_p,z_p,\bar{z}_p)\sim\lim\limits_{\epsilon\rightarrow0^+}-\frac{i}{2}\left[(\bar{z}-\bar{z}_p)^2\left(\frac{\Phi_{0,0}}{{(\Delta\tilde{z}_p)}^{2}}+\frac{\partial_{\bar{z}_p}S^+_1}{\Delta\tilde{z}_p}\right)+(\bar{z}-\bar{z}_p)\left(\frac{\Phi_{0,-1}}{{(\Delta\tilde{z}_p)}^{2}}+\frac{2S^+_1}{\Delta\tilde{z}_p}\right)\right.\nonumber\\
&\hspace{15mm}\left.+\frac{\Phi_{0,-2}}{{(\Delta\tilde{z}_p)}^{2}}+(t-t_p)\left\{(\bar{z}-\bar{z}_p)\left(\frac{\bm{\xi}\cdot S^+_1}{{(\Delta\tilde{z}_p)}^2}+\frac{\partial_{t_p} S^+_1}{{(\Delta\tilde{z}_p)}}\right)+\frac{\frac{1}{2}\Phi_{\frac{1}{2},-\frac{3}{2}}}{{(\Delta\tilde{z}_p)}^{2}}-\frac{\bar{{\bm{\xi}}}\cdot S^+_1}{(\Delta\tilde{z}_p)}\right\}\right](\mathbf{x}_p)
\end{align}
where the fields $\Phi_{h,\bar{h}}$ are yet undetermined local fields. Using the bosonic exchange property similarly as in the case of $S^+_0S^+_0$ OPE, we obtain several constraints that combined with the previously obtained ones completely determine all the $\Phi_{h,\bar{h}}$ fields (within an OPE). The results, consistent with the dimensions of the $\Phi_{h,\bar{h}}$ fields, are:
\begin{align*}
&\left({{\bm{\xi}}}\cdot S^+_1\right)\sim\left(\bar{{\bm{\xi}}}\cdot S^+_1\right)\sim\Phi_{\frac{1}{2},-\frac{3}{2}}\sim\Phi_{\frac{1}{2},\frac{1}{2}}\sim\Phi_{\frac{1}{2},-\frac{1}{2}}\sim0\hspace{5mm};\hspace{5mm}\Phi_{0,0}\sim K\\
&\Phi_{0,-2}(\mathbf{x}_1)\Psi(\mathbf{x}_2)\sim A{(\bar{z}_1-\bar{z}_2)}^2\Psi(\mathbf{x}_2)\hspace{5mm};\hspace{5mm}\Phi_{0,-1}(\mathbf{x}_1)\Psi(\mathbf{x}_2)\sim A{(\bar{z}_1-\bar{z}_2)}\Psi(\mathbf{x}_2)
\end{align*} 
where $K$ and $A$ are two constants and $\Psi$ is an arbitrary local field.

\medskip

Since $\left(P_{0,0}S^+_1\right)=-\frac{1}{2}\Phi_{\frac{1}{2},-\frac{3}{2}}\sim0$ and we assumed $S^+_1$ to be Lorentz quasi-primary, we have just shown that $S^+_1$ is a $1+2$D CC quasi-primary field. Moreover, ${{\bm{\xi}}}\cdot S^+_1\sim\bar{{\bm{\xi}}}\cdot S^+_1\sim0$ and, as we shall see, $S^+_1$ does not mix with any other field under Carrollian boost; these imply that $S^+_1$ transforms under an irrep of the spin-boost sub-algebra \eqref{eq:7}. Thus, under the global $1+2$D CC i.e. $\text{ISL}(2,\mathbb{C})$ transformations, $S^+_1$ transforms as a 2D CFT quasi-primary i.e. as \eqref{69}. Correlation function between two such $1+2$D CC quasi-primaries were derived in \cite{Bagchi:2016bcd} to be exactly of the same form as that of the 2D CFT quasi-primary two-point functions. So, the $\left\langle S_1^+S^+_1\right\rangle$ correlator should be:
\begin{align}
\left\langle S_1^+(t_1,z_1,\bar{z}_1)S^+_1(t_2,z_2,\bar{z}_2)\mathbf{I}(t_3,z_3,\bar{z}_3)\right\rangle\propto\lim\limits_{\epsilon\rightarrow0^+}\frac{{(\bar{z}_1-\bar{z}_2)}^2}{\left[{(z_1-z_2)-j\epsilon(t_1-t_2)}\right]^2}
\end{align}
If we calculate this correlator using the $S_1^+S^+_1$ OPE and the fact that $\left\langle S^+_1\right\rangle$=0, it is immediately seen that $A=0$ must hold.

\medskip

Thus, we note down the final form of the following OPEs:
\begin{align}
&S_0^+(\mathbf{x})S^+_0(\mathbf{x}_p)\sim0\hspace{5mm};\hspace{5mm} S_0^+(\mathbf{x})S^+_1(\mathbf{x}_p)\sim\lim\limits_{\epsilon\rightarrow0^+}-i\text{ }\frac{\bar{z}-\bar{z}_p}{{(\Delta\tilde{z}_p)}}\text{ }\partial_{t_p} S^+_1(\mathbf{x}_p)\nonumber\\
&S_1^+(\mathbf{x})S^+_0(\mathbf{x}_p)\sim\lim\limits_{\epsilon\rightarrow0^+}-\frac{i}{2}\left[\frac{(\bar{z}-\bar{z}_p)^2}{{(\Delta\tilde{z}_p)}}\text{ }\partial_{\bar{z}_p} S^+_0+\frac{\bar{z}-\bar{z}_p}{{(\Delta\tilde{z}_p)}}\text{ }S^+_0\right](\mathbf{x}_p)\\
&\left(S_1^+-\frac{t-t_p}{2}S^+_0\right)(\mathbf{x})S^+_1(\mathbf{x}_p)\sim\lim\limits_{\epsilon\rightarrow0^+}-\frac{i}{2}\left[\frac{(\bar{z}-\bar{z}_p)^2}{{(\Delta\tilde{z}_p)}^2}K+\frac{(\bar{z}-\bar{z}_p)^2}{{(\Delta\tilde{z}_p)}}\text{ }\partial_{\bar{z}_p} S^+_1+\frac{\bar{z}-\bar{z}_p}{{(\Delta\tilde{z}_p)}}\text{ }2S^+_1\right](\mathbf{x}_p)\nonumber
\end{align}
Clearly, $S^+_1$ is not a $1+2$D CC primary field when $K\neq0$.

\medskip

We now construct the OPEs involving the $T$ field following similar procedures. We shall not assume that $T$ is a Lorentz quasi-primary; this fact will emerge automatically.

\medskip

Keeping in mind that $S^+_0$ has $h=\frac{3}{2}$ , $\left(P_{1,-1}S^+_0\right)\sim{{\bm{\xi}}}\cdot S^+_0\sim\partial_tS^+_0\sim0$ and the assumption that $S^+_0$ is a Lorentz quasi-primary implying $\left(L_1S^+_0\right)=0$ from \eqref{67}, from the general form \eqref{70} of a $T\Phi$ OPE, we see that the singularities of the $TS^+_0$ OPE are completely fixed since we also assumed that no local field has $h<0$ :
\begin{align}
T(t,z,\bar{z})S^+_0(t_p,z_p,\bar{z}_p)\sim\lim\limits_{\epsilon\rightarrow0^+}-i\left[\frac{\frac{3}{2}S^+_0}{(\Delta\tilde{z}_p)^2}+\frac{\partial_{z_p}S^+_0}{\Delta\tilde{z}_p}\right](t_p,z_p,\bar{z_p})
\end{align}
Thus, $S^+_0$ is a $1+2$D CC primary field.

\medskip

Since $T$ also has integer spin $m=2$ , directly using the bosonic exchange property and keeping in mind that $\bar{\partial}^2S^+_0\sim\partial_tS^+_0\sim0$ and $\frac{1}{2}\partial\bar{\partial}S^+_0\sim\partial_tT$ from \eqref{24}, we readily obtain the following $S^+_0T$ OPE:
\begin{align}
iS^+_0(t,z,\bar{z})T(t_p,z_p,\bar{z}_p)\sim\lim\limits_{\epsilon\rightarrow0^+}\left[(\bar{z}-\bar{z}_p)\left(\frac{\partial_{t_p}T}{(\Delta\tilde{z}_p)}+\frac{\frac{3}{2}\partial_{\bar{z}_p}S^+_0}{{(\Delta\tilde{z}_p)}^2}\right)+\frac{\frac{3}{2}S^+_0}{{(\Delta\tilde{z}_p)}^2}+\frac{\frac{1}{2}\partial_{z_p}S^+_0}{(\Delta\tilde{z}_p)}\right](\mathbf{x}_p)
\end{align}
Comparison with \eqref{64} reveals the following properties of the $T$ field:
\begin{align}
{{\bm{\xi}}}\cdot T\sim\frac{3}{2}\bar{\partial}S^+_0\hspace{5mm}\text{ ; }\hspace{5mm}\bar{{{\bm{\xi}}}}\cdot T\sim-\frac{1}{2}{\partial}S^+_0\hspace{5mm}\text{ ; }\hspace{5mm}\left(P_{0,0}T\right)\sim-\frac{3}{2}S^+_0
\end{align} 
The third property ensures that $T$ is not even a $1+2$D CC quasi-primary field while the first two tell us that the three local fields $\partial S^+_0$, $T$ and $\bar{\partial}S^+_0=P$ transform under the three-dimensional reducible but indecomposible representation \eqref{71} (with $l=2,a=\frac{3}{2},b=-\frac{1}{2}$) of the spin-boost sub-algebra \eqref{eq:7}.

\medskip

The consistent ansatz for the $TT$ OPE is now written following the general form \eqref{70} and remembering that $T$ has $(h,\bar{h})=(2,0)$ (as said before, we do not assume $T$ to be a Lorentz quasi-primary):
\begin{align}
&T(t,z,\bar{z})T(t_p,z_p,\bar{z}_p)\sim\lim\limits_{\epsilon\rightarrow0^+}-i\left[\sum\limits^2_{n=1}\frac{\left(L_{n}T\right)}{{(\Delta\tilde{z}_p)}^{n+2}}+\frac{2T}{(\Delta\tilde{z}_p)^2}+\frac{\partial_{z_p}T}{\Delta\tilde{z}_p}\right.\nonumber\\
&\hspace{71mm}\left.-\frac{t-t_p}{2}\left\{\frac{\partial_{t_p}T}{(\Delta\tilde{z}_p)^2}+\frac{3\partial_{\bar{z}_p}S^+_0}{{(\Delta\tilde{z}_p)}^3}\right\}\right](\mathbf{x}_p)
\end{align} 
Using the bosonic exchange property and the fact that the local field $\left(L_{n}T\right)$ has $(h,\bar{h})=(2-n,0)$ , we find that:
\begin{align*}
\left(L_{1}T\right)\sim0\hspace{5mm};\hspace{5mm}\left(L_{2}T\right)\sim \frac{c}{2}
\end{align*} 
with $c$ being a constant.

\medskip

Next, we find that the singular terms in the following ansatz for the $TS_1^+$ OPE, obeying \eqref{70}, is already completely fixed: 
\begin{align}
T(t,z,\bar{z})S^+_1(t_p,z_p,\bar{z}_p)\sim\lim\limits_{\epsilon\rightarrow0^+}-i\left[\frac{S^+_1}{(\Delta\tilde{z}_p)^2}+\frac{\partial_{z_p}S^+_1}{\Delta\tilde{z}_p}-\frac{t-t_p}{2}\text{ }\frac{\partial_{t_p} S^+_1}{(\Delta\tilde{z}_p)^2}\right](t_p,z_p,\bar{z_p})
\end{align}
where we have used the facts that $S^+_1$ has $h=1$ , that it is assumed to be a Lorentz quasi-primary i.e. $\left(L_1S^+_1\right)\sim0$ and that no local field has $h<0$. Using the bosonic exchange property, we immediately find the $S^+_1T$ OPE below:
\begin{align}
&S_1^+(t,z,\bar{z})T(t_p,z_p,\bar{z}_p)\sim\lim\limits_{\epsilon\rightarrow0^+}-\frac{i}{2}\left[(\bar{z}-\bar{z}_p)^2\left(\frac{\bar{\partial}^2_pS^+_1}{{(\Delta\tilde{z}_p)}^{2}}+\frac{\bar{\partial}_pT}{\Delta\tilde{z}_p}\right)+(\bar{z}-\bar{z}_p)\text{ }\frac{2\bar{\partial}_pS^+_1}{{(\Delta\tilde{z}_p)}^{2}}+\frac{2S^+_1}{{(\Delta\tilde{z}_p)}^{2}}\right.\nonumber\\
&\hspace{28mm}\left.+(t-t_p)\left\{(\bar{z}-\bar{z}_p)\left(\frac{\frac{3}{2}\bar{\partial}_pS^+_0}{{(\Delta\tilde{z}_p)}^2}+\frac{\partial_{t_p}T}{{(\Delta\tilde{z}_p)}}\right)+\frac{\frac{3}{2}S^+_0}{{(\Delta\tilde{z}_p)}^{2}}+\frac{\frac{1}{2}\partial_pS^+_0}{(\Delta\tilde{z}_p)}\right\}\right](\mathbf{x}_p)
\end{align} 
Comparing with \eqref{63}, we find that $T$ has $\bar{h}=0$ as expected and $\left(j^{(+)}_0T\right)\sim0$ which along with $\left(L_1T\right)\sim0$ establishes that $T$ is a Lorentz quasi-primary field.

\medskip

In the above discussion, we did not find any other local field that mixes with $S^+_1$ under Carrollian boost. So, along with the fact that $\bm{\xi}\cdot S^+_1\sim0$ , we conclude that $S^+_1$ indeed transforms under a spin-boost irrep.

\medskip

We also note that in the above OPEs, no local field with $\Delta<0$ appear; so, the assumption that $\Delta\geq0$ for all local fields would be equivalent to the assumption that $h\geq0$ for all local fields.

\medskip

The mutual OPEs of the six fields $T_e$, $j^a_e$ and $P_i$ are read off these OPEs to be:
\begin{align}
&P_i(\mathbf{x})P_j(\mathbf{x}_p)\sim0\\
&T_e(\mathbf{x})P_i(\mathbf{x}_p)\sim\lim\limits_{\epsilon\rightarrow0^+}-i\left[\frac{\frac{3}{2}P_i}{(\Delta\tilde{z}_p)^2}+\frac{\partial_{z_p}P_i}{\Delta\tilde{z}_p}\right](\mathbf{x}_p)\\
&T_e(\mathbf{x})T_e(\mathbf{x}_p)\sim\lim\limits_{\epsilon\rightarrow0^+}-i\left[\frac{\frac{c}{2}}{(\Delta\tilde{z}_p)^4}+\frac{2T_e}{(\Delta\tilde{z}_p)^2}+\frac{\partial_{z_p}T_e}{\Delta\tilde{z}_p}\right](\mathbf{x}_p)\label{72}\\
&j^a_e(\mathbf{x})P_j(\mathbf{x}_p)\sim\lim\limits_{\epsilon\rightarrow0^+}-\frac{i}{2}\left(\frac{a-1}{2}-j\right)\text{ }\frac{P_{j+a}(\mathbf{x}_p)}{\Delta\tilde{z}_p}\label{74}\\
&T_e(\mathbf{x})j^a_e(\mathbf{x}_p)\sim\lim\limits_{\epsilon\rightarrow0^+}-i\left[\frac{j^a_e}{(\Delta\tilde{z}_p)^2}+\frac{\partial_{z_p}j^a_e}{\Delta\tilde{z}_p}\right](\mathbf{x}_p)\\
&j^a_e(\mathbf{x})j^b_e(\mathbf{x}_p)\sim\lim\limits_{\epsilon\rightarrow0^+}-\frac{i}{2}\left[\frac{K g^{ab}}{(\Delta\tilde{z}_p)^2}+\frac{(a-b)j^{a+b}_e}{\Delta\tilde{z}_p}\right](\mathbf{x}_p)\label{73}
\end{align}
with the non-zero terms of the `metric' $g^{ab}$ being $g^{+-}=g^{-+}=1$ and $g^{00}=-\frac{1}{2}$.

\medskip

These OPEs look exactly same as some well-known holomorphic 2D CFT OPEs; e.g. \eqref{72} is the 2D CFT $TT$ OPE giving rise to the Virasoro algebra with central charge $c$ ; \eqref{73} says that in that 2D CFT, $2j^a_e$ would be the Kac-Moody currents generating a $\text{sl}(2,\mathbb{R})$ current algebra at level $K$ ; $P_i$ fields would be Kac-Moody primaries of dimension $h=\frac{3}{2}$ , transforming under the two-dimensional representation of $\text{sl}(2,\mathbb{R})$.

\medskip

But we are not studying 2D CFT. More precisely, the technology utilized in 2D CFT is not readily applicable here. So, we should explicitly derive the mode-algebra from these OPEs. Below we demonstrate one such calculation:
\begin{align*}
\left[j^a_n\text{ },\text{ }P_k(\mathbf{x}_p)\right]&=[\frac{2}{2\pi j}\oint\limits_{C_u^\prime}d\hat{z}\text{ }\hat{z}^{n}\text{ }j^a_e(t,\hat{z},\hat{\bar{z}})\text{ },\text{ }P_k(\mathbf{x}_p)]\hspace{15mm}[\text{definition \eqref{75}}]\\
&=\frac{2}{2\pi j}\oint\limits_{C_u}d\hat{z}\text{ }\hat{z}^{n}\text{ }\hat{\mathcal{T}} j^a_e(t_p^+,\hat{z},\hat{\bar{z}})P_k(t_p,\vec{x}_p)-0\hspace{15mm}[\text{contour subtraction: Figure 1}]\\
&=\lim\limits_{\epsilon\rightarrow0^+}-i\left(\frac{a-1}{2}-k\right)\frac{1}{2\pi j}\oint\limits_{C_u}d\hat{z}\text{ }\hat{z}^{n}\text{ }\frac{P_{k+a}(\mathbf{x}_p)}{\hat{z}-z_p-j\epsilon0^+}\hspace{15mm}[\text{from OPE \eqref{74}}]\\
&=-i\left(\frac{a-1}{2}-k\right)z_p^nP_{k+a}(\mathbf{x}_p)
\end{align*}
Using now the mode-expansion for $P_k$ , given by \eqref{76}, on both sides and comparing coefficients of individual powers, one obtains the following commutator:
\begin{align*}
i\left[j^a_n\text{ },\text{ }P_{m,k}\right]=\left(\frac{a-1}{2}-k\right)P_{m+n,k+a}
\end{align*}
Below we list the complete mode-algebra with $n,m\in\mathbb{Z}$ ; $j,k\in\{-1,0\}$ and $a,b\in\{0,\pm1\}$, derived in a similar way:
\begin{align}
&i\left[P_{n,j}\text{ },\text{ }P_{m,k}\right]=0&&\text{(Abelian super-translations)}\\
&i\left[L_n\text{ },\text{ }P_{m,j}\right]=\left(\frac{n-1}{2}-m\right)P_{m+n,j}\\
&i\left[L_n\text{ },\text{ }L_{m}\right]=\left(n-m\right)L_{m+n}+\frac{c}{12}\left(n^3-n\right)\delta_{n+m,0}&&\text{(Holomrphic Virasoro algebra)}\\
&i\left[j^a_n\text{ },\text{ }P_{m,j}\right]=\left(\frac{a-1}{2}-j\right)P_{m+n,j+a}\\
&i\left[L_n\text{ },\text{ }j^a_m\right]=-mj^a_{m+n}\\
&i\left[j^a_n\text{ },\text{ }j^b_m\right]=(a-b)j^{a+b}_{m+n}+2Kg^{ab}n\delta_{n+m,0}&&\text{( $\overline{\text{sl}(2,\mathbb{R})}$ current algebra)}
\end{align}
Thus, the Lorentz i.e. $\text{sl}(2,\mathbb{C})$ symmetry, generated by the global super-rotation generators, gets infinitely enhanced to a $\text{Vir}\ltimes\hat{\overline{\text{sl}{(2,\mathbb{R}})}}$ symmetry governed by the holomorphic super-rotation generators and the generators of the diffeomorphism      \eqref{61}. Besides, the global Poincare translations get embedded in the abelian sub-algebra of the $z^{n+1}$ and $z^{n+1}\bar{z}$ super-translation generators. This result, barring the Virasoro central charge $c$ and the Kac-Moody algebra level $K$, completely agrees with the leading and the sub-leading conformally soft graviton symmetry algebra in Celestial holography \cite{Banerjee:2020zlg,Guevara:2021abz,Banerjee:2021dlm,Banerjee:2022wht}. As usual, the central charge $c$ and the level $K$ can not be fixed using the general symmetry principles alone; some model-specific, dynamical input is necessary to determine them.

\medskip

Similarly, had $\{\bar{T},S^-_0,S^-_1\}$ been treated as local fields, the symmetry algebra arising from the (anti-holomorphic sector) OPEs would be a semi-direct product of the $\overline{\text{Vir}}\ltimes\hat{{\text{sl}{(2,\mathbb{R}})}}$ algebra and an abelian algebra of the $\bar{z}^{n+1}$ and $z\bar{z}^{n+1}$ super-translation generators.

\medskip

Thus, just as in the Celestial CFT \cite{Banerjee:2022wht}, the choice of the generator-fields as the local fields that can appear in the OPEs determines which symmetry is manifest at the level of the $1+2$D Carrollian conformal OPEs.

\medskip

\section{Conclusions}\label{s6}
In this work, we present a study of the general quantum symmetric aspects of the $1+2$D Carrollian CFT (on flat Carrollian background) by adopting a first-principle field-theoretic approach. Along the way, we report how these field theoretic results can be connected to the physics of mass-less scattering in the $1+3$D bulk AFS and the Celestial holography, thus taking a step towards the formulation of an AFS/CarrCFT holographic correspondence.  

\medskip

We started by deriving the position-space Ward identities of a source-less $1+2$D CarrCFT and went on to show how the super-translation and super-rotation memory effects \cite{Strominger:2014pwa,Pasterski:2015tva}, originating from the gravitational radiation in the $1+3$D bulk AFS, emerge in the $1+2$D Carrollian CFT upon some simple manipulations of these Ward identities, manifested by the presence of a temporal step-function \cite{Strominger:2017zoo}.

\medskip

We then performed the temporal Fourier transformation (from $t$ to $\omega$) of the position-space $(t,z,\bar{z})$ Carrollian Ward identities that describe the memory effects. When the $\omega$ of the generators of those Ward identities were taken close to zero, the $(n+1)$-point Carrollian correlators in the $(\omega,z,\bar{z})$ space factorized into a factor of universal form and the $n$-point Carrollian correlators without the generator. Following the identification of the $(\omega,z,\bar{z})$ space correlators of the Carrollian conformal primaries with $\Delta=1$ and transforming under a spin-boost irrep, with the bulk AFS null-momentum space $S$-matrices \cite{Donnay:2022wvx}, the universal factor was recognized to contain both the leading \cite{Weinberg:1965nx,He:2014laa} and the sub-leading \cite{Cachazo:2014fwa,Kapec:2014opa} soft graviton theorems of the bulk AFS. Importantly, the pole(s) of this soft-factor at $\omega=0$ originated from the temporal step-function that bears the imprint of the memory effects.

\medskip

The generators of these Ward identities, that were constructed completely out of the Carrollian EM tensor components but independently of any bulk AFS metric components, were then interpreted as the leading and the sub-leading soft graviton fields. Moreover, we also provided the purely Carrollian construction of the EM tensor of the 2D Celestial CFT \cite{Kapec:2016jld,Cheung:2016iub}. From these constructions, we inferred that the Celestial (anti-)holomorphic EM tensor and the (positive)negative-helicity (energetically) sub-leading soft graviton fields are the 2D shadow transformations of each other, instead of assuming this fact as in \cite{Fotopoulos:2019tpe,Fotopoulos:2019vac,Banerjee:2020kaa}.

\medskip

We then moved on to find the quantum conserved charge operators that generate the extended BMS$_4$ transformations \cite{Barnich:2009se,Barnich:2010ojg,Barnich:2010eb,Barnich:2011mi} on the space of the quantum fields. We could find the finite charges generating all the super-rotations and the holomorphic and the anti-holomorphic but the mixed super-translations. All of these finite quantum charges are different from the expected Carrollian conformal Noether charges. 

\medskip

The defining property of the temporal step-function allowed us to directly relate the (covariant) time-ordered OPEs with the corresponding operator commutation relations via complex contour integrals without any need to go through a 2D CFT-like radial quantization procedure, similarly as in $1+1$D CarrCFT \cite{Saha:2022gjw}. It also led to a $j\epsilon$-form of the Carrollian conformal Ward identities and the OPEs that helped us completely establish the definitions of the quantum conserved charges. As opposed to the $\theta$-prescription, the $j\epsilon$-prescription facilitates a direct use of the powerful algebraic properties of the OPEs, like associativity and commutativity.

\medskip

Meanwhile, following \cite{Banerjee:2020zlg,Polyakov:1987zb}, we observed that the Carrollian conformal Ward identity of the field $S^+_{1e}$ (extracted from $S^+_1$) which had been identified with the (bulk) positive-helicity sub-leading energetically soft graviton, can be recast into a form resembling the holomorphic $\overline{\text{sl}(2,\mathbb{R})}$ Kac-Moody Ward identities, by decomposing $S^+_{1e}$ into the three generating currents each with $h=1$ but different $\bar{h}$. The global transformations corresponding to this alleged $\overline{\text{sl}(2,\mathbb{R})}$ Kac-Moody symmetry was found to be the three anti-holomorphic Lorentz transformations. 

\medskip

We now summarize the actions of the (symmetry) generator fields in the $1+2$D CarrCFT:
\begin{itemize}
\item The $S^+_0$ field, identified with the positive-helicity leading soft graviton \cite{Banerjee:2018fgd,He:2014laa}, embeds two quantum fields $P_{-1}$ and $P_0$, both with $h=\frac{3}{2}$. Their modes respectively generate the $z^{a+1}$ and the $z^{a+1}\bar{z}$ super-translations that include the four Poincare translations (for $a=-1,0$). The field $S^-_0$ which is the 2D shadow transformation of $S^+_0$, consists of the modes generating the $\bar{z}^{a+1}$ and the $z\bar{z}^{a+1}$ super-translations.
\item The $T$ field consists of two parts: $\partial\bar{\partial}S^+_0$ and the field $T_e$ which is identified with the holomorphic EM tensor of the Celestial CFT \cite{Kapec:2016jld,Cheung:2016iub}. The modes of $T_e$ generate the holomorphic super-rotations. The modes of the similarly defined field $\bar{T}_e$, embedded into $\bar{T}$, generate the anti-holomorphic super-rotations as anticipated. The ($\bar{T}_e$) $T_e$ Ward identity resembles the (anti-)holomorphic Virasoro one. 
\item The $S^\pm_1$ field is identified with the conformally soft \cite{Donnay:2018neh,Puhm:2019zbl} $\pm$ve-helicity sub-leading soft graviton that consists of $S^\pm_0$ and $S^\pm_{1e}$ identified with the energetically sub-leading soft graviton. By construction, $S^\pm_1$ ($S^\pm_{1e}$) turned out to be the 2D shadow transformation of $\bar{T}$ or $T$ ($\bar{T}_e$ or $T_e$). $S^\mp_{1e}$ embeds three quantum currents ($\bar{j}^a_e$) $j^a_e$ whose Ward identities take the exactly same form as the (anti-)holomorphic $\text{sl}(2,\mathbb{R})$ Kac-Moody Ward identities.   
\end{itemize}

\medskip

But, before reading too much into the strong resemblance of the above Ward identities to the ones appearing in 2D relativistic CFTs, we need to remember that most of these Carrollian generator fields have $h$ and $\bar{h}$ both non-zero, unlike in 2D (anti-)holomorphic CFT. So, it would be more appropriate to extract the quantum symmetry algebra, i.e. the (charge-)algebra generated by the modes, from the (allowed) mutual OPEs of the six fields $S^\pm_0$, $S^\pm_1$, $T$ and $\bar{T}$. Recalling that    all the fields involved in an OPE must be local and that an OPE is associative, we argued that among the two sets $\{T,S^+_1,S^+_0\}$ and $\{S^-_1,\bar{T},S^-_0\}$, only one can be treated as a set of local fields for the purpose of forming consistent mutual OPEs.

\medskip

In this work, we chose to treat $T$, $S^+_1$ and $S^+_0$ as the local fields, i.e. we would analyze the holomorphic sector OPEs. Assuming that $S^+_0$ and $S^+_1$ are both Lorentz quasi-primary local fields, inspired by a similar assumption in Celestial holography \cite{Fotopoulos:2019tpe} and that no local field in the holomorphic sector possesses $h<0$, we were able to completely determine the pole-singularities of the mutual OPEs of $T$, $S^+_1$ and $S^+_0$ following the procedures of \cite{Zamolodchikov:1989mz,Saha:2022gjw}. Namely, we started from the general OPE structures of these three fields, read off from the corresponding Ward identities, and then appealed to only the bosonic exchange properties (commutativity) of these OPEs. 

\medskip

In particular, we did not require any hint from the physics of the bulk AFS mass-less scattering. Using only the general Carrollian symmetry arguments and the general algebraic properties of the OPEs was enough to fix the singular parts of these OPEs. While we did not say anything about the regular parts of the OPEs, it is easily noted, using the analogy with the usual 2D CFT \cite{Belavin:1984vu} through our discussion in section \ref{s4}, that they consist of the descendant local fields whose correlators can be completely determined from the correlators of the parent primary fields.

\medskip

Extracting the forms of the mutual OPEs of the six fields $T_e$, $j^a_e$ (or $S^+_{1e}$) for $a\in\{0,\pm1\}$ and $P_i$ for $i\in\{-1,0\}$ from the holomorphic sector OPEs, we could finally conclude using the $\text{OPE}\longleftrightarrow\text{Commutator}$ prescription developed in section \ref{s4}, that the corresponding modes indeed generate a $\text{Vir}\ltimes\hat{\overline{\text{sl}(2,\mathbb{R})}}$ symmetry algebra along with an abelian super-translation ideal, perfectly agreeing with the Celestial holographic conclusions \cite{Banerjee:2020zlg,Guevara:2021abz,Banerjee:2021dlm,Banerjee:2022wht}.

\medskip

Since we traced the $1+2$D CarrCFT roots of only the leading and sub-leading soft-graviton theorems \cite{Weinberg:1965nx,Cachazo:2014fwa}, an obvious next step is to find out how the whole $w_{1+\infty}$ tower of symmetries \cite{Strominger:2021lvk} can emerge from the Carrollian conformal physics. In Celestial holography, the $w_{1+\infty}$ symmetry follows completely from the leading, the sub-leading and the sub-subleading soft graviton theorems \cite{Guevara:2021abz,Banerjee:2021cly}. So, presumably we should first try to construct a Carrollian conformal field that would correspond to the sub-subleading soft graviton field. We hope to report on this in a very near future.

\medskip 

\acknowledgments
The author would like to thank Arjun Bagchi for helpful discussions at various stages of this work and also for his valuable comments on the initial version of the draft. A.S. is financially supported by the PMRF fellowship, MHRD, India.

\end{document}